\documentclass[useAMS,usenatbib]{mn2e}

\usepackage{amsmath}
\usepackage{amsfonts}
\usepackage{amssymb}
\usepackage[usenames,dvipsnames]{xcolor}
\usepackage{graphicx}
\usepackage{epstopdf}
\usepackage{epsfig}
\DeclareGraphicsExtensions{.pdf,.png,.jpg}
\graphicspath{{./Figures/}}
\usepackage{tabularx}
\usepackage{rotating}
\usepackage{lscape}
\usepackage{bold-extra}
\usepackage{appendix}
\usepackage{listings}
\usepackage{float}
\usepackage{pdfpages}
\usepackage{natbib}

\def\micron{$\mu$m }

\usepackage{url}
\usepackage{hyperref}
\usepackage{enumerate}
\usepackage{setspace}
\let\oldsqrt\sqrt
\def\sqrt{\mathpalette\DHLhksqrt}
\def\DHLhksqrt#1#2{%
\setbox0=\hbox{$#1\oldsqrt{#2\,}$}\dimen0=\ht0
\advance\dimen0-0.2\ht0
\setbox2=\hbox{\vrule height\ht0 depth -\dimen0}%
{\box0\lower0.4pt\box2}}

\title[Formation history of massive cluster galaxies]{The formation history of massive cluster galaxies as revealed by CARLA}
\author[E.A.Cooke et al.]{E. A. Cooke$^{1}$\thanks{e-mail: Elizabeth.Cooke@nottingham.ac.uk}, N. A. Hatch$^{1}$, A. Rettura$^{2,3}$,  D. Wylezalek$^{4}$, A. Galametz$^{5}$, \newauthor D. Stern$^{2}$,  M. Brodwin$^{6}$, S. I. Muldrew$^{7}$, O. Almaini$^{1}$, C. J. Conselice$^{1}$, \newauthor P. R. Eisenhardt$^{2}$, W. G. Hartley$^{8}$, M. Jarvis$^{9,10}$, N. Seymour$^{11}$, S. A. Stanford$^{12}$ \\
$^{1}$School of Physics and Astronomy, University of Nottingham, University Park, Nottingham NG7 2RD, UK\\
$^{2}$Jet Propulsion Laboratory, California Institute of Technology, MS 169-234, Pasadena, CA 91109, USA\\
$^{3}$Infrared Processing and Analysis Center, California Institute of Technology, MS 220-6, Pasadena, CA 91125, USA\\
$^{4}$Department of Physics and Astronomy, Johns Hopkins University, 3400 N. Charles St, Baltimore, MD 21218, USA\\
$^{5}$Max-Planck-Institut fuer Extraterrestrische Physik, Giessenbachstrasse, D-85748 Garching, Germany\\
$^{6}$UMKC Department of Physics and Astronomy, 257 Flarsheim Hall, 5110 Rockhill Road, Kansas City, MO 64110, USA\\
$^{7}$Department of Physics and Astronomy, University of Leicester, University Road, Leicester LE1 7RH, UK\\
$^{8}$ETH Zurich, Institute for Astronomy, Wolfgang-Pauli-Strasse 27, 8093 Zurich, Switzerland\\
$^{9}$Astrophysics, University of Oxford, Denys Wilkinson Building, Keble Road, Oxford OX1 3RH, UK\\
$^{10}$Physics Department, University of the Western Cape, Bellville, South Africa\\
$^{11}$International Centre for Radio Astronomy Research, Curtin University, Perth, Australia\\
$^{12}$Physics Department, One Shields Avenue, University of California, Davis, CA 95616, USA\\
}

\begin{document}

\date{Accepted 2015 June 23.  Received 2015 May 28; in original form 2015 March 17}

\pagerange{\pageref{firstpage}--\pageref{lastpage}} \pubyear{2015}

\maketitle

\label{firstpage}

\begin{abstract}
We use a sample of 37 of the densest clusters and protoclusters across $1.3 \le z \le 3.2$ from the Clusters Around Radio-Loud AGN (CARLA) survey to study the formation of massive cluster galaxies. We use optical $i'$-band and infrared 3.6\,\micron and 4.5\,\micron images to statistically select sources within these protoclusters and measure their median observed colours; $\langle i'-[3.6] \rangle$. We find the abundance of massive galaxies within the protoclusters increases with decreasing redshift, suggesting these objects may form an evolutionary sequence, with the lower redshift clusters in the sample having similar properties to the descendants of the  high redshift protoclusters. 
We find that the protocluster galaxies have an approximately unevolving observed-frame $i'-[3.6]$ colour across the examined redshift range. We compare the evolution of the  $\langle i'-[3.6] \rangle$ colour of massive cluster galaxies with simplistic galaxy formation models. 
Taking the full cluster population into account, we show that the formation of stars within the majority of massive cluster galaxies occurs over at least 2\,Gyr, and peaks at $z \sim 2$-3. From the median $i'-[3.6]$ colours we cannot determine the star formation histories of individual galaxies, but their star formation must have been rapidly terminated to produce the observed red colours. Finally, we show that massive galaxies at $z>2$ must have assembled within 0.5\,Gyr of them forming a significant fraction of their stars. This means that few massive galaxies in $z>2$ protoclusters could have formed via dry mergers. 

\end{abstract}


\section{Introduction}
In the local Universe, most massive cluster galaxies are old and have little-to-no ongoing star formation. 
They form a very homogenous, slowly-evolving population, exhibiting similar, red colours. 
When viewed in colour-magnitude space, these massive, old galaxies form a characteristic ``red sequence". Such red sequences of galaxies are nearly ubiquitous in low redshift clusters, and persist out to $z\sim1.5$ \citep[e.g.][]{Blakeslee2003,Holden2004,Mei2006,Eisenhardt2008}.  
Red sequences have commonly been used to examine the formation history of massive cluster galaxies. The colour, slope and low scatter of the red sequence within clusters are consistent with early-type cluster galaxies forming concurrently in a short burst of star formation at high redshift ($z>2$) and passively evolving thereafter \citep[e.g.][]{Bower1992,Eisenhardt2007}, although there are indications that further star formation occurs within galaxies towards the outskirts of the cluster \citep{Ferre-Mateu2014}. 

The red sequences of low redshift clusters indicate a high formation redshift, though it is difficult to determine the exact epoch and history of galaxy formation using their galaxy colours. The colours of galaxies that have been passively evolving for more than a few billion years are very similar \citep{Kauffmann2003}. Thus it is difficult to differentiate between formation redshifts if the 
time between the galaxy's formation and observation is several Gyr. 
Observing galaxies closer to their formation period makes it possible to measure galaxy ages more accurately. 
By measuring the colours of galaxies within high redshift clusters we can determine the exact epoch and history of formation and break the degeneracies between single collapse models and those which include extended periods of galaxy growth \citep[e.g.][]{Snyder2012}.

The scatter in colour of early-type cluster galaxies at $z>1$ is low and consistent with passive evolution for ${z\lesssim2.3}$ \citep[e.g.][]{Stanford1998,Blakeslee2003,Mei2006,Lidman2008}. However, when examining the full cluster population, it is no longer possible to model the galaxy formation history with a single formation timescale, but rather there is a scatter in the inferred ages \citep[e.g.][]{Eisenhardt2008,Kurk2009}. Clusters at $z>1$ were much more active than they are today; they exhibited significant ongoing star formation \citep{Snyder2012,Brodwin2013,Zeimann2013,Alberts2014}, merging between galaxies \citep{Mancone2010,Lotz2013} and increased AGN activity \citep{Galametz2010a,Martini2013}. The red sequence of clusters and protoclusters was much less populated at $z>1$ than today \citep[e.g.][]{Kodama2007,Hatch2011a,Rudnick2012}, which means some of the progenitors of local red sequence galaxies would have bluer colours and lie below the red sequence at $z>1$. Thus, when tracing the evolution of just the galaxies that already lie on the red sequence, high redshift studies are prone to progenitor bias \citep{vanDokkum2001}. To robustly trace the evolution of cluster galaxies it is important to study all the progenitors; those that are already passive at high redshift, and those that only become passive at a later time.

To trace the early formation history of massive cluster galaxies, we have undertaken a large survey of clusters, pushing the study of galaxy colours to even higher redshifts. Our sample extends from $z=1.3$ out to $z=3.2$, covering the full timescale of massive cluster galaxy formation measured from previous works \citep[e.g.][]{Blakeslee2003}. We have taken our cluster sample from the Clusters Around Radio-Loud AGN \citep[CARLA;][]{Wylezalek2013} survey. CARLA is a $400$-hour Warm \emph{Spitzer Space Telescope} programme designed to locate and investigate clusters of galaxies at $z>1.3$. Radio Loud Active Galactic Nuclei (RLAGN) preferentially reside in dense environments \citep[e.g.][ and references therein]{Hatch2014}, so the CARLA survey targeted the environment of 419 powerful RLAGN lying at $1.3<z<3.2$ and having a 500\,MHz luminosity $\ge10^{27.5}$\,W\,Hz$^{-1}$. \citet{Wylezalek2013} showed that 55\% of these RLAGN are surrounded by significant excesses of galaxies with red $[3.6]-[4.5]$ colours that are likely associated with the RLAGN and therefore are likely to be high redshift clusters and protoclusters.

The luminosity functions of the CARLA cluster galaxies were investigated in \citet{Wylezalek2014}, showing that 
passive galaxy evolution models were consistent with their measured $m_*$ up to $z\sim3.2$. In this paper we investigate the formation epoch and star formation history of the cluster galaxies. We observe $37$ of the densest CARLA fields in the $i'$ band and calculate the average observed $i'-[3.6]$ colour of the cluster galaxies. We avoid progenitor bias by measuring the colours of all $M_*>10^{10.5}$\,M$_{\odot}$ cluster galaxies to estimate the overall average colour evolution of massive cluster galaxies.  Although using average colours loses information about the individual cluster galaxies, if they are formed as concurrent bursts, then averaging will reduce photometric errors. If the galaxy population is more diverse, then by modelling that diversity we can compare to the average colour, which still contains information about the population. The scatter in colours can also be used to reduce the effect of degeneracies between models.

In Section \ref{sec:method} we describe our data and control fields, as well as the methodology of measuring cluster galaxy properties. Section \ref{sec:results} presents our key results, the implications of which are discussed in Section \ref{sec:discussion}. Our conclusions are summarised in Section \ref{sec:conclusions}. 
We assume a $\Lambda$CDM cosmology with $h = 0.70$, $\Omega_M = 0.3$ and $\Omega_\Lambda = 0.7$ throughout, unless stated otherwise. 
Magnitudes are expressed in the AB system.

\begin{table*}
\begin{centering}
\begin{tabular}{ l c r c c c c c c }
\hline
Field & RA & \multicolumn{1}{c}{DEC} & z & Instrument & Exp. time (s) & $5 \sigma$ $i'$ depth & Seeing (arcsec) & Density$^{a}$ \\ \hline \hline
J085442.00$+$57572 	& 08:54:42.00 & 57:57:29.16 	& 1.317 & ACAM   & 6600  & 24.59 & 1.00 & 4.1 \\
J110020.21$+$09493 	& 11:00:20.16 & 09:49:35.00 	& 1.321 & GMOS-S & 2405  & 24.86 & 0.65 & 3.3 \\
J135817.60$+$57520 	& 13:58:17.52 & 57:52:04.08 	& 1.373 & ACAM   & 8400  & 24.95 & 0.89 & 6.2 \\
MRC0955$-$288 		& 09:58:04.80 & $-$29:04:07.32 	& 1.400 & GMOS-S & 2645  & 25.02 & 0.58 & 2.7 \\
7C1756$+$6520 		& 17:57:05.28 & 65:19:51.60 	& 1.416 & ACAM   & 10205 & 24.26 & 1.73 & 4.4 \\
6CE1100$+$3505 		& 11:03:26.40 & 34:49:48.00 	& 1.440 & ACAM   & 9600  & 25.19 & 1.12 & 6.3 \\
TXS 2353$-$003 		& 23:55:35.87 & $-$00:02:48.00 	& 1.490 & ACAM   & 6000  & 24.99 & 0.81 & 5.6 \\
J112914.10$+$09515 	& 11:29:14.16 & 09:51:59.00 	& 1.519 & GMOS-S & 2645  & 24.78 & 0.44 & 6.3 \\
MG0122$+$1923 		& 01:22:30.00 & 19:23:38.40 	& 1.595 & ACAM   & 7200  & 24.83 & 0.79 & 3.2 \\
7C1753$+$6311 		& 17:53:35.28 & 63:10:49.08 	& 1.600 & ACAM   & 6000  & 25.08 & 0.74 & 4.5 \\
BRL1422$-$297 		& 14:25:29.28 & $-$29:59:56.04 	& 1.632 & GMOS-S & 2689  & 24.87 & 0.62 & 3.5 \\
J105231.82$+$08060 	& 10:52:31.92 & 08:06:07.99 	& 1.643 & GMOS-S & 2645  & 25.04 & 0.56 & 5.0 \\
TNJ0941$-$1628 		& 09:41:07.44 & $-$16:28:03.00 	& 1.644 & GMOS-S & 2645  & 25.03 & 0.58 & 4.7 \\
6CE1141$+$3525 		& 11:43:51.12 & 35:08:24.00 	& 1.780 & ACAM   & 10187 & 24.89 & 1.59 & 4.5 \\
6CE0905$+$3955 		& 09:08:16.80 & 39:43:26.04 	& 1.883 & ACAM   & 7200  & 25.09 & 0.89 & 4.1 \\
MRC1217$-$276 		& 12:20:21.12 & $-$27:53:00.96 	& 1.899 & GMOS-S & 2645  & 24.96 & 0.69 & 3.5 \\
J101827.85$+$05303 	& 10:18:27.84 & 05:30:29.99 	& 1.938 & ACAM   & 7200  & 25.19 & 0.81 & 5.0 \\
J213638.58$+$00415 	& 21:36:38.64 & 00:41:53.99 	& 1.941 & ACAM   & 6300  & 25.00 & 0.69 &  5.7\\
J115043.87$-$00235 	& 11:50:43.92 & $-$00:23:54.00 	& 1.976 & GMOS-S & 2645  & 25.04 & 0.62 & 2.7 \\
J080016.10$+$402955.6 	& 08:00:16.08 & 40:29:56.40 	& 2.021 & ACAM   & 6600  & 25.16 & 0.93 & 6.3 \\
J132720.98$+$432627.9 	& 13:27:20.88 & 43:26:27.96 	& 2.084 & ACAM   & 7200  & 25.03 & 0.75 & 4.1 \\
J115201.12$+$102322.8 	& 11:52:01.20 & 10:23:22.92 	& 2.089 & ACAM   & 7200  & 25.18 & 1.13 & 4.2 \\
TNR 2254$+$1857 	& 22:54:53.76 & 18:57:03.60 	& 2.154 & ACAM   & 9000  & 25.41 & 1.10 & 5.6 \\
J112338.14$+$052038.5 	& 11:23:38.16 & 05:20:38.54 	& 2.181 & GMOS-S & 2645  & 25.00 & 0.64 & 4.7 \\
J141906.82$+$055501.9 	& 14:19:06.72 & 05:55:01.92 	& 2.293 & ACAM   & 7800  & 24.56 & 1.28 & 4.5 \\
J095033.62$+$274329.9 	& 09:50:33.60 & 27:43:30.00 	& 2.356 & ACAM   & 7200  & 24.92 & 1.31 & 4.4 \\
4C 40.02 		& 00:30:49.00 & 41:10:48.00 	& 2.428 & ACAM   & 8700  & 25.26 & 0.94 & 4.2 \\
J140445.88$-$013021.8 	& 14:04:45.84 & $-$01:30:21.88 	& 2.499 & GMOS-S$^{*}$ & 2640 & 25.08 & 0.38 & 2.7 \\
J110344.53$+$023209.9 	& 11:03:44.64 & 02:32:09.92 	& 2.514 & ACAM   & 7800  & 24.76 & 1.33 & 4.1 \\
TXS 1558$-$003 		& 16:01:17.28 & $-$00:28:46.00 	& 2.520 & GMOS-S & 2405  & 24.94 & 0.91 & 2.4 \\
J102429.58$-$005255.4 	& 10:24:29.52 & $-$00:52:55.43 	& 2.555 & GMOS-S & 2645  & 25.14 & 0.53 & 4.5 \\
J140653.84$+$343337.3 	& 14:06:53.76 & 34:33:37.44 	& 2.566 & ACAM   & 7800  & 25.19 & 0.73 & 3.3 \\
6CSS0824$+$5344 	& 08:27:59.04 & 53:34:14.88 	& 2.824 & ACAM   & 7200  & 24.65 & 1.17 & 4.1 \\
J140432.99$+$072846.9 	& 14:04:32.88 & 07:28:46.96 	& 2.864 & ACAM   & 8400  & 25.06 & 1.02 & 4.1 \\
B2 1132$+$37 		& 11:35:06.00 & 37:08:40.92 	& 2.880 & ACAM   & 7200  & 25.19 & 0.98 & 5.9 \\
MRC 0943$-$242 		& 09:45:32.88 & $-$24:28:50.16 	& 2.922 & GMOS-S & 2645  & 24.83 & 0.44 & 4.7 \\ 
NVSS J095751$-$213321 	& 09:57:51.36 & $-$21:33:20.88 	& 3.126 & GMOS-S & 2645  & 25.01 & 0.61 & 2.7 \\ \hline
\end{tabular}
\caption{Details of $i'$ band observations for the 37 CARLA fields observed with ACAM on the WHT and GMOS-S on Gemini South. \newline
$^{a}$ Density is the number of standard deviations from the average field density of red IRAC sources, as calculated in \citet{Wylezalek2014}. \newline
$^{*}i'$ imaging of J140445.88$-$013021.8 was taken with the new Hamamatsu CCDs on GMOS-S, with reduced fringing effects. }
\label{table:iimages}
\end{centering}
\end{table*}

\section{Method} \label{sec:method}
In this Section we describe our methodology and datasets. We have two cluster samples: a high redshift ($1.3<z<3.2$) sample from the CARLA survey (Section \ref{sec:CARLA}), and a low and intermediate redshift sample ($0.1<z<1.8$) taken from the IRAC Shallow Cluster Survey (ISCS, described in Section \ref{sec:ISCS}). We also utilise a control field sample, from the UKIDSS Ultra Deep Survey (Section \ref{sec:controlfield}), in order to statistically subtract field contaminants from our cluster samples. Our selection of high redshift galaxies and cleaning of foreground interlopers is described in Section \ref{sec:sampleselection}. Our method in calculating the colours of (proto)cluster galaxies is then described in Section \ref{sec:measuring_colours}. 

\subsection{High redshift cluster sample: CARLA} \label{sec:CARLA}
\subsubsection{Data}
Infrared (IR) data at 3.6\,\micron and 4.5\,\micron for the CARLA sample were taken during Cycles 7 and 8 using the \emph{Spitzer} Infrared Array Camera \citep[IRAC;][]{Fazio2004}.  The imaged fields are $5.2 \times 5.2$\,arcmin$^2$, with the 4.5\,\micron data being slightly deeper; 95\% completeness is reached at magnitudes of $[3.6]=22.6$ and $[4.5]=22.9$. For a full description of the IR observations and data reduction see \citet{Wylezalek2013}. 

We complement our existing \emph{Spitzer} dataset with $i'$ band imaging. 
The $i'$ and $3.6$\,\micron bands bracket the $4000$\,{\AA} break at the redshifts covered by the CARLA survey, and allow direct comparison to previous work at lower redshifts \citep{Eisenhardt2008}. The extra $i'$ band also allows us to refine the selection of cluster member galaxies, providing more detail in order to study the evolution of clusters more thoroughly than before. 

Optical $i'$ band data were obtained for 37 of the densest CARLA fields with the auxiliary-port camera (ACAM) on the $4.2$\,m William Herschel Telescope (WHT) in La Palma and the Gemini Multi-Object Spectrograph South instrument \citep[GMOS-S;][]{Hook2004} on Gemini-South in Chile. 
These fields were selected as the CARLA targets which contained the highest densities of sources with red IRAC $[3.6]-[4.5]$ colours 
that were visible at the latitudes of each telescope. 
Densities of most of these fields are $4$-$6\sigma$ denser than an average blank field (see Table \ref{table:iimages}), as calculated in \citet{Wylezalek2014}, with a few at $2$-$3\sigma$ due to higher density fields not being observable. 
Fields with bright stars ($m_{i'}\le10$) in the field of view were rejected to avoid saturation and bleeding in the images. Fields with known low-redshift clusters within the field of view were also rejected to avoid biasing our measurements of overdensities. 

Twenty-three fields were imaged with ACAM during the period 2013 September-2014 December. The field of view of ACAM is circular, with a diameter of 8.3\,arcmin and pixel scale 0.25\,arcsec\,pixel$^{-1}$.  

The remaining 14 fields were imaged with GMOS-S. The majority of the GMOS-S data was taken using the EEV detectors between February and April 2014, though one field (J140445$-$013021.8) was imaged in June 2014 with the new  Hamamatsu CCDs. GMOS-S covers an area of $5.5\times5.5$\,arcmin$^2$ with a pixel scale of 0.146\,arcsec\,pixel$^{-1}$. Exposure times were adapted to take varying seeing into account, in order to obtain a consistent depth across all fields (see Table \ref{table:iimages}).

\subsubsection{Data reduction} 
The $i'$ band images were reduced using the publicly available {\sc{THELI}} software \citep{Erben2005,Schirmer2013}. 
The data were de-biased and flatfielded using a superflat created from median-combining all images taken on the same night. A fringing model was created and subtracted and the sky background was subtracted. 
The GMOS-S data taken with the old EEV detectors exhibited bad fringing (typically $\sim67\%$ of the background), whereas the data taken with the Hamamatsu CCDs has significantly reduced fringing effects ($<1\%$ of the background). 

Within {\sc{THELI}}, astrometric solutions for the images were derived using {\sc{SCAMP}} \citep{Bertin2006} with catalogues from the Sloan Digital Sky Survey (SDSS), or the Two Micron All Sky Survey (2MASS) where fields were not covered by the SDSS. Each field was mean-combined using {\sc{SWarp}} \citep{Bertin2002} within the  {\sc{THELI}} software. 
The reduced images were flux calibrated by comparing unsaturated stars in each field to the SDSS where possible. Two WHT fields were not covered by SDSS and were flux calibrated with unsaturated stars in SDSS-covered exposures taken immediately before and after the observations. The GMOS-S fields not covered by SDSS were flux calibrated using standard star observations. We compared the different calibration methods for fields with both SDSS coverage and standard star observations and find that calibrations with standard stars differ from calibrations with SDSS by $\le0.05$\,mag. Details of the $i'$ band data for each of the 37 CARLA fields are given in Table \ref{table:iimages}.

\subsubsection{Source extraction} 
Source extraction was performed using SExtractor \citep{Bertin1996} in dual image mode, using the 4.5\,\micron images for detection and performing photometry on the 3.6\,\micron images. IRAC fluxes were measured in $4$\,arcsec apertures and corrected to total fluxes using aperture corrections of 1.42 and 1.45 for 3.6\,\micron and 4.5\,\micron respectively \citep{Wylezalek2013}. 
At the RA and DEC of each source detected at 4.5\,$\mu$m, $i'$-band fluxes were measured using the IDL APER routine. These fluxes were measured in either 2.5\,arcsec or 3.2\,arcsec diameter apertures, depending on the full-width half-maximum (FWHM) of the image. The aperture sizes were chosen to be $\sim2.5\times$ the seeing and were a compromise between including as much flux as possible, and avoiding blending. 
The larger aperture was used for images with seeing $> 1.15$\,arcsec (see Table \ref{table:iimages}). These $i'$ fluxes were then corrected to total flux using correction factors (typically 1.15 and 1.04 for ACAM and GMOS-S data respectively) measured from the growth curves of unsaturated stars in the images. 

Image depths, shown in Table \ref{table:iimages}, were calculated by measuring the flux in $\sim100,000$ random apertures, with the aperture size dependent on the FWHM of each image (as above). The median $1\sigma$ depth of the WHT data is $i' = 26.79$\,mag, and for the Gemini data is $i'=26.75$\,mag. Due to the similarity in depth of all the fields, the overall median depth of $i'=26.76$\,mag ($=1\sigma_{med}$) was used for all fields. 

Colours were calculated from aperture-corrected $i' -$ aperture-corrected $3.6$\,\micron magnitudes. 
The colours derived from the control field (see Section \ref{sec:controlfield}) were calculated in exactly the same way as the (proto)cluster sample. The distribution of colours of all galaxies in the (proto)cluster fields are consistent with the control field so there are no systematic errors in our colour measurements.

\subsection{Control field: UDS and SpUDS} \label{sec:controlfield}
We utilise the UKIDSS Ultra Deep Survey (UDS; PI O. Almaini) to statistically subtract contamination from fore- and background galaxies in the cluster fields. The UDS is a deep 0.8\,deg$^2$ near-infrared survey, which overlaps part of the Subaru/\emph{XMM-Newton} Deep Survey \citep[SXDS;][]{Furusawa2008}. 

Galaxies in the UDS have photometric redshift information derived in \citet{Hartley2013} using 11 photometric bands. The $K$-band selected catalogue incorporates $U$-band data from the Canada-France-Hawaii Telescope (Foucaud et al., in preparation), optical $BVRi'z'$ photometry from the SXDS, $JHK$ photometry from the 8th data release (DR8) of the UDS, and  \emph{Spitzer} Ultra Deep Survey 3.6 and 4.5\,\micron data (SpUDS; PI J. Dunlop). 
Photometric redshifts were determined 
by fitting spectral energy distribution (SED) templates to the photometric data points. The resulting dispersion is $\Delta z$/($1 + z$)$ = 0.031$.

The SpUDS survey is a 1\,deg$^2$ Cycle 4 \emph{Spitzer} Legacy program that encompasses the UDS field. We use the SpUDS $3.6$ and $4.5$\,\micron catalogues of \citet{Wylezalek2013}, which were extracted from the public mosaics in the same way as for the CARLA survey. Catalogues were created using SExtractor in dual-image mode, using the $4.5$\,\micron image as the detection image. The SpUDS data reach 3$\sigma$ depths of $\sim24$\,mag at both 3.6 and 4.5\,$\mu$m, but here we use only sources down to the shallower depth of the CARLA data. 

The $i'$ band data covering the 0.8\,deg$^2$ UDS was obtained as part of the SXDS and resampled and registered onto the UDS $K$-band image \citep[see][]{Cirasuolo2010}. 
The $5\sigma$ limiting depth is $i'=27.2$\,mag, but we use only sources down to the shallower depth of the CARLA fields. Any source with $i'$ magnitude fainter than the median depth is limited to $1\sigma_{med}$. Fluxes were measured in the same way as for the CARLA sample, using positions from the $4.5$\,\micron catalogue and measuring fluxes using the IDL APER routine with $2.5$\,arcsec diameter apertures.

\subsection{Selection of high redshift galaxies} \label{sec:sampleselection}
The cluster membership of the galaxies in the CARLA fields is not yet known, therefore we determine the average $i'-[3.6]$ colours of the cluster members by statistically subtracting the back- and foreground galaxies. Statistical subtraction is most accurate when the cluster members are the dominant population in the sample; however, for the high redshift CARLA clusters, the number of interloping galaxies outweighs the number of cluster members. We therefore make a spatial cut and two colour cuts to pre-select galaxies that are most likely to be cluster members.

\subsubsection{Spatial selection}
To select the highest fraction of (proto)cluster galaxies to interlopers, we only consider sources within 1\,arcmin of the central RLAGN for each CARLA field. \citet{Wylezalek2013} showed that the galaxy density was highest within this region. At these redshifts, 1\,arcmin is $\sim500$\,kpc in physical coordinates. 
In co-moving coordinates this corresponds to $\sim 1.8$\,Mpc radius at $z\sim3$ and $\sim1.2$\,Mpc radius at $z=1.3$. 
This decreasing co-moving radius with time traces the expected collapse of the cluster. Although protoclusters always collapse in the co-moving reference frame, in physical and angular coordinates the size of the protocluster remains approximately constant across $1.3<z<3.2$ as gravity is almost balanced by the Hubble expansion \citep[for a full explanation see][]{Muldrew2015}. Whilst our 1\,arcmin aperture can only capture a fraction of these protoclusters, it encloses approximately the same fraction at all redshifts between $z=1.3$ and $z=3.2$. 
Assuming the CARLA clusters all have approximately the same $z=0$ mass (see Section \ref{sec:overdensity}), this means that we expect to select the same fraction of the (proto)clusters at each redshift. 

\subsubsection{Colour cuts at $z>1.3$}
\begin{figure}
\centering
\includegraphics[scale=0.5]{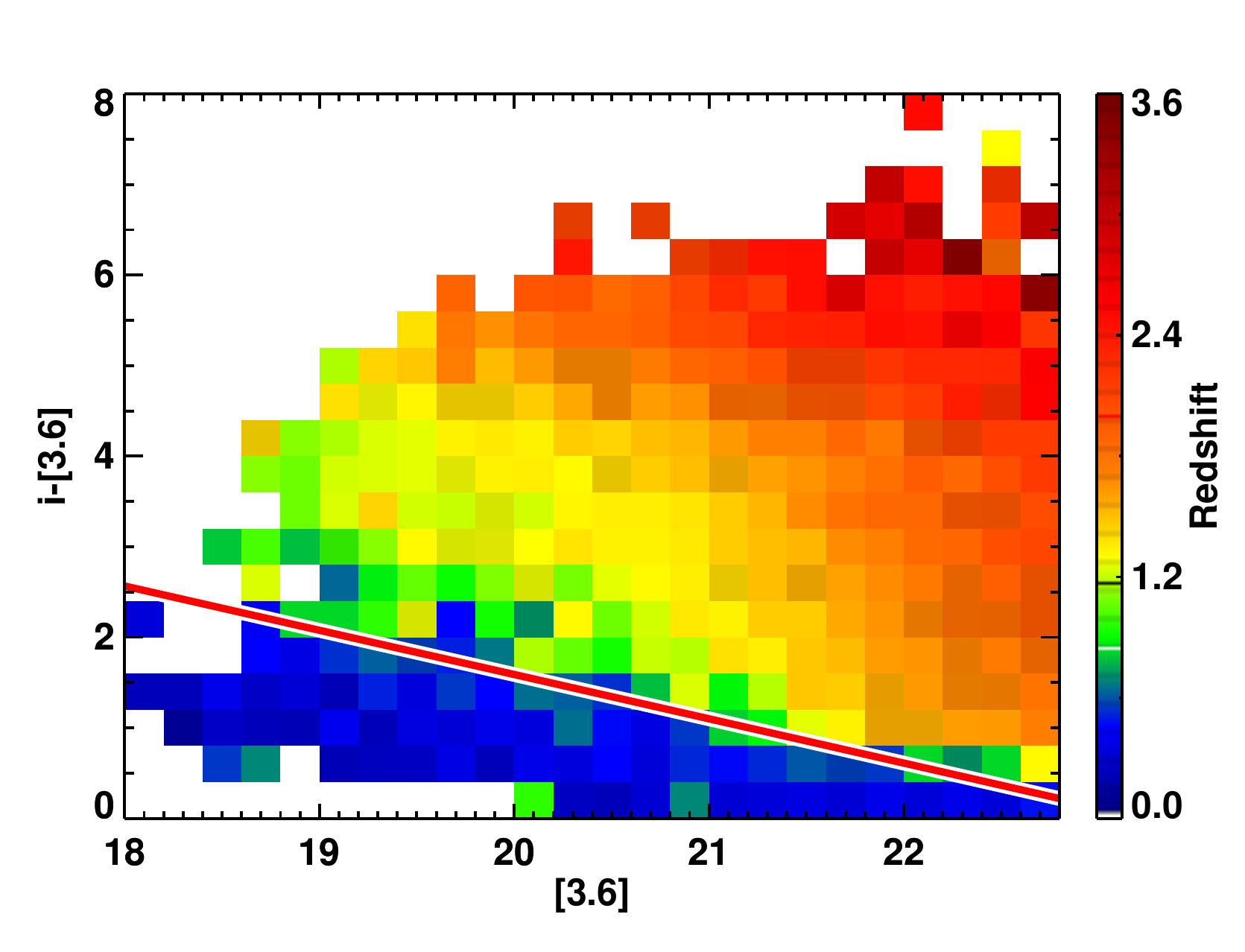}
\caption{Distribution of UDS sources by redshift in colour-magnitude space. Each pixel's colour represents the mean redshift of all the sources within that pixel. We employ a colour-magnitude cut, shown by the red line, to remove bright foreground contaminants, whilst retaining 99\% of our IRAC-selected sources. }
\label{fig:lowz}
\end{figure}

Without spectroscopic measurements we cannot ascertain true cluster membership; however, with the available photometry we can remove low redshift foreground contaminants. 
The \emph{Spitzer} IRAC colour cut $[3.6]-[4.5]>-0.1$ \citep{Papovich2008} was used to select sources at $1.3 < z < 3.2$. Hereafter we refer to these mid-infrared colour-selected sources as ``IRAC-selected sources". 
This colour cut effectively selects galaxies at $z>1.3$, with only $10$-$20\%$ contamination from foreground sources \citep{Muzzin2013}. Potential contaminants include strongly star-forming galaxies at $0.2<z<0.5$ and powerful AGN at all redshifts. To remove these bright interlopers, as well as other bright foreground sources, we apply a further cut of $i'-[3.6] > -0.5 \times [3.6]+11.4$, shown in 
Figure \ref{fig:lowz} by the red line. This line was derived from UDS data, using the photometric redshift information to determine where foreground contaminants are most likely to lie in colour-magnitude space. From the UDS, contours of the probability of a source lying at $z<1.3$ were derived. This cut is a linear fit to the contour corresponding to a $\ge 80\%$ likelihood of a source lying at $z<1.3$.  
This cut removes the brightest foreground contaminants while retaining 99\% of the IRAC-selected sources, likely to lie at $z>1.3$. 
Throughout the rest of this paper we refer to the IRAC-selected sources which have colours above these cuts as our ``high redshift sample". 
After applying these colour cuts, the CARLA fields are 1.5-2 times the density of the average field. Given the fact that we are observing a very deep cylinder (from $z\sim1.3$), and a typical protocluster is at most $\sim40$\,Mpc (co-moving diameter) deep \citep{Muldrew2015}, an overdensity level of 2 times the field is quite extreme, meaning these structures are highly likely to be forming clusters. 

\begin{figure*}
\centering
\includegraphics[scale=0.5]{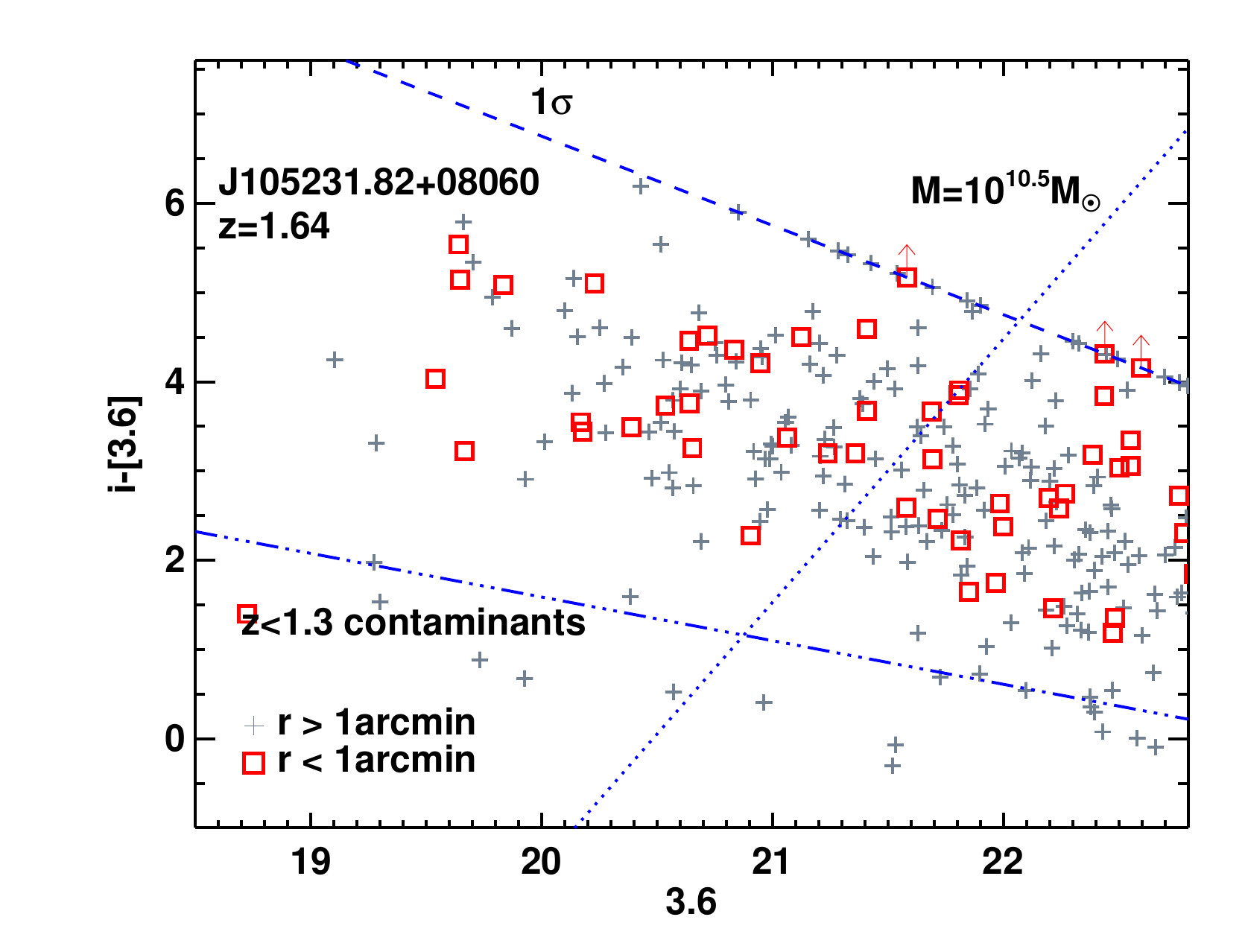} 
\includegraphics[width=0.4\textwidth]{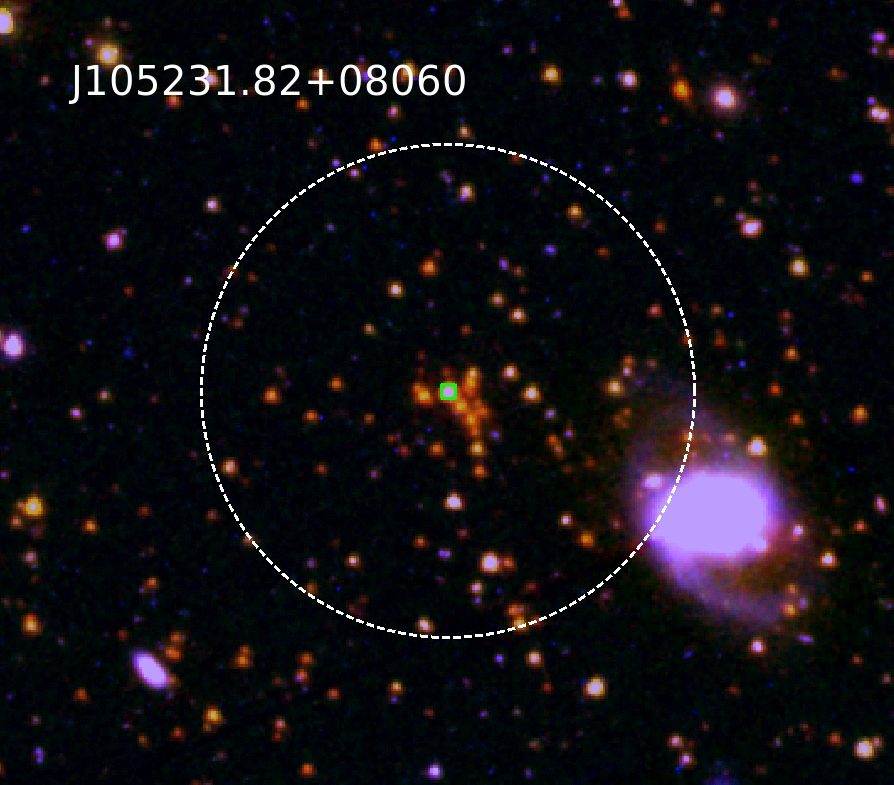}
\includegraphics[scale=0.5]{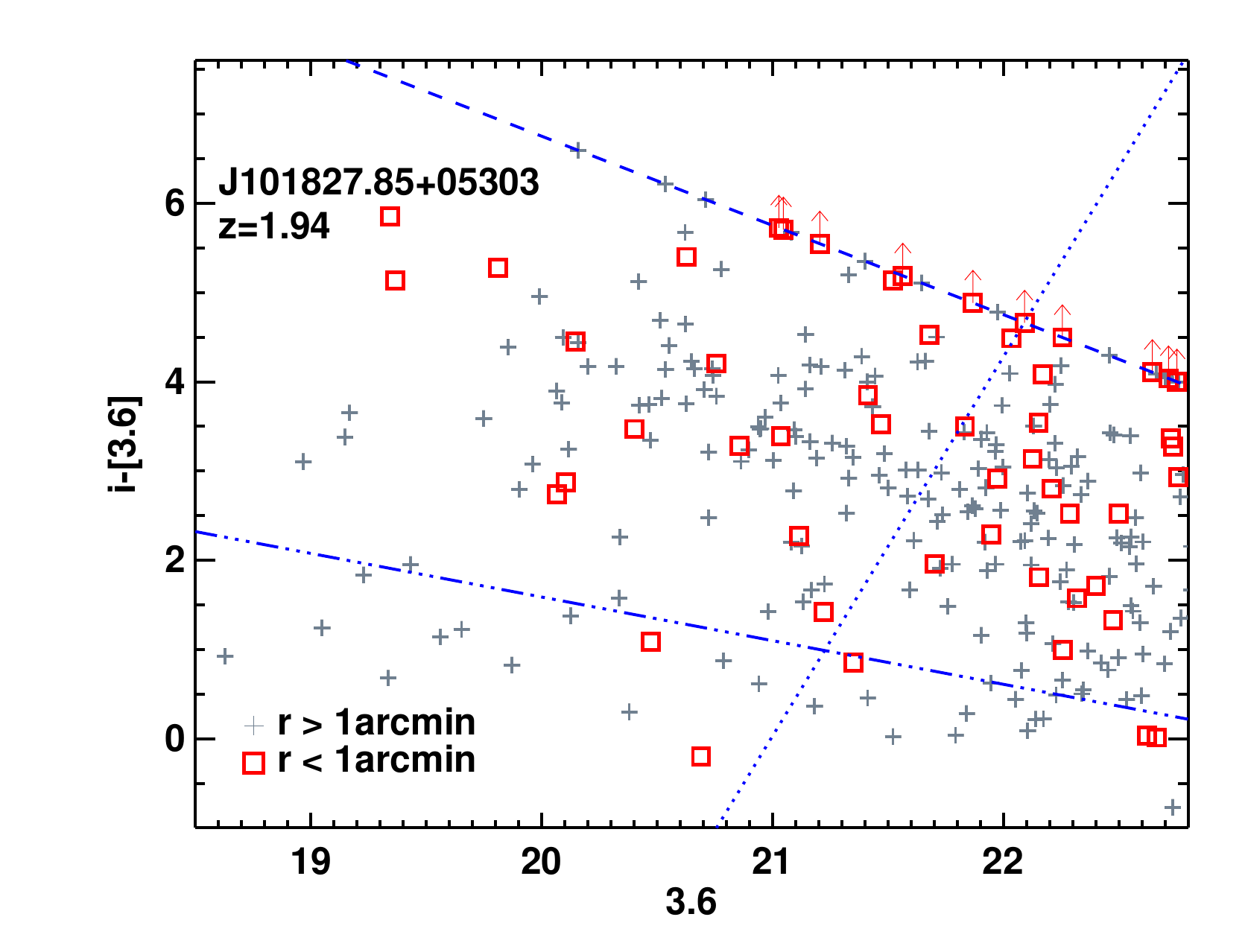} 
\includegraphics[width=0.4\textwidth]{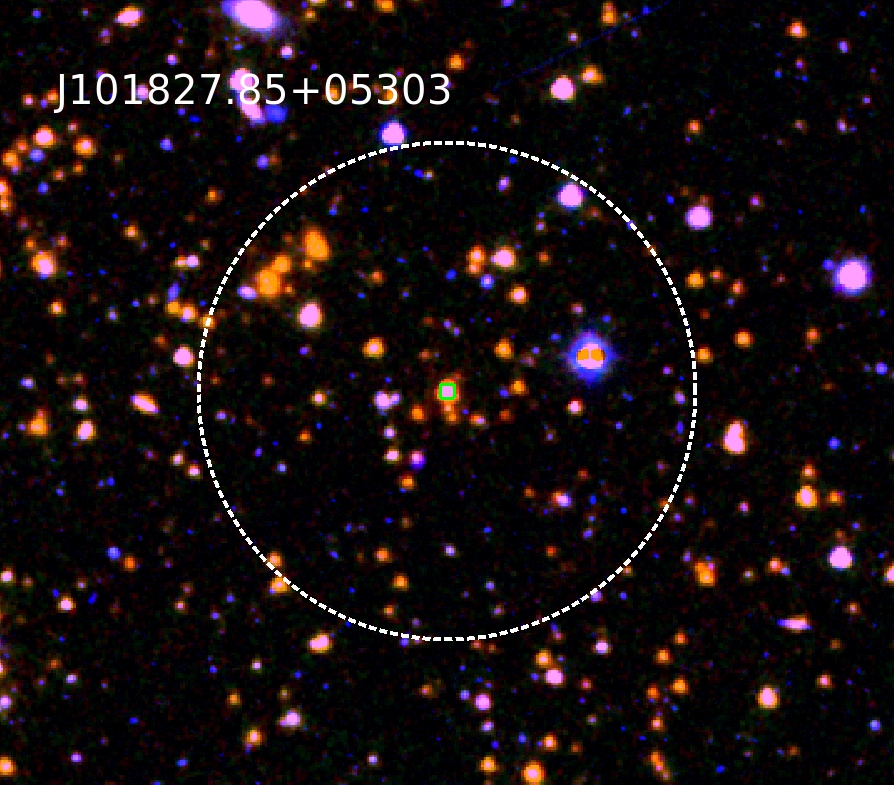}
\caption{\emph{Left:} $i'-[3.6]$ colour-magnitude diagrams for J105231.82$+$08060 at $z=1.64$ and J101827.85$+$05303 at $z=1.94$, two of the 37 CARLA fields. J105231.82$+$08060 appears to show a dense core of red galaxies, perhaps indicating that the red sequence is being populated, whereas the red sources in J101827.85$+$05303 are more spread out across the field. Red squares show sources within 1\,arcmin of the RLAGN. Grey plus symbols show those sources lying further than 1\,arcmin from the RLAGN, which are more likely to be contaminants. The blue dashed lines show our $1\sigma$ $i'$ median depth. Sources fainter than the $1\sigma$ median depth are set to $1\sigma_{med}$ and shown as lower limits. The dotted blue lines show a $M_*>10^{10.5}$\,M$_{\odot}$ mass cut and the dash-triple dotted blue lines show the cut used to remove low redshift contaminants. The RLAGN in these fields are too bright to fit on this scale and are not shown. No sources have been removed through statistical subtraction here. The colour-magnitude diagrams of the other 35 CARLA targets are shown in Appendix \ref{Append:CMDs}. \emph{Right:} $i'$,$[3.6]$,$[4.5]$ three-colour images of J105231.82$+$08060 and J101827.85$+$05303, showing the 1\,arcmin radius apertures, within which cluster galaxies are selected. The central RLAGN is marked with a green square.}
\label{fig:3colimages}
\end{figure*}

\subsubsection{Mass cuts} \label{sec:masscuts}
As the CARLA survey covers a wide redshift range, from 2\,Gyr to 7\,Gyr after the Big Bang, we are likely to detect more low-mass galaxies at low redshifts than at high redshifts. This may bias our results by giving a bluer average colour at low redshift simply because we can probe further down the mass function than at high redshift. In order to better compare these clusters across redshift, we select galaxies with stellar masses of $M_*>10^{10.5}$\,M$_{\odot}$ at each redshift\footnote{Note that taking a $10^{10.5}$\,M$_{\odot}$ mass cut does not fully account for progenitor bias \citep{Mundy2015}. The lower mass progenitors of $M_*=10^{10.5}$\,M$_{\odot}$ galaxies at low redshift will not be selected by our mass cut at high redshift. This means that there will be additional galaxies that enter the sample at low redshift that are not detected at high redshift. These low mass objects will typically have bluer colours, which may cause our measured colours to be progressively bluer at lower redshifts. An evolving mass cut, or a constant number density selection would provide a more accurate measure of the evolution of the $\langle i'-[3.6]\rangle$ colour, however these cuts are dependent on the galaxy evolution model adopted and are beyond the capabilities of the current data.}. 
We approximate stellar mass using the $[3.6]$ magnitude and $i'-[3.6]$ colour as a mass proxy. For redshifts $0.1<z<3.2$ in steps of $\Delta z=0.1$, a line in the $(i'-[3.6])$-$[3.6]$ plane was determined for a $10^{10.5}$\,M$_{\odot}$ galaxy using \citet{BC03} models. We used stellar population models with exponentially declining star formation models following ${\rm SFR}\propto e^{-t/ \tau }$ with $\tau $ of 0.01, 0.1, 0.5, 1 and 10\,Gyr. For all models we assume solar metallicity and that the stars are formed with the initial mass function of Chabrier (2003).  We obtain SEDs of these models at a variety of ages since the onset of star formation, ranging between 0.5 and 12\,Gyrs in 0.5\,Gyr steps (but not allowing the model to be older than the age of the Universe). The best-fit line to these models then formed the mass cuts for each redshift bin. 
These mass cuts were applied to each (proto)cluster, according to its redshift, and are shown as dotted lines in Figure \ref{fig:AppendCMDs}. For clusters with $z<1.3$, an evolving $[3.6]$ magnitude limit was also applied, to avoid faint, low mass galaxies entering the sample. This magnitude limit was calculated, using the same models as above, as the faintest possible magnitude that a $10^{10.5}$\,M$_{\odot}$ galaxy could have at each redshift.

\subsection{Colours of (proto)cluster galaxies} \label{sec:measuring_colours}
In Figure \ref{fig:3colimages}, we show the colour-magnitude diagrams of two of the CARLA fields, J105231.82$+$08060 (imaged with GMOS-S) and J101827.85$+$05303 (imaged with ACAM) at $z=1.64$ and $z = 1.94$, along with their $i',[3.6],[4.5]$ three-colour images.  Red squares show sources within 1\,arcmin of the RLAGN, and grey plus symbols show those sources lying further than 1\,arcmin from the RLAGN, which are likely to contain a higher fraction of field contaminants. The blue dashed lines show our $1\sigma$ $i'$ median depth. 
Due to the depth of our $i'$ data we cannot probe the faint red population, although at these redshifts the red sequence is depleted at faint magnitudes \citep[e.g.][]{Papovich2010}. The faint red sources shown as limits are also likely to be cluster members. 
The dotted blue lines in Figure \ref{fig:3colimages} show the mass cut used to select galaxies with $M_*>10^{10.5}$\,M$_{\odot}$, as described in Section \ref{sec:masscuts}. The colour-magnitude diagrams of the remaining CARLA clusters are shown in Appendix \ref{Append:CMDs}. 

\begin{figure}
\centering
\includegraphics[scale=0.5]{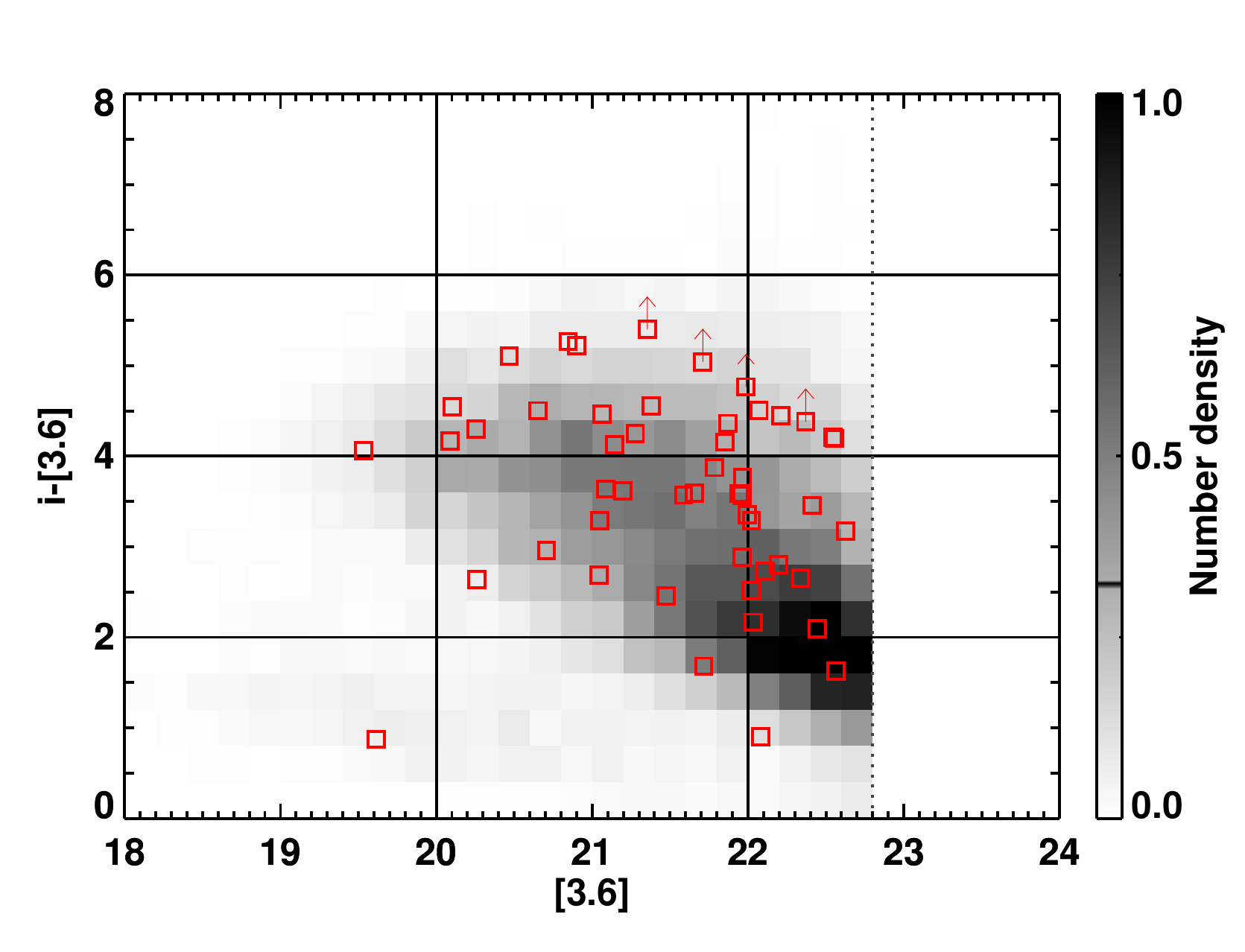}
\caption{Example of the grid cells used in calculating the average $i'-[3.6]$ colour of each cluster. In each cell, $\mu^{\rm{UDS}}_{cell} \pm \sigma^{\rm{UDS}}_{cell}$ sources from each cluster field were removed iteratively before calculating the average colour, in order to statistically remove the field sources in each CARLA (proto)cluster candidate region. This grid is as fine as possible, whilst still ensuring a significant number of sources in each cell. To guide the eye, shading shows the normalised number density distribution of IRAC-selected UDS sources. The depth of the 3.6\,\micron images is shown by the dotted line. Also overlaid are red points showing the colour-magnitude diagram of J213638.58$+$00415 at $z=1.94$. See Appendix \ref{Append:CMDs} for full details of the colour-magnitude diagrams.}
\label{fig:grid}
\end{figure}

We measure the median $i'-[3.6]$ colour of the CARLA cluster galaxies (used in Section \ref{sec:results}) by dividing the colour-magnitude diagram of the clusters into grid cells (see Figure \ref{fig:grid}) and statistically subtracting the expected number of field galaxies in each grid cell before taking the median colour of the remaining galaxies. The full method is as follows: 

We use the UDS data to derive the average number of sources expected from field contamination, and their expected distribution in  $i'-[3.6]$ vs. $[3.6]$ colour-magnitude space.  Colour-magnitude diagrams were calculated for 401 randomly located 1\,arcmin radius regions in the UDS. The colour-magnitude diagrams of the 401 random UDS regions were then divided into twelve grid cells (Figure \ref{fig:grid}) and the mean number of sources in each cell was measured ($\mu^{\rm{UDS}}_{cell}$). 
The errors on these numbers were taken as the standard deviation of the number of sources per cell. These were normalised such that $\sum \sigma^{\rm{UDS}}_{cell} = \sigma^{\rm{UDS}}_{total}$, so that overall $\mu^{\rm{UDS}}_{total} \pm \sigma^{\rm{UDS}}_{total}$ were used, although distributed according to the total population\footnote{$\sigma^{\rm{UDS}}_{total}$ was measured from the number distribution of all sources in the 401 random regions, before applying the grid.}. All these values were calculated after applying the appropriate mass cut for the CARLA field being investigated, so the field and cluster were treated in the same way throughout. 

In order to statistically remove field contaminants, in each (proto)cluster field 
randomly-selected sources were removed from each cell before calculating the median $i'-[3.6]$ colour of the remaining sources. The number of randomly-selected sources removed each time was taken from a Gaussian centred on $\mu^{\rm{UDS}}_{cell}$ with a width of $\sigma^{\rm{UDS}}_{cell}$, helping to deal with sample variance in interloper galaxies. 
This was repeated for 1001 iterations 
to give an overall median colour. 
The mean colours were calculated in the same way, though sources with  $i'$ band magnitudes fainter than $1\,\sigma_{med}$ were set equal to $1\,\sigma_{med}$ value (shown as lower limits in Figure \ref{fig:3colimages}). The mean $i'-[3.6]$ colours typically differ by $\sim0.13$\,mag compared with the median colours and at most differ by $\sim0.34$\,mag. The median colours are typically redder than the means, with seven exceptions ($>80\%$ are redder). 
We use the median colours hereafter in order to avoid biasing our results, although we emphasise that there is good agreement between the mean and median colours. 

The median and mean $[3.6]-[4.5]$ colours were measured similarly, dividing the $([3.6]-[4.5])$-$[3.6]$ colour-magnitude diagrams into cells and statistically removing field contamination. The median $[3.6]-[4.5]$ colours are used throughout.

\subsection{Low \& intermediate redshift cluster sample} \label{sec:ISCS}
In order to compare the results of the high redshift CARLA clusters in this paper to lower redshift clusters, we use photometric catalogues from the IRAC Shallow Cluster Survey \citep[ISCS;][]{Eisenhardt2008}, covering the Bo{\"o}tes region of the NOAO Deep Wide-Field Survey \citep[NDWFS;][]{Jannuzi1999}. In \citet{Eisenhardt2008} 335 cluster and group candidates were identified spanning $0< z \lesssim 2$. These form a low and intermediate redshift cluster comparison sample. We also include in our sample two higher redshift clusters discovered in the same sky region by the IRAC Distant Cluster survey (IDCS; the deeper IRAC extension of the ISCS), at $z=1.75$ \citep{Stanford2012, Brodwin2012, Gonzalez2012} and $z=1.89$ \citep{Zeimann2012}. 

The NDWFS Johnson $I$ magnitudes \citep{Eisenhardt2008} were converted to SDSS $i'$ magnitudes using the $R-I$ colours:
\begin{equation}
i' = 0.004+0.46(R-I)+I
\end{equation}
This conversion was derived using \citet{BC03} models with both exponentially declining star formation models following a star formation rate ${\rm (SFR)}\propto e^{-t/ \tau }$ with $\tau $ of 0.01, 0.1, 0.5, 1\,Gyr and simple stellar population models where stars form in a single burst at high redshift and passively evolve thereafter. A linear equation was then fit to the model galaxy $(R-I)$ and $(i'-I)$ colours.

For the low and intermediate redshift sample, galaxies were selected if they reside within 1\,arcmin of the cluster centre and have a \emph{Spitzer} IRAC colour of $[3.6]-[4.5] \le 0$ for clusters with $0<z<1.3$, thus removing contaminants at higher redshifts. 

Selecting galaxies within a constant 1\,arcmin radius of the cluster centre at $z<1.3$ corresponds to an increasingly smaller fraction of the (proto)cluster towards lower redshift \citep{Muldrew2015}. This effect is small \citep[at most an 8\% decrease in the area observed between $z=1.3$ and $z=0.5$][]{Muldrew2015}, however it may bias our selection towards the very core of the lowest redshift clusters, and potentially bias our colours to those of the most massive cluster galaxies, with the reddest colours, due to the SFR-density relation. This effect is unlikely to cause a bias in our results for two reasons: first, we are only selecting the most massive cluster galaxies in each cluster, which are likely to be in the central cluster regions anyway. Secondly, \citet{Eisenhardt2008} used a constant physical radius for their cluster galaxy selection. This would have the opposite effect: selecting a larger fraction of the lower redshift clusters. Our results for the ISCS clusters agree with the results found in \citet{Eisenhardt2008} and thus are unlikely to be biased by our choice of aperture size. 

Clusters at $z>1.3$ were treated in exactly the same way as the 37 CARLA clusters, as described above. Mass cuts of $M_*>10^{10.5}$\,M$_{\odot}$ were taken for all clusters, as described in Section \ref{sec:masscuts}. 

\subsubsection{Testing the method}
\begin{figure}
\centering
\includegraphics[scale=0.5]{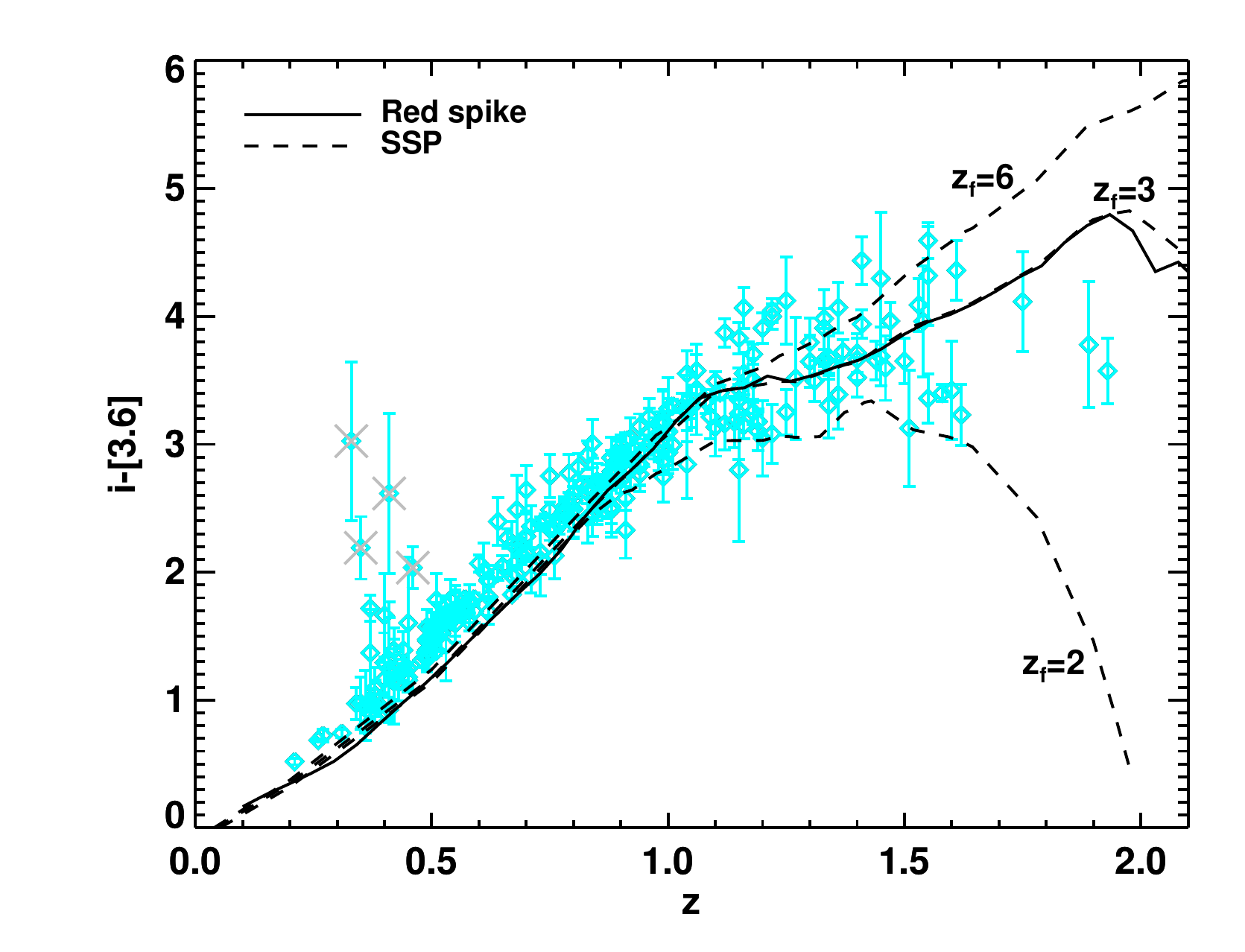}
\caption{Average  $i'-[3.6]$ colours for the ISCS clusters, measured in the same way as for the 37 CARLA clusters. Overlaid is a ``red spike" model (solid line), where stars are formed in a 0.1\,Gyr burst beginning at $z_f=3$, as in \citet{Eisenhardt2008}. Also shown are simple stellar population (SSP) models, where stars form in a delta burst at $z_f=2$, 3, 6 (dashed lines). There is very little difference in the resulting colours of the red spike model and a SSP with $z_f=3$. The measured colours agree with a formation redshift for these cluster galaxies of $z_f\sim3$, in agreement with \citet{Eisenhardt2008}, however note the degeneracies between the different models at $z\lesssim 1$. Points overlaid with grey crosses show low redshift clusters whose colours are affected by secondary overdensities at higher redshift, which have colours of $i'-[3.6]\sim4$.}
\label{fig:Bootes}
\end{figure}

The average $i'-[3.6]$ colours for the ISCS clusters are shown in Figure \ref{fig:Bootes}. Although no cluster membership information is used in this study, the average cluster galaxy colours agree well with those found in \citet{Eisenhardt2008} who used photometric and spectroscopic redshifts to determine cluster membership. The trend of increasing $i'-[3.6]$ colour with redshift agrees with a formation redshift for these cluster galaxies of $z_f\sim3$, showing larger scatter in the colours at higher redshift, as found in \citet{Eisenhardt2008}. This proves that the statistical subtraction method used in this paper to measure average cluster galaxy colours can replicate the results found when cluster membership information is taken into account. Four clusters lie significantly off the $z_f=3$ trend at $z\sim0.4$ (shown with light grey crosses in Figure \ref{fig:Bootes}). The colour-magnitude diagrams for these clusters were visually inspected and found to show secondary structures at higher redshift, which have colours of $i'-[3.6]\sim4$. Since we cannot separate out and remove these potential higher redshift clusters using our method, we instead remove these four clusters from the ISCS sample. At $z<1$ the ISCS data lies slightly above the model. This is due to the mass cut we employ, which selects just the most massive galaxies in order to be consistent with the higher redshift data. This slight offset is expected from \citet{Eisenhardt2008}, where the most luminous (massive) cluster galaxies were systematically redder than simple stellar population models due to the mass-metallicity relation.

\section{Results} \label{sec:results}
\begin{figure*}
\centering
\includegraphics[scale=0.45]{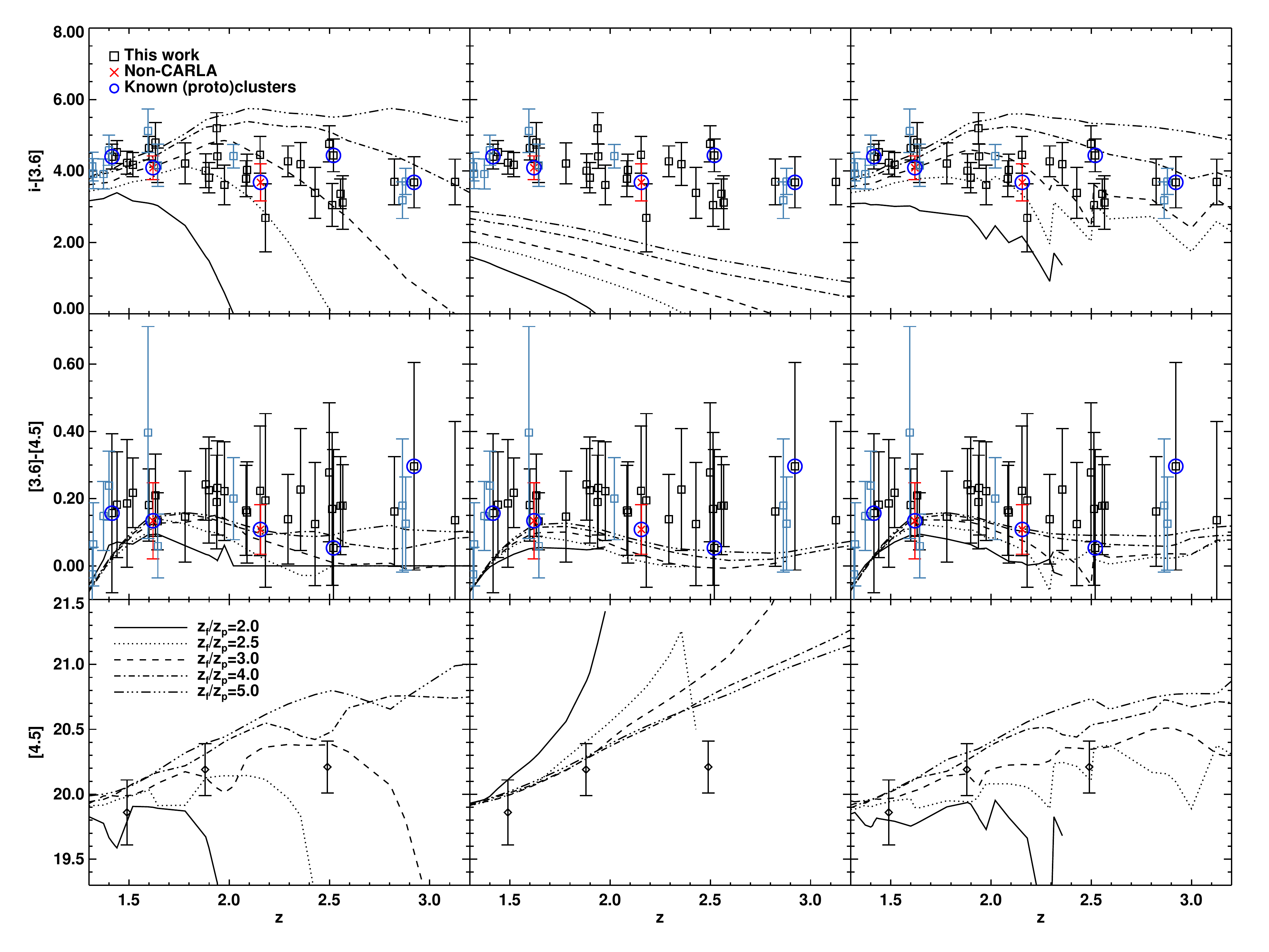}
\caption{\emph{Left:} SSP models. \emph{Centre:} CSP models: exponentially declining models with an e-folding time of 1\,Gyr. \emph{Right:} mSSP models: multiple, normally distributed, bursts of star formation with $z_{peak}=z_p$ and FWHM $=1$\,Gyr. Each column shows the predicted $i'-[3.6]$, $[3.6]-[4.5]$ and $m^*_{[4.5]}$ for $z_f$ ($z_p$ for mSSP) $=2.0,2.5,3.0,4.0,5.0$. Overlaid are the data points from the CARLA clusters (black squares), as well as two non-CARLA (proto)clusters from previous studies (red crosses). Confirmed (proto)clusters are circled in dark blue. \emph{Top:} The $i'-[3.6]$ colours. The points show the median and errorbars represent the standard deviation of the median from 1001 iterations plus a 10\% error in the fluxes. \emph{Middle:} The $[3.6]-[4.5]$ colours. Points show the median source colour and the error on the median. A 20\% error on the flux measurements is not shown, but would add 0.3\,mag to the error bars. \emph{Bottom:} The $m^*_{[4.5]}$ data points in the highest density bin from \citet{Wylezalek2014}. In the top two panels, grey-blue points show those (proto)clusters which lie significantly off the trend found in Figure \ref{fig:overdensity}. }
\label{fig:3x3}
\end{figure*}

The median $i'-[3.6]$ and $[3.6]-[4.5]$ colours of the CARLA clusters are plotted as a function of redshift in Figure \ref{fig:3x3}. For each field, the redshift of the central RLAGN is used as the cluster redshift. 
We also plot the colours for two spectroscopically confirmed clusters, for comparison:  CLG\,0218.3$-$0510 at $z=1.62$ \citep{Tanaka2010,Papovich2010} and the protocluster around the Spiderweb radio galaxy, PKS\,1138$-$262 at $z=2.156$ \citep{Pentericci2000}. Their colours were measured as described in Section \ref{sec:measuring_colours}.  
In the bottom panels of Figure \ref{fig:3x3} we plot the characteristic $[4.5]$ magnitude, $m^*_{[4.5]}$, measured by \citet{Wylezalek2014}. 
\citet{Wylezalek2014} studied the luminosity functions of CARLA clusters within three density bins. Since most of the CARLA clusters in our present study are more than $4\sigma$ denser than the average field, we use the $m^*_{[4.5]}$ derived for the highest density bin used in the \citet{Wylezalek2014} study.

To determine the galaxy formation history of the clusters we compare the average $i'-[3.6]$, $[3.6]-[4.5]$ colours and $m^*_{[4.5]}$ values to three simplistic models (see Figure \ref{fig:cartoon}): simple burst models (SSP; Section \ref{sec:SSP}), exponentially declining models (CSP; Section \ref{sec:CSP}) and multiple burst models (mSSP; Section \ref{sec:mSSP}). 
We generate model galaxies using the publicly available model calculator, EzGal \citep{EzGal}, with \citet{BC03} models\footnote{We have also tested our models using \citet{Maraston2005} models (see Appendix \ref{Append:Maraston}) but find that the \citet{BC03} models provide a better fit to our data.} normalized to match the observed $m^*$ of galaxy clusters at $z\sim0.82$, $[4.5]=19.82$ (AB) \citep{Mancone2012}. 
The scatter in the average $[3.6]-[4.5]$ colours is very large, $\sim0.2$\,mag, meaning it is difficult to constrain a formation history using these colours. They are consistent with all the models we examine in the following Sections and are not discussed further.

\subsection{Simple Stellar Population model} \label{sec:SSP}
\subsubsection{Model description}
The first model we examine is a single simple stellar population (SSP), where galaxies form in an instantaneous burst (hereafter referred to as a delta burst) at $z_f$ and passively evolve thereafter (see the top panel of Figure \ref{fig:cartoon}). Such a model is commonly used in the literature to estimate the formation epoch of cluster galaxies 
and provides a good fit to the $z<1.5$ data (see Figure \ref{fig:Bootes}). 

In the left hand column of Figure \ref{fig:3x3} we compare the SSP models with a range of formation redshifts to the average $i'-[3.6]$ colours of the CARLA clusters, as well as the average $[3.6]-[4.5]$ colour and the characteristic magnitudes $m_{[4.5]}^*$. 
Whereas the SSP model with $z_f=3$ agrees with the $m_{[4.5]}^*$ values well at all redshifts, no SSP with a single formation redshift is able to match the $i'-[3.6]$ colour data. For the CARLA clusters at $z\lesssim2.5$ a formation redshift of $z_f\sim2.5$-$3$ seems to fit the average $i'-[3.6]$ colours well. For clusters at higher redshifts, however, the measured average colour seems to imply a higher formation redshift. 
This means that, either
the basic SSP model is a poor representation of the galaxy formation history of clusters above $z\sim1.3$; or the CARLA clusters selected at $z\sim3$ formed earlier than those at $z\sim1.5$, and thus they do not all lie on one evolutionary sequence. We explore which of these scenarios is likely to be the case in the following Section.

\subsubsection{Are CARLA clusters an evolutionary sequence?} \label{sec:overdensity}
The CARLA survey imaged over 400 high redshift RLAGN across the entire sky. From this survey, over 230 RLAGN appear to be located in regions denser than 2$\sigma$ above average \citep{Wylezalek2013}.  Finally, from these $\sim230$ (proto)cluster candidates we selected the 37 most overdense candidates in every redshift bin (see the grey points in the left hand panel of Figure \ref{fig:overdensity}). 
Therefore, through our selection method, we have isolated some of the most overdense (proto)clusters across $1.3\le z \le 3.2$. 
According to hierarchical structure formation, the most overdense regions at $z=3$ should evolve into the most overdense regions at $z=1.3$ (and subsequently $z=0$), albeit with a large amount of scatter \citep{Chiang2013}. 
So we expect the CARLA (proto)clusters in this study to form an approximate evolutionary sequence, with the high redshift protoclusters being the statistical ancestors of the lower redshift clusters in our sample. 

To test this hypothesis, we examine the galaxy growth within the CARLA clusters. With the red crosses in Figure \ref{fig:overdensity}, we show the overdensity (left panel) and abundance (right panel) of $M_*>10^{10.5}$\,M$_{\odot}$ galaxies within 1\,Mpc of the RLAGN. 
These red points show a trend of increasing overdensity and abundance towards lower redshift, with a Spearman's rank correlation coefficient of $r=-0.72$. This is a highly significant trend with a p-value of $5.5\times10^{-7}$. 

This trend was not artificially introduced by our mass cuts, which evolve with redshift. We tested this by randomly reassigning the redshifts to our 37 CARLA clusters, which in turn randomised the mass cut line taken for each cluster. This was iterated 1000 times and the trend of increasing overdensity with redshift re-examined. 
The randomized samples produce a very weak correlation of median value $r=-0.29$, which is not significant (median p$=0.08$). Therefore, randomising the redshifts (and therefore the mass cut) for each cluster does not produce a significant trend with redshift.

Furthermore, this trend is not due to massive galaxies entering the cluster from the outskirts region, because our 1\,arcmin apertures contain the same fraction of the (proto)clusters at all redshifts between $z=3.2$ and $z=1.3$. \citet{Chiang2013} show that most of cluster collapse occurs at $z>1$ when viewed in the co-moving reference frame. However, in physical units, the cluster's effective radius stays relatively stable until $z\le1$ because gravity is almost balanced by the Hubble expansion \citep{Muldrew2015}. Our 1\,arcmin radius ($\sim 0.5$\,Mpc physical) apertures track the same fraction of the (proto)clusters across the $1.3<z<3.2$ epoch. Thus, the trend in Figure \ref{fig:overdensity} is not caused by cluster collapse, but rather is due to galaxy growth within the (proto)clusters.

In hierarchical cosmology it takes time for massive galaxies to assemble, therefore we use the abundance of massive galaxies as a proxy for cluster maturity. The increase in massive galaxy abundance therefore suggests an increase in cluster maturity. To test this hypothesis we compare the trend in Figure \ref{fig:overdensity} to the expected galaxy growth within semi-analytic models. 
We use the \citet{Guo2011} semi-analytic model built upon the Millennium Dark Matter Simulation \citep{Springel2005}. A full description of the models and identification of (proto)cluster members is provided in \citet{Cooke2014} and \citet{Muldrew2015}. In brief, we devolve 1938 clusters with $z=0$ halo masses of $>10^{14}$\,$h^{-1}$\,M$_{\odot}$ back in time and trace their member galaxies. At each output redshift we count the number of progenitor galaxies with $M_*>10^{10.5}$\,$h^{-1}$\,M$_{\odot}$. 
The solid black line in Figure \ref{fig:overdensity} shows the evolution of the number of $M_*>10^{10.5}$\,M$_{\odot}$ (proto)cluster galaxies from these models, normalised to the least squares fit to the data. 
Although the detailed physics of the semi-analytic models is uncertain, the general trend is in good agreement with the data. This provides compelling evidence that the growth in abundance of massive galaxies within these CARLA clusters suggests that they are likely to form an evolutionary sequence: the high redshift protoclusters could be the statistical ancestors of the lower redshift clusters in this sample.

\subsubsection{SSPs cannot explain high redshift colours}
We have shown that the increase in abundance of massive galaxies within the CARLA clusters follows the expected trend of galaxy growth within forming clusters. We therefore suggest that these CARLA clusters lie on an approximate evolutionary sequence, i.e. the lower redshift clusters have the expected properties of the descendants of the higher redshift protoclusters in our sample. Therefore the colour data in Figure \ref{fig:3x3} must be fit by a single formation model. 
However, although a single SSP of any $z_f>2$ fits the colours of $z<1.5$ cluster galaxies, at high redshift, we cannot fit one formation epoch to all the data across $1.3<z<3.2$. 
This implies that cluster galaxies did not form concurrently at high redshift, but rather a more complex formation history is required. 

We also note that the majority of these CARLA clusters are to be richer than the high redshift ISCS clusters. This is discussed further in Section \ref{sec:compwISCS}.

\subsection{Composite Stellar Population model} \label{sec:CSP}

\subsubsection{Model description}
In order to try to fit the unevolving colour in the data, we next examine a composite stellar population (CSP), where each of the galaxies undergo an exponentially decaying SFR ($SFR \propto \exp{-t/\tau}$) starting at $z_f$, with an e-folding timescale $\tau$. This is represented by the upper-middle panel of Figure \ref{fig:cartoon}. 
CSP models were examined with $\tau=$0.1, 1 and 10\,Gyr. All galaxies are assumed to have formed concurrently. The short e-folding time of $\tau=0.1$\,Gyr gives similar results to the SSP models, and the $\tau=10$\,Gyr models cannot produce $i'-[3.6]$ colours redder than 1.5. The CSP models with $\tau=1$\,Gyr are shown in the centre column in Figure \ref{fig:3x3}. 
CSP models with $z_f>2$ fit the $m_{[4.5]}^*$ values at low redshift, however they cannot explain the bright magnitudes at $z=2.5$. 
Although this model succeeds in producing a flatter $i'-[3.6]$ colour trend with redshift, the colours are still too blue to fit the CARLA data. This means that massive cluster galaxies could not have formed their stars gradually in one long period of star formation unless there is a large amount of dust attenuation.

\subsubsection{Dust extinction} \label{sec:dust}
Dust attenuation in the cluster galaxies will cause their colours to appear redder. Adding dust to the CSP models would make the models redder. 
Because the models get bluer at higher redshift, in order to fit the flat colour trend of the data, we require a varying amount of dust extinction ($A_V$) with redshift. 
Assuming the Calzetti extinction law, to match the CSP models\footnote{In order to match the SSP models, we would require a varying amount of dust extinction with redshift, with $A_V \sim 1.8$ at $z=3$, and $A_V \le 0.7$ at $z=1.3$. This amount of dust extinction in passively evolving galaxies is unlikely, due to the lack of on-going star formation.}, we would require $A_V \sim1.8$ at $z=3$, $A_V \sim 1.3$ at $z=2$ and $A_V \sim 1.1$ at $z=1.3$. This level of dust extinction is not extreme for these redshifts \citep{Garn2010,Cooke2014}.

A number of recent studies have found large numbers of dusty, star-bursting galaxies in high-redshift (proto)clusters \citep[e.g.][]{Santos2014,Santos2015,Dannerbauer2014}. These large numbers do not necessarily mean that the dusty star-forming population represent the majority of the cluster population. Indeed, despite the increase in star-formation rates, \citet{Papovich2012} found that the majority of cluster galaxies in the central regions are passive. 

The colours plotted in Figure \ref{fig:3x3} are the median values for each cluster. This means that a large fraction of the galaxies would need to be dusty in order to affect the overall median colour we measure. 
Up to 10\% of UDS sources (with all our selection criteria and cuts applied) are detected at 24\,$\mu$m. This suggests that the fraction of galaxies in our CARLA sample that are extremely dusty, star-forming galaxies is less than 10\%, and therefore are unlikely to affect our measured median colour.

\subsubsection{CSP cannot explain cluster colours without dust}
The CSP models shown here do not produce colours which are red enough to explain the observed $\langle i'-[3.6] \rangle$ data. In order to match the data, we require a significant fraction of the cluster population to be dusty, highly star-forming galaxies, and have an average dust attenuation that increases with redshift. Previous studies have shown that a significant fraction of the massive cluster population are likely to be passive (at least up to $z\sim1.6$), so it is unlikely that these CSP models are correct for this massive cluster galaxy population. 
Furthermore, significant dust extinction would bring further discrepancy between the models and the values of $m_{[4.5]}^*$.

\subsection{Multiple Simple Stellar Populations model} \label{sec:mSSP}
We have shown that the epoch of massive cluster galaxy formation has to be extended, but the CSP models which extend the period of star formation cannot fit the data unless we incorporate a significant amount of dust. In order to produce the observed red $\langle i'-[3.6] \rangle$ colours at $z=3$, at least some of the cluster population must already be passive at high redshift. 

\subsubsection{Model description}
\begin{figure}
\centering
\includegraphics[scale=0.5]{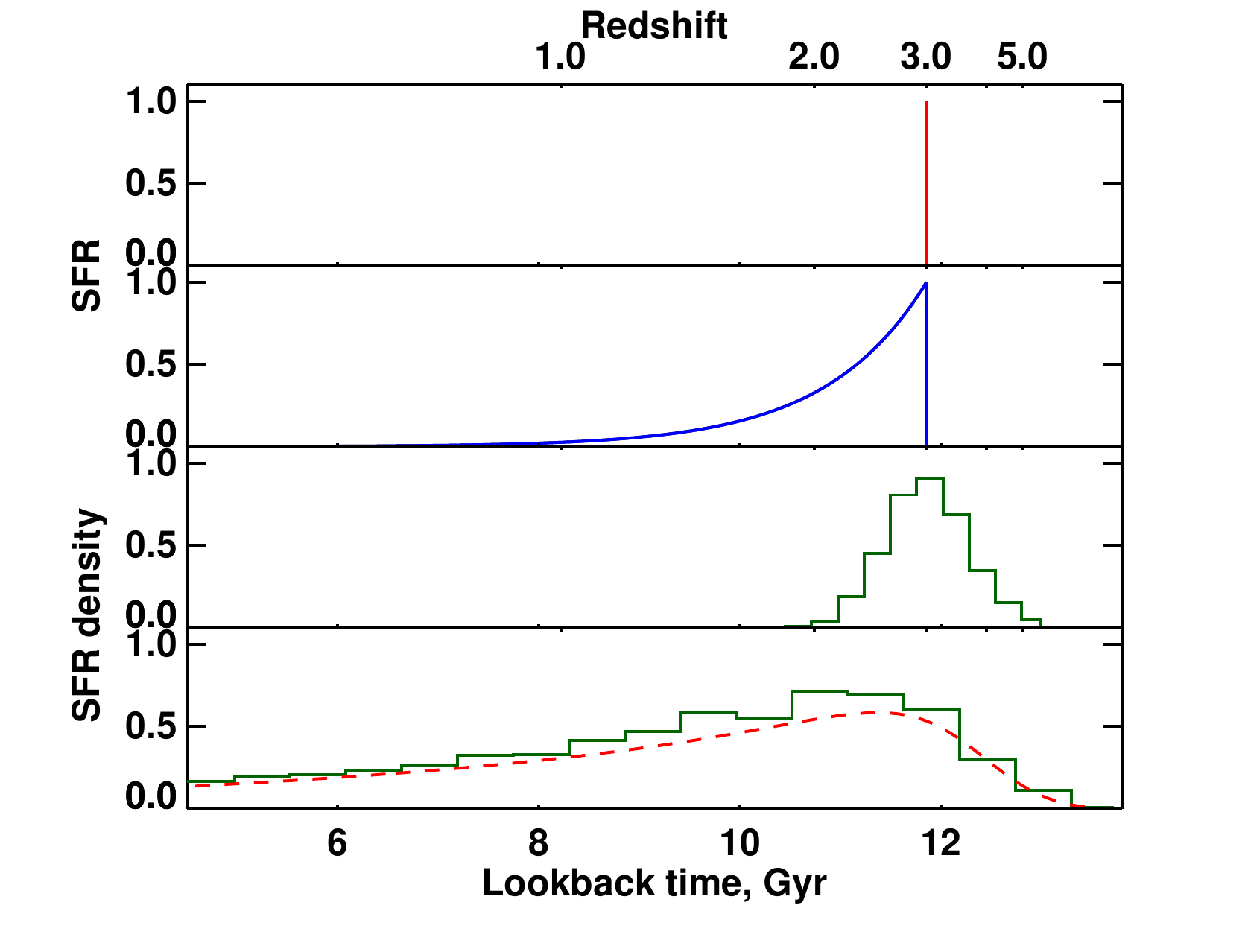}
\caption{Cartoon representation of the three models described in Sections \ref{sec:SSP}-\ref{sec:mSSP}, as well as a model based on the star formation rate density of the Universe, described in Section \ref{sec:SFRdens}. \emph{Top:} SSP model: stars form in a single delta burst at $z_f=3$ and passively evolve thereafter. \emph{Upper middle:} CSP model: the galaxy undergoes an exponentially decaying SFR ($SFR \propto \exp{-t/\tau}$) starting at $z_f=3$, with $\tau = 1$\,Gyr. \emph{Lower middle:} mSSP model: galaxies are formed in multiple bursts of star formation, normally distributed in time around $z_{peak}=3$ with FWHM$=1$\,Gyr. The green histogram illustrates the relative fraction of galaxies that are formed in each time interval. \emph{Bottom:} Multiple burst model where the distribution of galaxies follows the cosmic star formation rate density, see Section \ref{sec:SFRdens} for details. As above, the histogram illustrates the relative fraction of galaxies that are formed in each time interval. }
\label{fig:cartoon}
\end{figure}

In order to produce passive $M>10^{10.5}$\,M$_{\odot}$ galaxies by $z=3$, we again model galaxy formation as single bursts of star formation. To extend the period of cluster galaxy formation, and produce an approximately unevolving colour trend with redshift, we use multiple simple stellar populations (mSSP), where cluster galaxies are formed in individual, short bursts, with their formation redshifts distributed in time so that the total cluster population forms over the course of a few Gyr. This is illustrated in the lower-middle panel of Figure \ref{fig:cartoon}. The green histogram represents the relative fraction of galaxies being formed. 

We create mSSP models with 16 model galaxies\footnote{On average there are 16 cluster galaxies within 1\,arcmin of the RLAGN in each of the 37 CARLA fields, once field contaminants are statistically removed.}, which each form their stars in delta-bursts, with their formation redshifts normally distributed in time around a peak at redshift $z_{peak}$. The normal distribution's FWHM is set to 1\,Gyr. We then take the median colour of these model galaxies at each redshift. The right hand column of Figure \ref{fig:3x3} shows the mSSP models for the $i'-[3.6]$, $[3.6]-[4.5]$ and $m_{[4.5]}^*$ values with different $z_{peak}$. Including multiple bursts of star formation flattens the expected $i'-[3.6]$ colour over redshift and provides good agreement with the CARLA data. The $[3.6]-[4.5]$ and $m_{[4.5]}^*$ values are also consistent with $z_{peak}>2.5$. 
We have also tested this model with varying FWHM and find that the FWHM has to be $>0.9$\,Gyr in order to provide a good fit to the CARLA data.

\subsubsection{mSSP model provides a good description of the data}
The mSSP model describes the observed $i'-[3.6]$ colours of these CARLA clusters well, and also agrees with the $m_{[4.5]}^*$ values. 
We conclude that a more extended period of burst-like galaxy formation, spanning at least 1\,Gyr, is required to explain the colours of the CARLA cluster galaxies. We have modelled these galaxies as forming in single bursts, but due to the scatter in our data we cannot constrain the individual star formation histories of the cluster members. The median $i'-[3.6]$ colours mean that the galaxies must have ceased their star formation rapidly in order to produce red colours. 
This bursty appearance could also be produced with a variety of star formation histories, so long as the star formation is rapidly terminated. Investigating these formation histories is beyond the scope of these data and we just examine the most basic, burst models.

\begin{figure}
\centering
\includegraphics[scale=0.5]{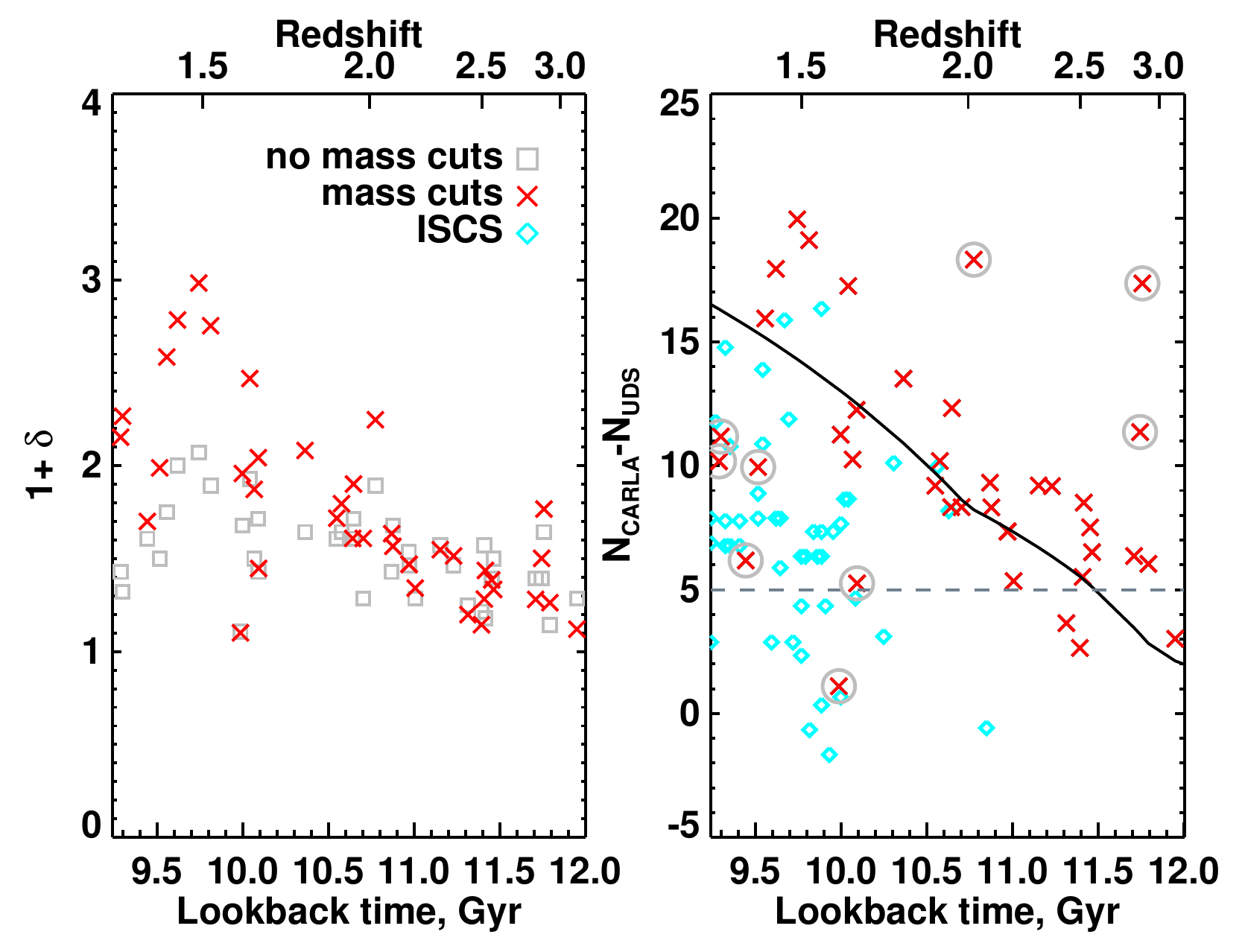}
\caption{Left: $1+\delta = N_{\rm{CARLA}}/N_{\rm{UDS}}$ for the clusters as a function of time. Grey points show the original densities, without taking any mass cuts. Red points show the densities after taking a $10^{10.5}$\,M$_{\odot}$ mass cut. There is a significant trend of increasing density over time. Clusters which do not lie on this trend are unlikely to be the progenitors/descendants of those on the trend. Right: $N_{\rm{CARLA}}-N_{\rm{UDS}}$. This shows how the number of galaxies within 1\,arcmin of the RLAGN with M$>10^{10.5}$\,M$_{\odot}$ increases with time. The black line shows the expected evolution of the number of massive (proto)cluster galaxies from semi-analytic models. Blue points show the ISCS clusters, discussed in Section \ref{sec:compwISCS}. Highlighted with grey circles are clusters that lie significantly off the expected trend for the CARLA clusters and thus are unlikely to form part of the evolutionary sequence.}
\label{fig:overdensity}
\end{figure}

\section{Discussion} \label{sec:discussion}

\subsection{Clusters undergo extended periods of galaxy formation}
We have examined three model star formation histories: a single stellar population, an exponentially declining SFR, and multiple bursts of star formation distributed normally around a peak period at $z_{peak}$. 
We find that SSP models (left hand column of Figure \ref{fig:3x3}) are unable to account for the red $i'-[3.6]$ colours of cluster galaxies at $z>2.5$ and the flat colour trend we find at $z>1.3$ (assuming that these clusters represent one evolutionary sequence; see Section \ref{sec:overdensity}). 
By examining the colours of cluster galaxies at $z>1.3$ we are able to distinguish the cluster formation histories and have shown that the epoch of galaxy formation in clusters has to be extended; a single formation redshift is not sufficient to produce the colour trend we observe.

We have shown that the $\langle i'-[3.6] \rangle$ colours of these cluster galaxies agree well with a model in which they formed in multiple short bursts over approximately 2\,Gyr, peaking at $z\sim3$. This is consistent with a model where different populations of galaxies form in individual bursts at different times, building up the galaxy population over time, rather than in one, short burst. This model is similar to the composite model from \citet{Wylezalek2014} used to explain the luminosity functions of CARLA clusters. 
Although we claim that the cluster galaxies formed over an extended period of time, our data are not sufficient to further constrain the galaxy formation history. A number of extended galaxy formation models could fit these data.

\begin{figure}
\centering
\includegraphics[scale=0.5]{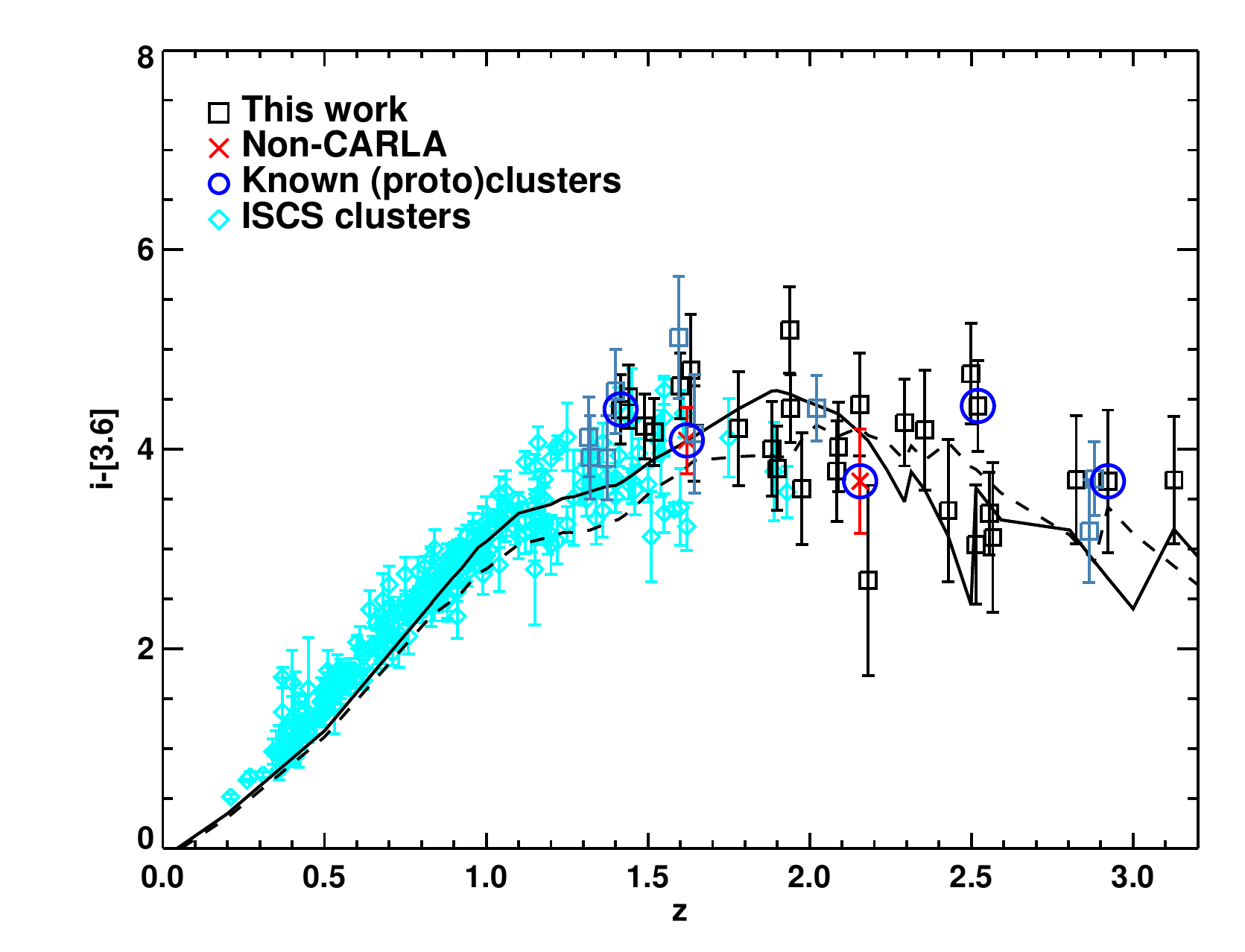}
\caption{The points are the same as in Figure \ref{fig:3x3}, also showing the ISCS clusters at lower redshift as cyan diamonds. Overlaid are the mSSP model with multiple bursts of star formation normally distributed around $z_{peak}=3$, with FWHM$=$1\,Gyr (solid line), and a multiple-burst model following the cosmic star formation history (dashed line). These extended galaxy formation models provide the best fit to our CARLA clusters at high redshift, and also account for the colours observed at lower redshift. The scatter in our data is too large to allow any further distinction between the two extended models.}
\label{fig:mSSPBootes}
\end{figure}

In order to further analyse the formation history of massive cluster galaxies, we must adopt a model.
Our following results do not strongly depend on the exact form of this extended model, but we choose to follow recent literature \citep[e.g.][]{Snyder2012} in assuming the cluster galaxies follow the star formation rate density trend of the Universe.

We expand our mSSP model to follow the cosmic star formation rate by producing 500 model galaxies, each formed in a single short burst, distributed in time according to the star formation history of the Universe from \citet{Hopkins2006}:
\begin{equation} \label{eq:SFRdens}
\rho_* = \frac{(a+bz)h}{1+(z/c)^d}
\end{equation}
where $\rho_*$ is the star formation rate density, $a=0.017$, $b=0.13$, $c=3.3$, $d=5.3$, and $h=0.7$. The bottom panel of Figure \ref{fig:cartoon} illustrates this multiple-burst model. The red dashed line shows the overall shape of the cosmic star formation rate, the green histogram indicates the relative fraction of galaxies that are formed in each time interval. 

In Figure \ref{fig:mSSPBootes} we show this model, as well as the mSSP model (Section \ref{sec:mSSP}), with multiple bursts of star formation around $z_{peak}=3$ (normally distributed bursts across $\sim2$\,Gyr). This Figure illustrates that we do not have sufficient data to distinguish between different extended models. Both models can also account for the colours of lower redshift clusters, providing a consistent explanation for the formation of all massive cluster galaxies.

\subsection{Formation timescale of massive galaxies} \label{sec:SFRdens}
In this Section we examine the period of time between high redshift cluster galaxies forming their stars and assembling into $M_*>10^{10.5}$\,M$_{\odot}$ objects. In hierarchical merging models \citep[e.g.][]{DeLucia2006} galaxies form in small entities and subsequently merge. Therefore there may exist a long time delay between the period of star formation and their assembly epoch. 
If galaxies merge with little gas and no significant star formation (a ``dry merger''), then the resulting massive galaxies will appear red. If the merger included a lot of gas \citep[i.e. a ``wet merger'', or if the galaxies formed via monolithic collapse,][]{Eggen1962}, and induced further star formation, then the resulting galaxies will have bluer colours. 
Thus we can use our data to estimate the time between star formation and assembly into $M_*>10^{10.5}$\,M$_{\odot}$ galaxies.

To do this we form 500 model galaxies distributed in redshift according to the cosmic star formation rate density (Equation \ref{eq:SFRdens}) and calculate an average $i'-[3.6]$ colour at each redshift. 
To simulate galaxies growing in mass through dry mergers and entering our sample only after a certain period of time, we impose a restriction whereby galaxies are only included in our sample after a time delay $\Delta t$. 

\begin{figure}
\centering
\includegraphics[scale=0.5]{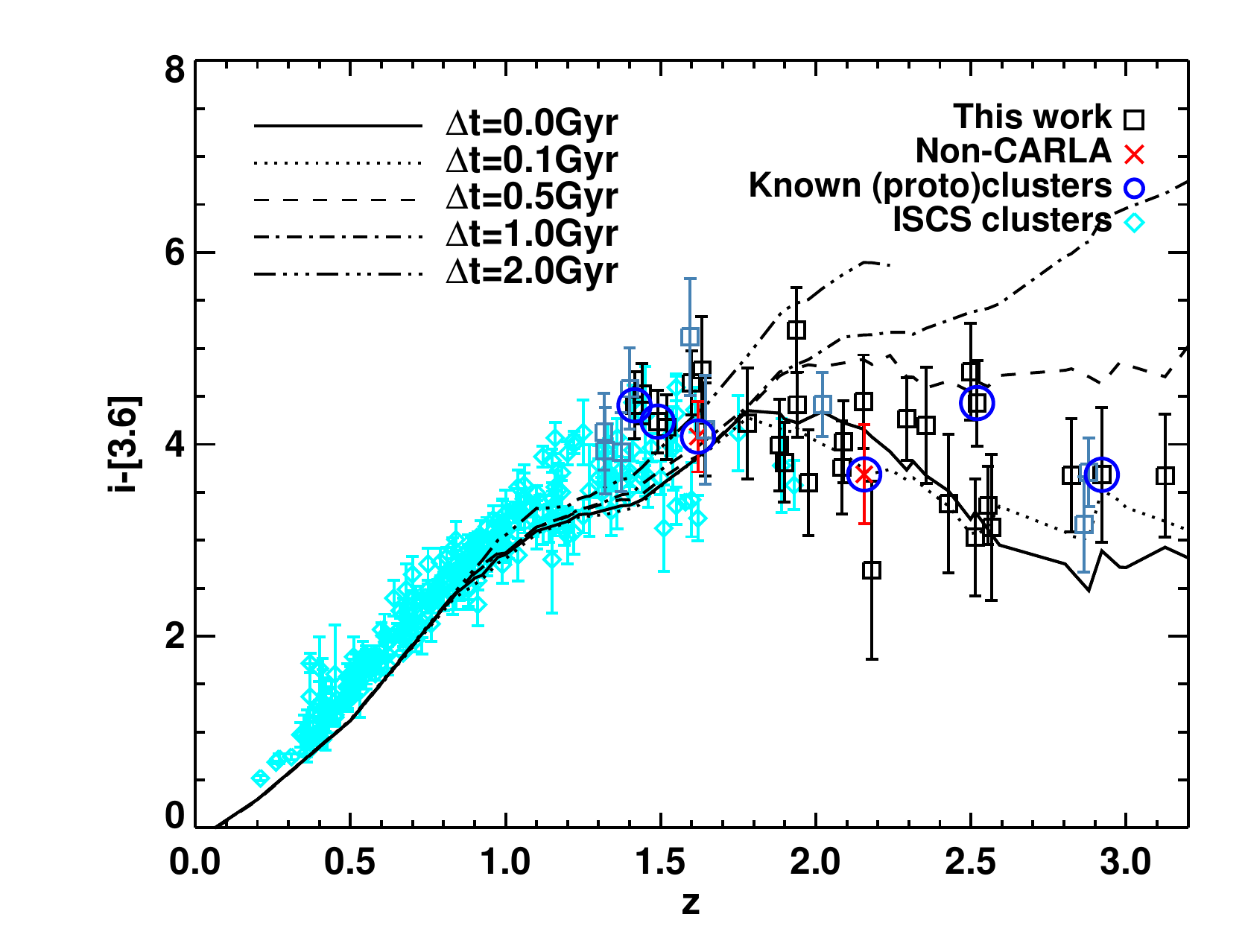}
\caption{The expected average $i'-[3.6]$ colours of clusters when the component galaxies are formed in single bursts, distributed according to the star formation rate density of the Universe. The points are the same as in Figure \ref{fig:mSSPBootes}. The models here include a time delay $\Delta t$ between galaxies forming and being included in our selection. Different lines correspond to different values of $\Delta t$. Values of $\Delta t$ larger than $\sim0.5$\,Gyr produce average cluster colours that are too red to explain the observations at $z>2$. 
}
\label{fig:SFRdens_dt}
\end{figure}

Figure \ref{fig:SFRdens_dt} shows this model with different values of $\Delta t$. The CARLA data at $z>2$ are only consistent with a maximum time delay of $\Delta t \approx 0.5$\,Gyr. 
This short delay between galaxies forming their stars and growing massive enough to enter our sample is in agreement with studies of the luminosity function of galaxies at high redshift, which show that the bright end of the luminosity function is established within 5\,Gyr of the Big Bang \citep[e.g.][]{dePropris2003,Andreon2006,Muzzin2008,Mancone2010,Wylezalek2014}.

These results do not depend on the exact form of the cluster's assembly history. We have tested different assembly histories (the best-fit normally distributed model from Section \ref{sec:mSSP} and a model with the same form as the cosmic star formation rate density, but shifted to higher redshifts) and found no qualitative difference in these results. Individual galaxies must still have assembled within $0.5$\,Gyr of formation of the majority of their stars. 

In summary, $z>2$ massive ($M_*>10^{10.5}$\,M$_{\odot}$) cluster galaxies must have assembled within $0.5$\,Gyr of forming their stars. This could have happened in a number of different ways, such as: formation through a single massive burst; merging into massive galaxies soon after they formed their stars; undergoing a merging event which triggered a massive starburst which dominated the observed colours of the galaxy thereafter. 

At $z<2$, there can be a long delay (several Gyr) between galaxies forming their stars and assembling into massive galaxies. Dry galaxy merging is likely to become a much more important route by which massive galaxies form at ${z\lesssim2}$. 

\subsection{Cessation of star formation within massive cluster galaxies} \label{sec:SFRdens_zend}
Massive galaxies at $z=0$ are passive and contain old stellar populations which suggest that they finished forming stars at $z>2$ \citep[e.g.][]{Bower1992}. 
The cosmic star formation model forms stars up to the present-day. 
In this Section we test whether a cut-off in galaxy formation at higher redshifts provides a better fit to the data.

\begin{figure}
\centering
\includegraphics[scale=0.5]{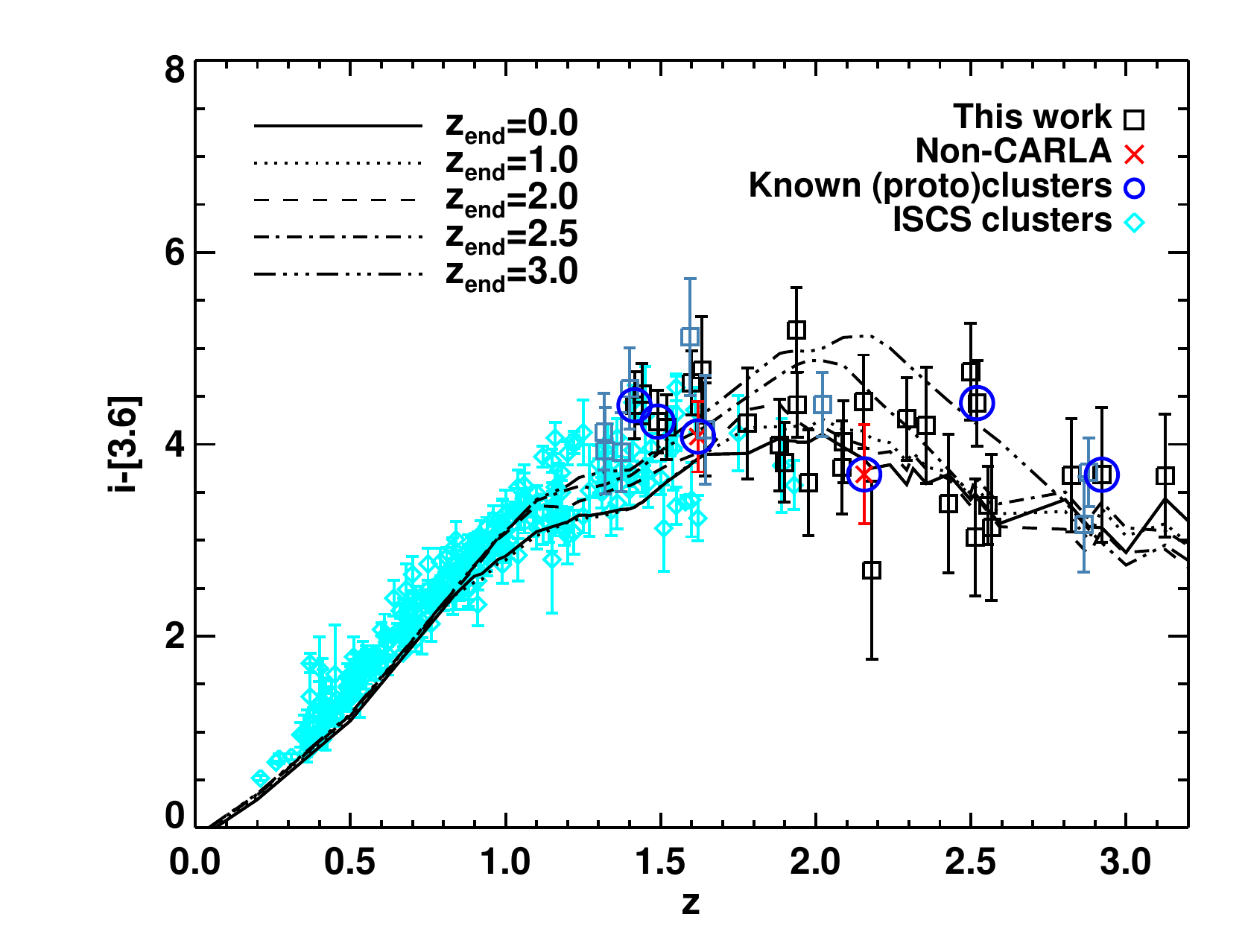}
\caption{The same as Figure \ref{fig:SFRdens_dt}, but the different lines here correspond to different cut-offs in star formation, i.e. star formation is allowed to continue down to $z_{end}$, following the cosmic star formation rate density, and then terminated. The choice of $z_{end}$ has little effect on the expected colours, suggesting that the cosmic star formation rate density itself is sufficient, requiring no cut-off. 
}
\label{fig:SFRdens_zend}
\end{figure}

To test when massive galaxy formation ceased in clusters, we form 500 model galaxies following the cosmic star formation rate density (Equation \ref{eq:SFRdens}), down to a defined redshift $z_{end}$, i.e. with no more star formation occuring in massive galaxies at $z<z_{end}$. Throughout this Section, we use $\Delta t =0$. 
Figure \ref{fig:SFRdens_zend} shows the average $i'-[3.6]$ colour of the model galaxies with various different values of $z_{end}$. 
Higher values of $z_{end}$ predict slightly redder colours at $1<z<2.5$, however the scatter in our data does not allow us to quantify whether a termination of star formation at any particular $z_{end}$ is required. 
The reddening of galaxy colours is entirely due to the peak epoch of star formation occurring at $2\le z\le 3$ and few stars forming in massive galaxies thereafter. 
To determine when star formation in massive cluster galaxies ceased, we require measurements of the individual star formation rates of the cluster members. The average $i'-[3.6]$ colours alone do not contain enough information.

\subsection{Is an extended galaxy formation history consistent with previous work?}
Previous studies have modelled the formation of cluster galaxies as a single concurrent event. Lower redshift data agree well with these models \citep[e.g.][]{Blakeslee2003,Mei2009}. In this Section we test whether an extended period of galaxy formation is consistent with the observational data from previous work. 

\begin{figure*}
\centering
\includegraphics[scale=0.5,page=1]{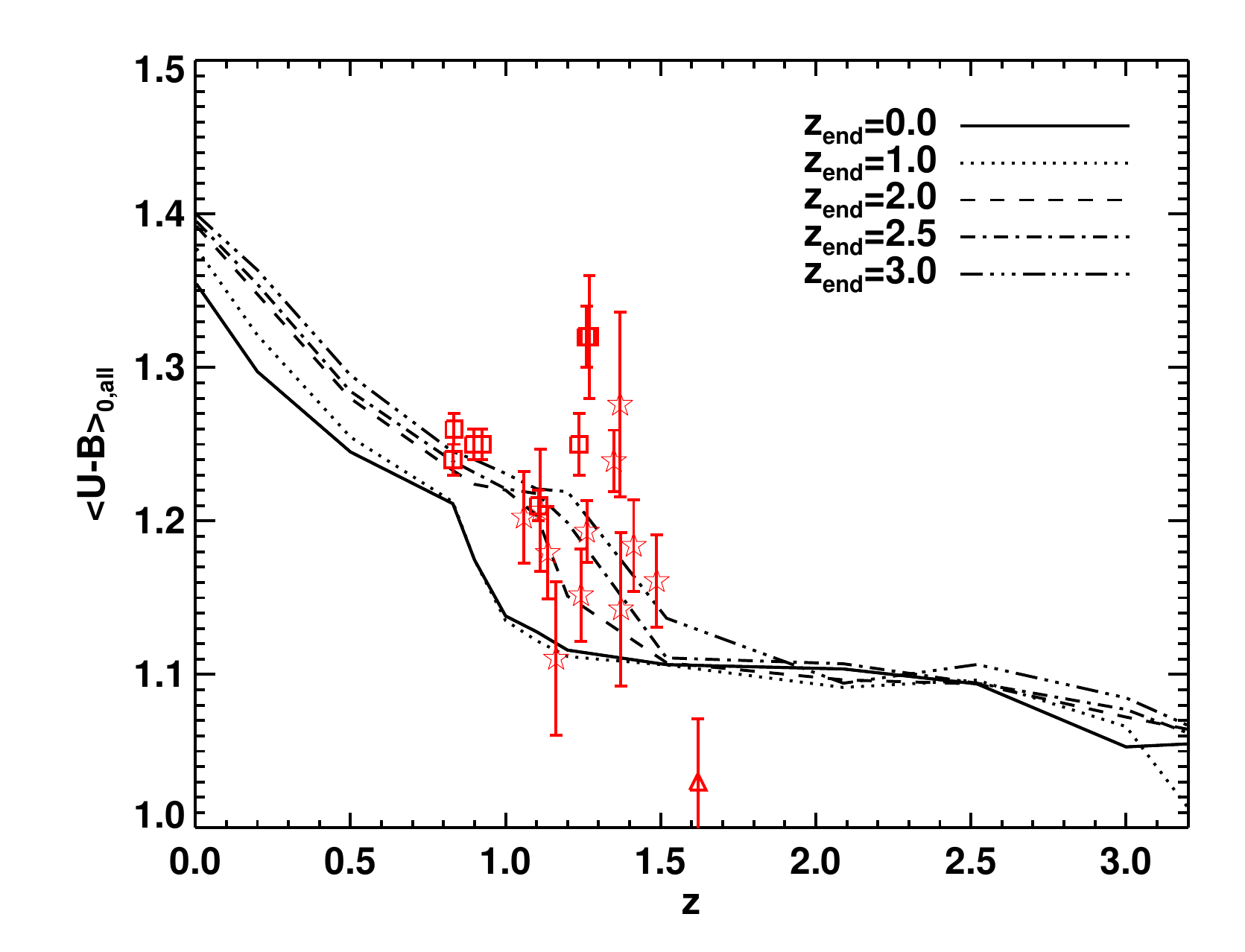}
\includegraphics[scale=0.5,page=2]{modelUBpredicts.pdf}
\caption{Predicted evolution of rest-frame $(U-B)_0$ colours for model galaxies forming as described in Section \ref{sec:SFRdens} with different values of $z_{end}$. \emph{Left:} The average colours for all cluster galaxies. \emph{Right:} The scatter in the $(U-B)_0$ colours for cluster galaxies. Overplotted in red are results from previous studies: \citet{Mei2009} (squares), \citet[][]{Blakeslee2003} (diamonds), \citet{Papovich2010} (triangle), \citet[][]{Snyder2012} (stars). Three studies at low redshift are shown by the red crosses; these are data taken from \citet{Bower1992,Ellis1997,vanDokkum1998}, adapted to rest-frame $(U-B)_0$ by \citet{Mei2009}.
}
\label{fig:UBcolscat}
\end{figure*}

In Figure \ref{fig:UBcolscat} we examine the trend of the rest-frame $(U-B)_0$ average colour and scatter predicted from the extended model from Section \ref{sec:SFRdens_zend}, following the cosmic star formation density, with different cut-off redshifts, $z_{end}$. Overlaid in red in Figure \ref{fig:UBcolscat} are findings from previous studies \citep{Bower1992,Ellis1997,vanDokkum1998,Blakeslee2003,Mei2009,Papovich2010,Snyder2012} at $0<z<1.6$.

The $\langle U-B \rangle_{0}$ colour and scatter depend on the adopted value of $z_{end}$. Stopping galaxy formation at higher $z_{end}$ decreases the scatter and reddens the expected $(U-B)_0$ colours. 
Our model colours and scatters were calculated taking the whole cluster population into account, whereas the data points were measured from red sequence galaxies only. Therefore the data are expected to have a redder colour and smaller scatter than the models, however almost all massive cluster galaxies at $z<1.5$ exhibit red colours \citep[e.g.][]{Kajisawa2006b}. The models with $z_{end} \sim1$-2 show good agreement with these previous results, although there is large scatter in the data, suggesting that our simple burst model following the cosmic star formation rate density of the Universe (with some reasonable $z_{end} \sim 1$-2) provides a possible explanation for the formation history of massive cluster galaxies.

\subsection{Cluster mass} \label{sec:compwISCS}
The luminosity functions of CARLA clusters are significantly different from the ISCS clusters, with the CARLA clusters having brighter $m^*_{[4.5]}$ values than those from the ISCS \citep{Wylezalek2014}. The lowest density bin examined in \citet{Wylezalek2014} showed results more consistent with the ISCS sample, suggesting that the lower-richness CARLA clusters are more similar to the clusters from ISCS. In this paper we only examine the densest CARLA clusters and in Figure \ref{fig:overdensity} we confirm that members of this subset are richer than the $z>1.3$ ISCS clusters. 

In the right hand panel of Figure \ref{fig:overdensity} we plot the number of massive ($M_*>10^{10.5}$\,M$_{\odot}$) galaxies in the CARLA fields. We also show the excess number of galaxies in the ISCS clusters at $z>1.3$. 
The majority of the high redshift ISCS clusters have systematically fewer massive galaxies than the CARLA sample. This indicates that the CARLA clusters are more overdense than the ISCS sample and are therefore likely to be more massive clusters and protoclusters.

\citet{Brodwin2007} found that the correlation function of ISCS clusters indicates they reside in dark matter haloes of $\sim10^{13.9}$\,M$_{\odot}$, and will evolve into clusters of  $2$-$3\times10^{14}$\,M$_{\odot}$ by $z=0$. Figure \ref{fig:overdensity} suggests that the majority of the 37 CARLA clusters in this study will collapse to become more massive clusters ($\gtrsim 5\times 10^{14}$\,M$_{\odot}$) by the present day.

\citet{Brodwin2013} suggest that the formation history of massive cluster galaxies depends on the overall mass of the cluster in which they reside. 
They found that the lower-mass ISCS clusters were undergoing a major epoch of merging and galaxy formation peaked around $z\sim1.4$, and predicted that the formation epoch would peak at higher redshifts for more massive clusters. Our CARLA data are consistent with a peak formation and assembly epoch of $z\sim2$-3 for the more massive CARLA clusters, in agreement with a higher assembly epoch for more massive clusters. 
With the ${\langle i'-[3.6] \rangle}$ data, however, we find that the cluster density does not affect the measured average colour. This is perhaps due to the broad range of wavelengths covered by the observed $i'-[3.6]$ colour at these redshifts. 

In summary, the CARLA fields studied here are unlikely to be the progenitors of the lower redshift ISCS clusters. They are likely to evolve into more massive clusters ($\gtrsim 5\times 10^{14}$\,M$_{\odot}$). However, we do not think this will affect the results of Figures \ref{fig:mSSPBootes}-\ref{fig:SFRdens_zend} as both populations exhibit similar average $i'-[3.6]$ colours.

\subsection{Caveats}

\subsubsection{CARLA sample are not yet confirmed}
Most of the CARLA clusters studied here are not yet spectroscopically confirmed, which may affect our conclusions on the evolution of clusters. 
These fields were selected as the most dense CARLA fields, which are significantly denser than the average field at a $\gtrsim 4\sigma$ level, and thus are likely to contain protoclusters. 
Also, in Figure \ref{fig:overdensity}, the trend of increasing density towards lower redshifts suggests that most of our sample are indeed (proto)clusters. 
Some fields in our sample have clear evidence for a forming red sequence (see Figure \ref{fig:AppendCMDs}) and three of our sample are spectroscopically confirmed structures: 7C1756$+$6520 \citep{Galametz2010a}, TXS\,1558$-$003 \citep{Hayashi2012}, and MRC\,0943$-$242 \citep{Venemans2007}. 
These three clusters, as well as the two non-CARLA clusters from \citet{Papovich2010} and \citet{Pentericci2000}, follow the same flat trend in colour as the unconfirmed clusters at all redshifts. 
This all provides strong evidence that the majority of the CARLA fields in this study are likely to be (proto)clusters. Further spectroscopic studies are required to confirm this.

\subsubsection{AGN}
The presence of AGN may cause redder colours in our cluster sample. The fraction of AGN in clusters is known to be enhanced compared to the field \citep[e.g.][]{Galametz2010a}, which may affect the IRAC bands. 
Our use of median colours throughout should prevent small numbers of AGN significantly affecting our measured average colours. 

\subsubsection{Blending}
The FWHM of the \emph{Spitzer} 3.6 and 4.5\,\micron data is $\sim1.7$\,arcsec. This means that source fluxes may be affected by blending with nearby sources, particularly in crowded fields. The $i'$ data, although having a small FWHM, may also experience some blending. If blending occurs between galaxies at similar redshifts, i.e. between cluster members, our conclusions will be unaffected, as we measure median colours of clusters throughout. Blending with fore- or background sources may cause inaccuracies in the measured colours. Further data with better resolution is required to gain more accurate measurements of galaxy colours.

\section{Conclusions} \label{sec:conclusions}
We have used a sample of 37 clusters and protoclusters across $1.3 \le z \le 3.2$ from the CARLA survey of high-redshift clusters to study the formation history of massive cluster galaxies. These fields are the densest regions in the CARLA survey, and as such are likely to be the sites of formation for massive clusters. We have used optical $i'$-band and infrared 3.6\,\micron and 4.5\,\micron images to statistically select sources likely to lie within these (proto)clusters and examined their average observed $i'-[3.6]$ colours. The abundance of massive galaxies within these (proto)clusters increases with decreasing redshift, suggesting these CARLA (proto)clusters form an evolutionary sequence, with the lower redshift clusters in the sample having similar properties to the descendants of the  high redshift protoclusters. This sequence allows us to study how the properties of their galaxy populations evolve as a function of redshift. By comparing the abundance of massive galaxies in these CARLA (proto)clusters to those of $z>1.5$ ISCS clusters we have shown that the CARLA sample are likely to collapse into more massive clusters, typically $\gtrsim 5\times 10^{14}$\,M$_{\odot}$.

We have compared the evolution of the average colour of massive cluster galaxies with simple galaxy formation models. 
Taking the full cluster population into account, we have shown that cluster galaxies did not all form concurrently, but rather formed over the course of a few Gyr. 
The overall colour evolution is consistent with the stars in each galaxy forming in a single burst, although more complex individual star formation histories that are rapidly truncated may produce this effect. 
This galaxy formation history is consistent with galaxies within different groups of the (proto)cluster forming concurrently, but the whole cluster population building up over a longer period of time. Overall this produces an approximately unevolving average observed $i'-[3.6]$ colour for cluster galaxies at $z=1.3$ to $z\sim3$.

In summary, our main conclusions are as follows: 
\begin{enumerate}
\item The average colours of massive cluster galaxies are relatively flat across $1.3<z<3.2$. It is not possible to describe the formation of these galaxies with a burst model at a single formation redshift. Cluster galaxies formed over an extended period of time. 
\item The formation of the majority of massive cluster galaxies is extended over at least 2\,Gyr, peaking at $z \sim 2$-3. From the average $i'-[3.6]$ colours we cannot determine the star formation histories of individual galaxies, but their star formation must have been rapidly terminated to produce the observed colours. 
\item Massive galaxies at $z>2$ must have assembled within 0.5\,Gyr of them forming a significant fraction of their stars. This means that few massive galaxies in $z>2$ clusters could have formed via dry mergers. 
\end{enumerate}

\section*{Acknowledgements}
The authors would like to thank Anthony Gonzalez for useful comments and suggestions. 
Thank you also to the CARLA team for producing the survey on which this paper is based. 
We thank the anonymous referee for their careful review and helpful comments, which improved the content of the paper. 
We are grateful to Fiona Riddick, Cecilia Fari\~{n}a, Raine Karjalainen, James McCormac and Berto Gonz\'{a}lez for all their help and support with the observations at the WHT. 

EAC acknowledges support from the STFC. 
NAH is supported by an STFC Rutherford Fellowship. 
The work of DS was carried out at Jet Propulsion Laboratory, California Institute of Technology, under a contract with NASA. 
SIM acknowledges the support of the STFC consolidated grant (ST/K001000/1). 
NS is the recipient of an ARC Future Fellowship. 

Based on observations made with the William Herschel Telescope under programme IDs W/2013b/10, W/2014a/6 and SW/2013b/34, and the Gemini Observatory under programme ID GS-2014A-Q-45. 
The William Herschel Telescope operates on the island of La Palma by the Isaac Newton Group in the Spanish Observatorio del Roque de los Muchachos of the Instituto de Astrof\'{i}sica de Canarias. 

The Gemini Observatory is operated by the Association of Universities for Research in Astronomy, Inc., under a cooperative agreement with the NSF on behalf of the Gemini partnership: the National Science Foundation (United States), the National Research Council (Canada), CONICYT (Chile), the Australian Research Council (Australia), Minist\'{e}rio da Ci\^{e}ncia, Tecnologia e Inova\c{c}\~{a}o (Brazil) and Ministerio de Ciencia, Tecnolog\'{i}a e Innovaci\'{o}n Productiva (Argentina). 

This work is based on observations made with the Spitzer Space Telescope, which is operated by the Jet Propulsion Laboratory, California Institute of Technology under a contract with NASA. Support for this work was provided by NASA through an award issued by JPL/Caltech. 

\bibliographystyle{mn2e}
\bibliography{protoclustersbib}

\appendix

\section{Colour magnitude diagrams} \label{Append:CMDs}
The $i'-[3.6]$ vs. $[3.6]$  colour-magnitude diagrams for the remaining 35 CARLA fields (not showing Figure \ref{fig:3colimages}) are shown in Figure \ref{fig:AppendCMDs}. There is a large scatter in the colours of sources, suggesting that each of these clusters still has continuing star formation. 

\begin{figure*}
\centering
\includegraphics[scale=0.3]{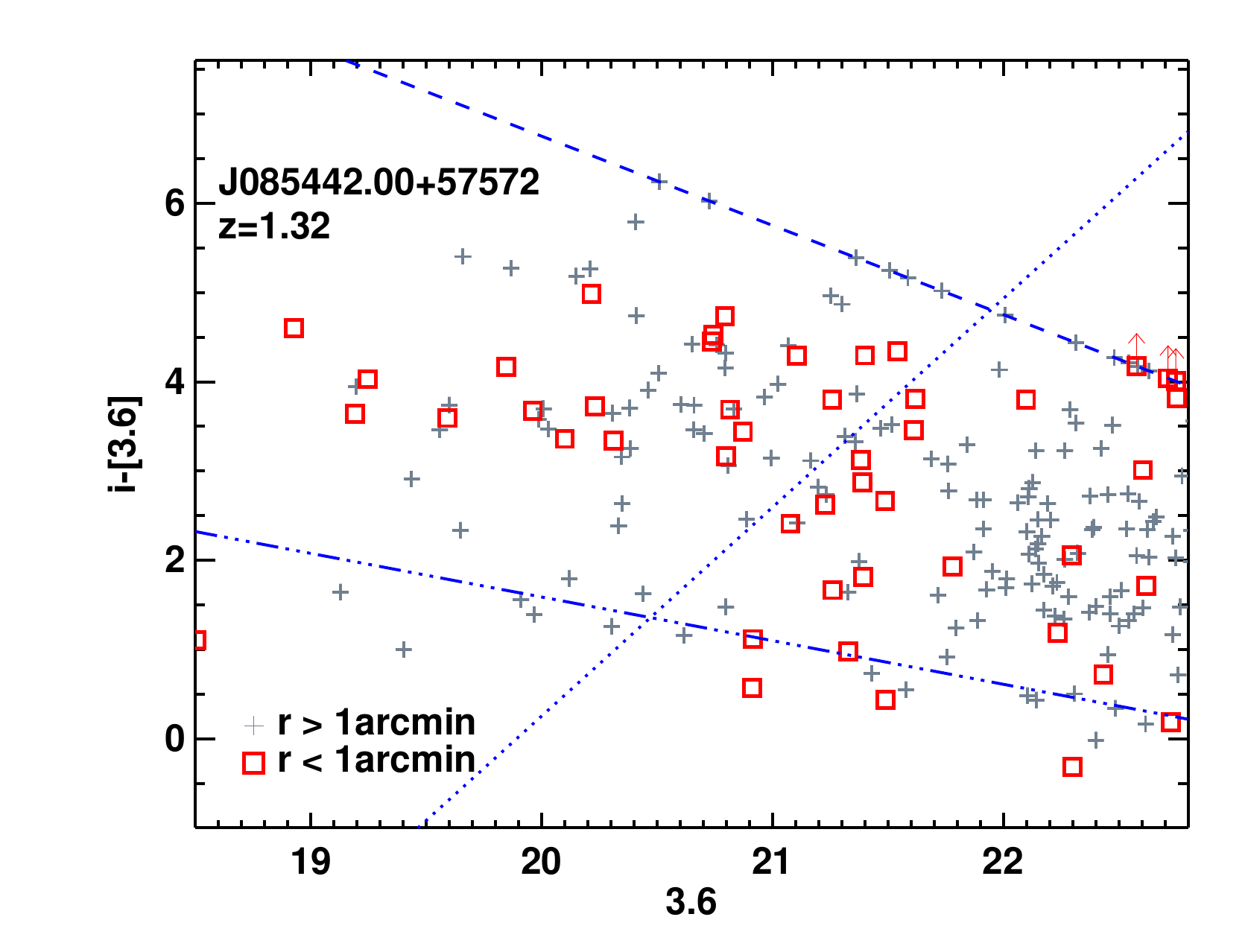}
\includegraphics[scale=0.3]{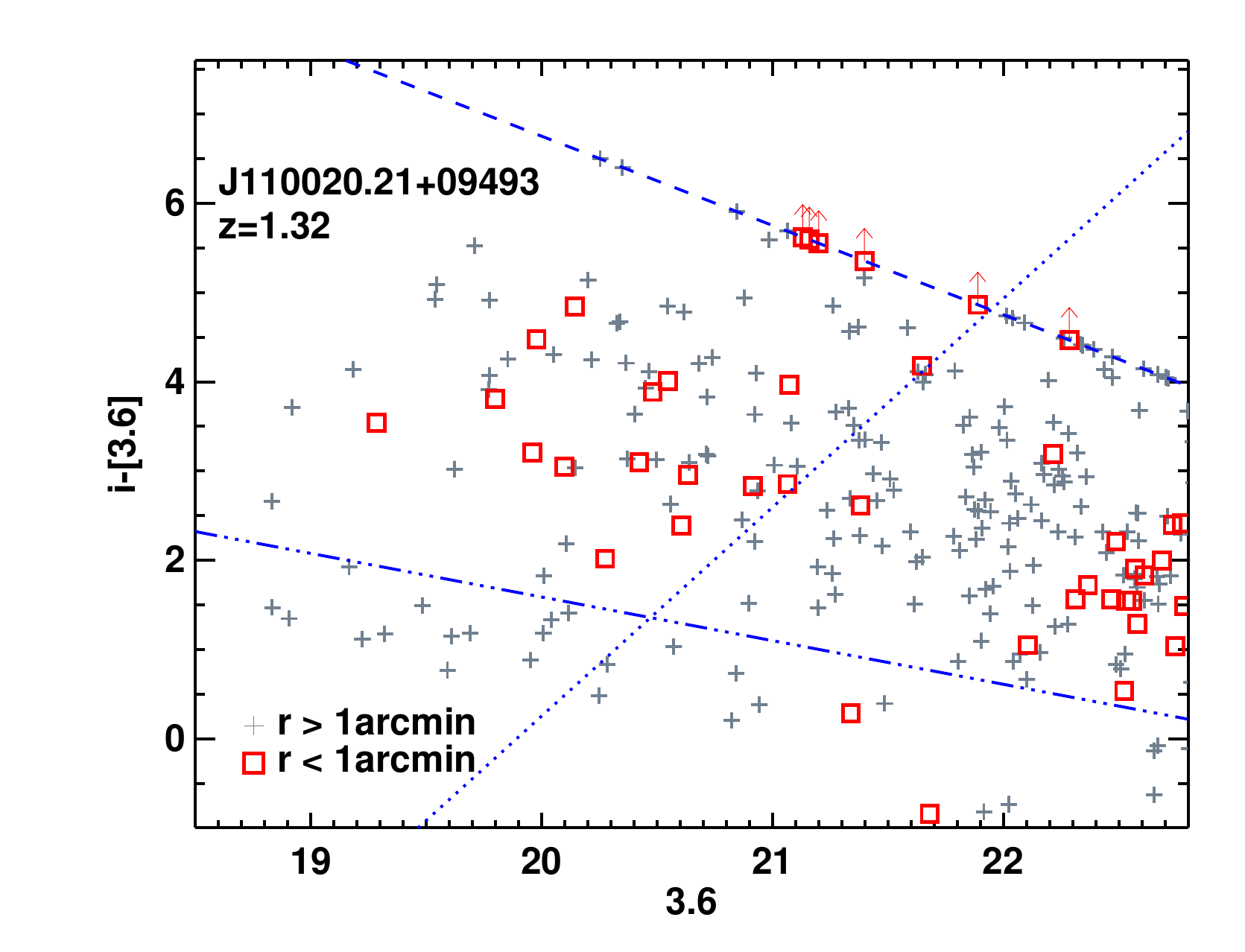}
\includegraphics[scale=0.3]{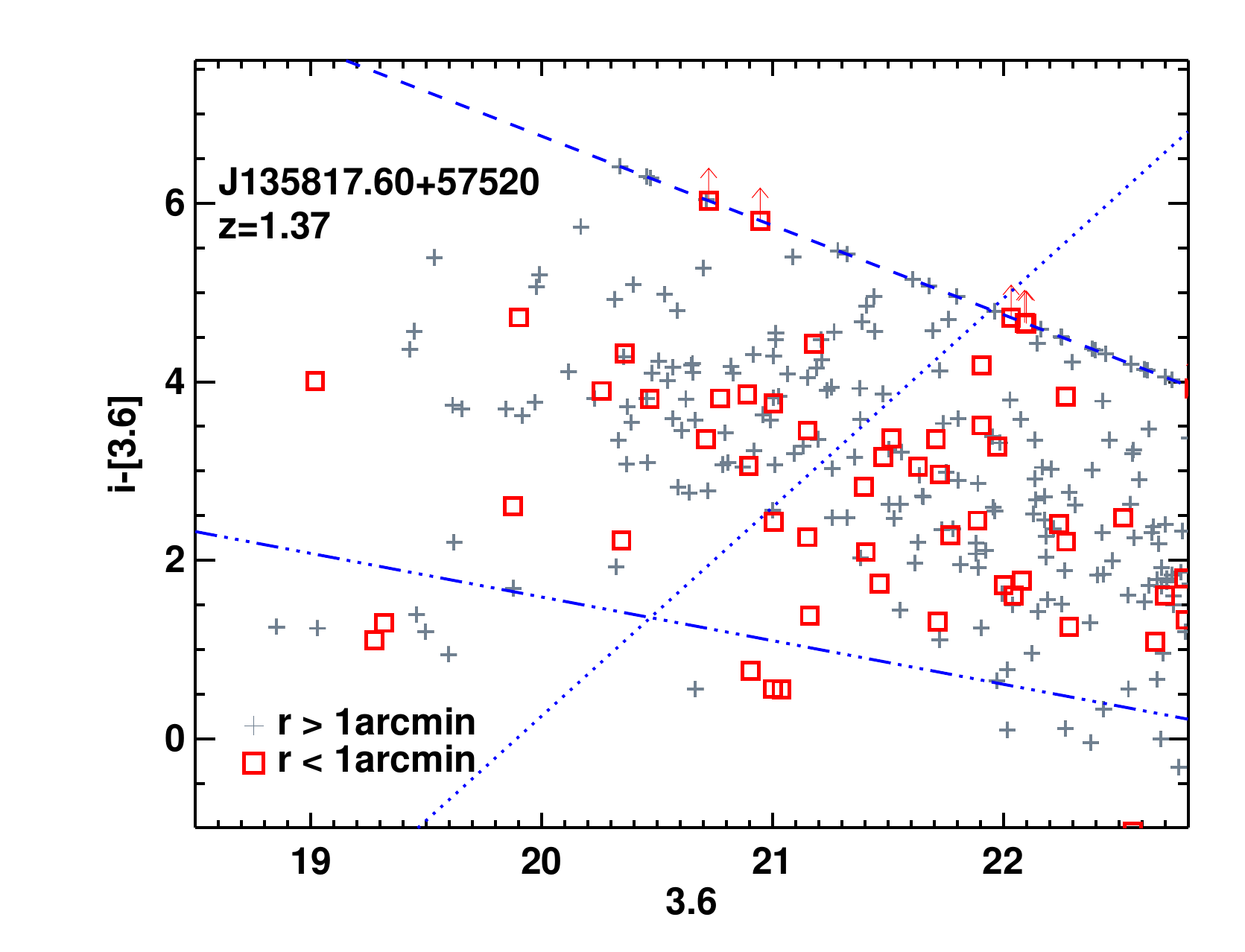} 
\includegraphics[scale=0.3]{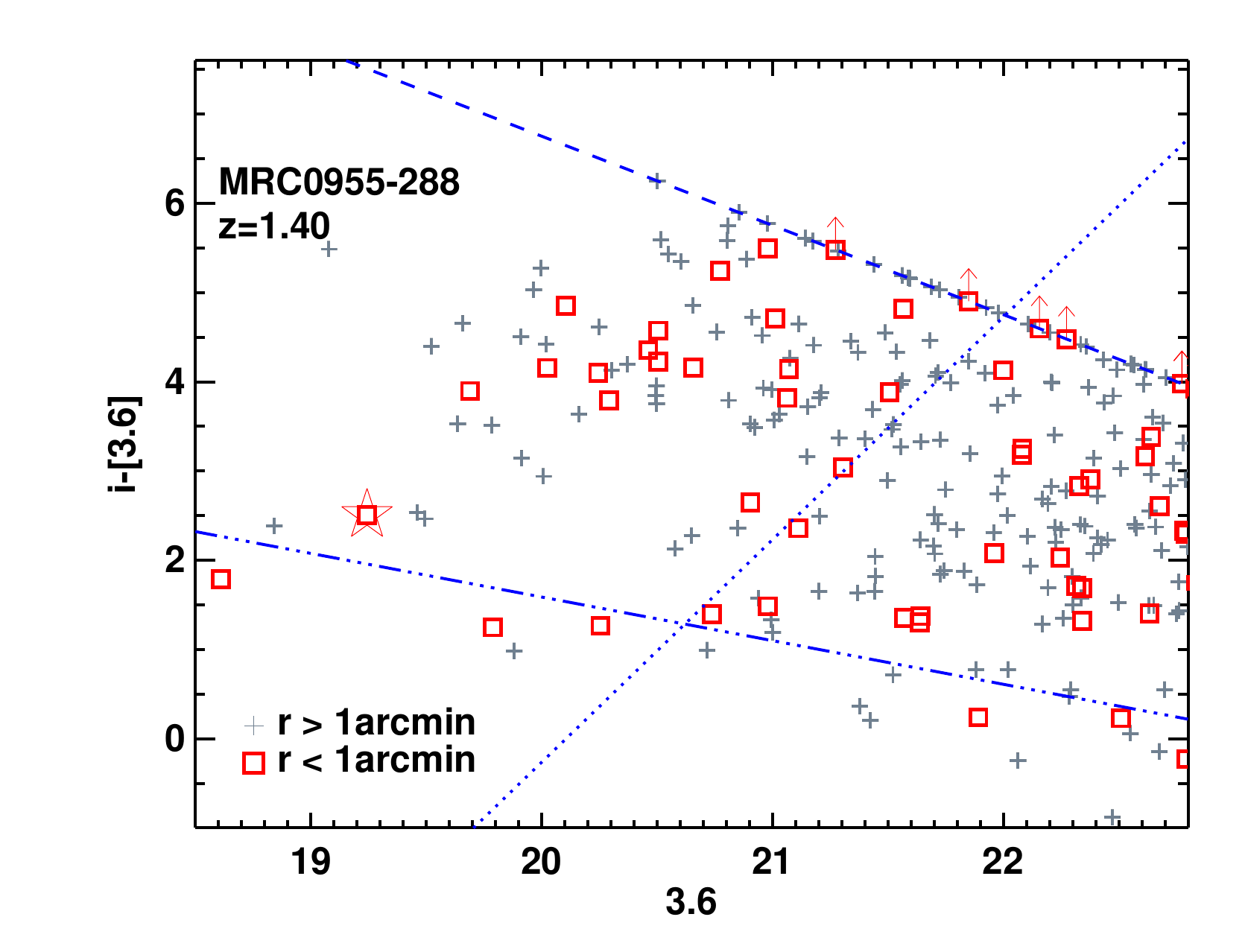}
\includegraphics[scale=0.3]{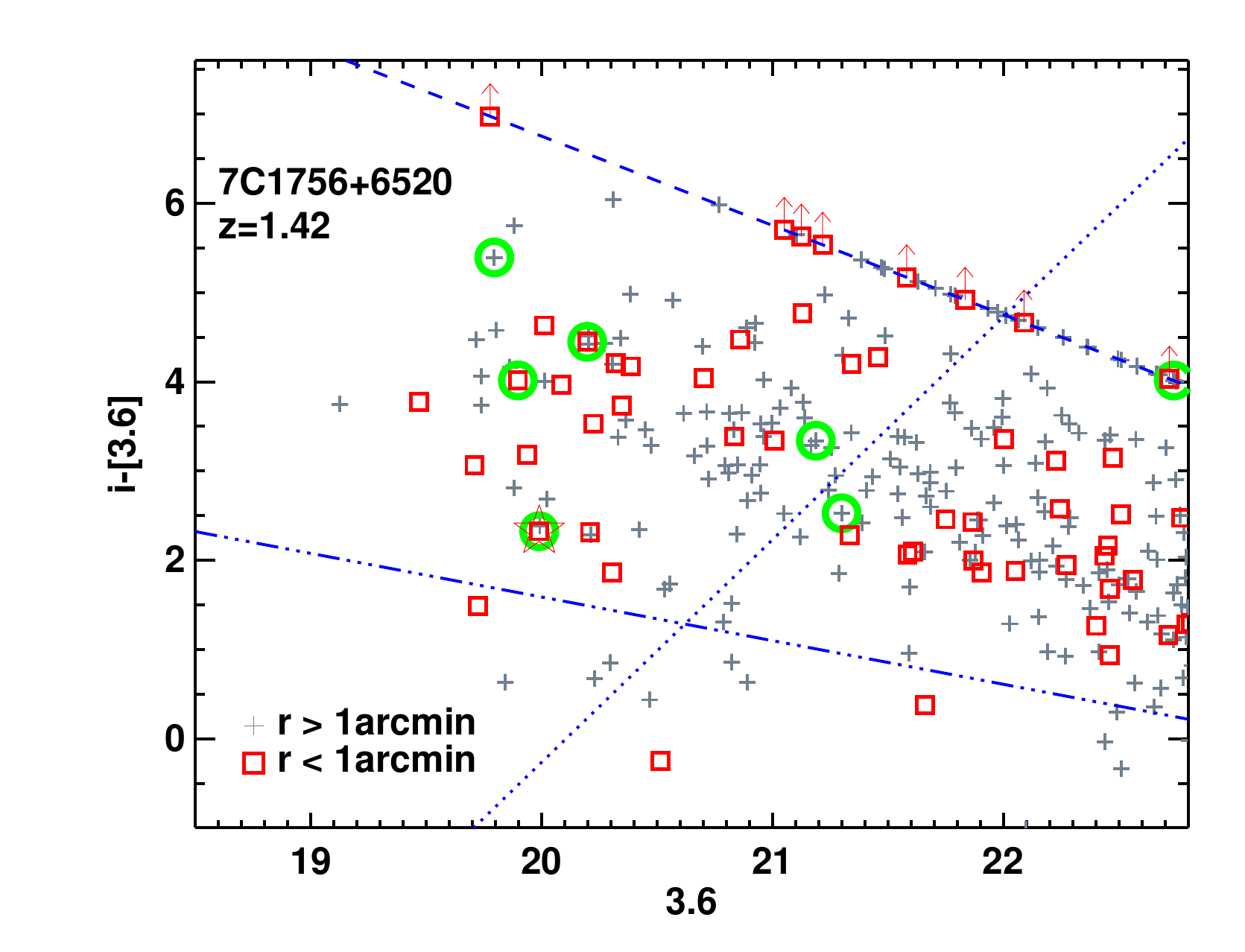}
\includegraphics[scale=0.3]{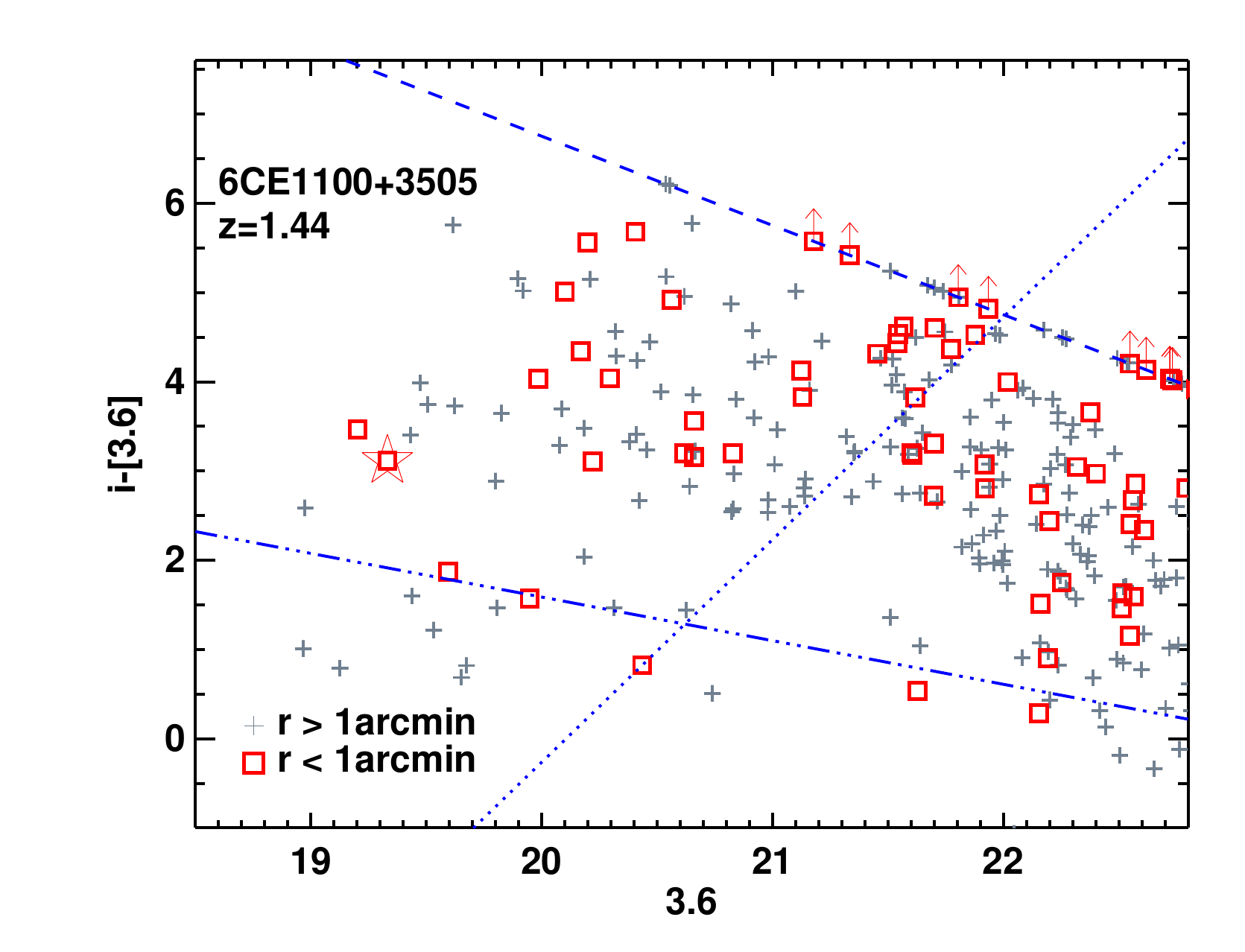} 
\includegraphics[scale=0.3]{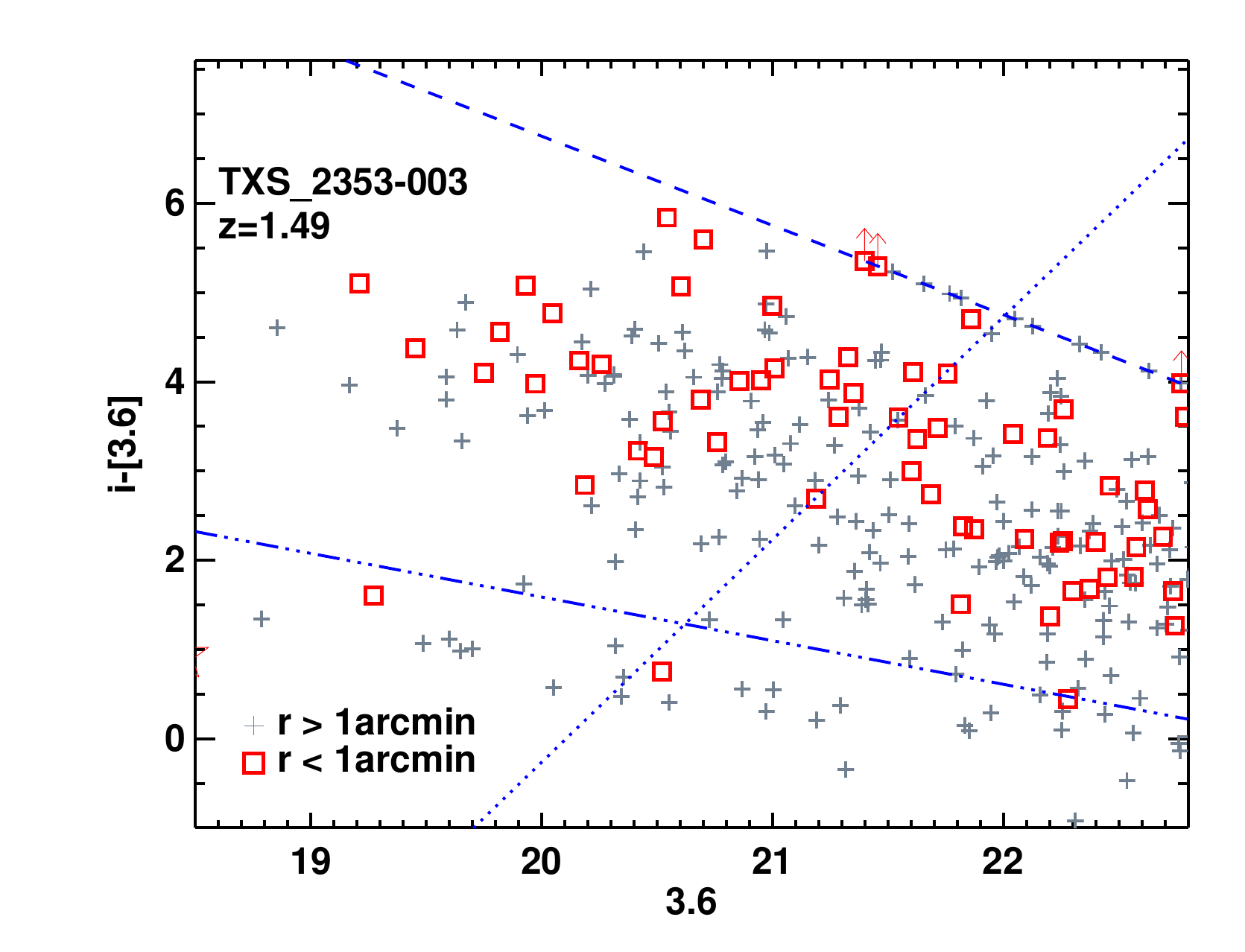}
\includegraphics[scale=0.3]{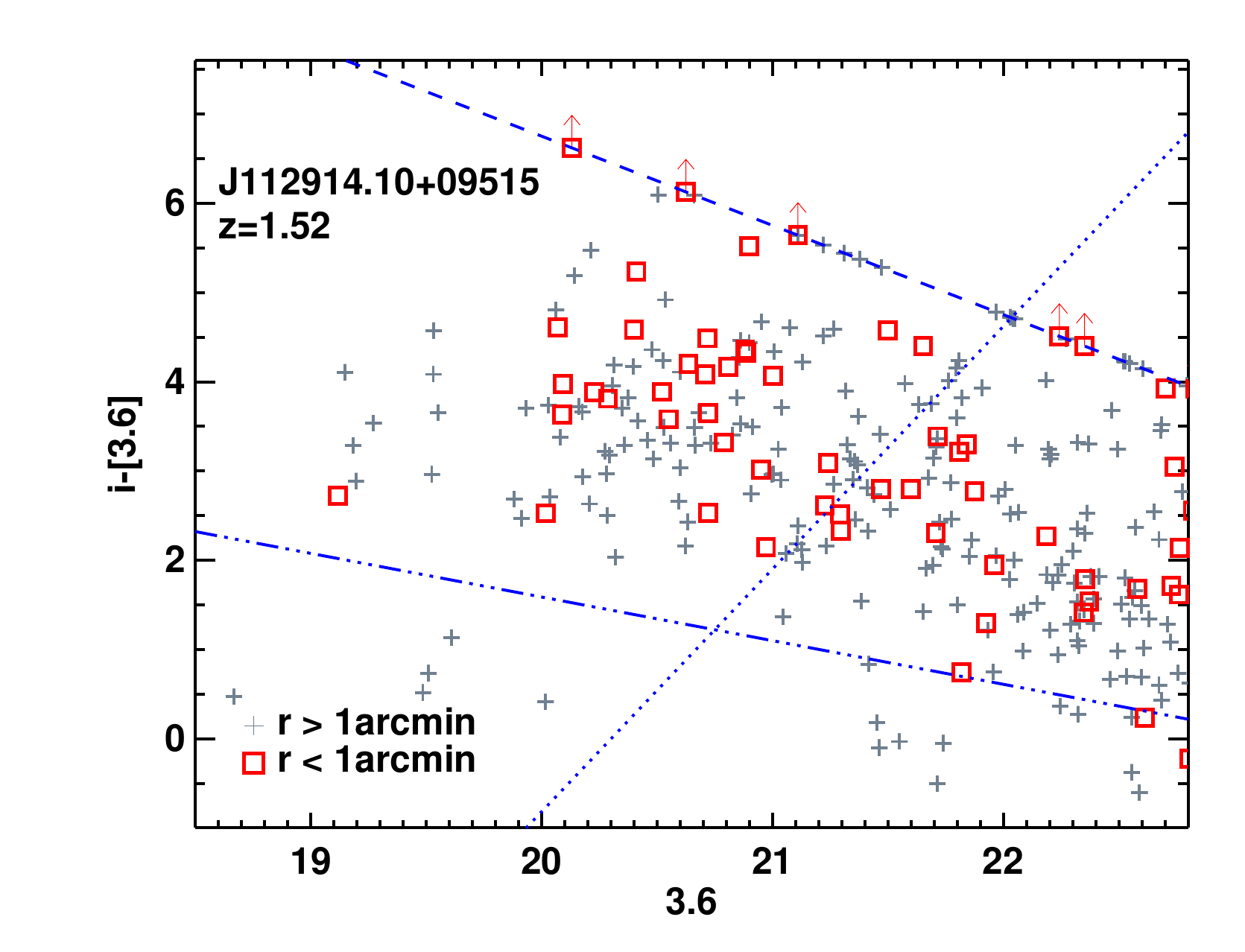}
\includegraphics[scale=0.3]{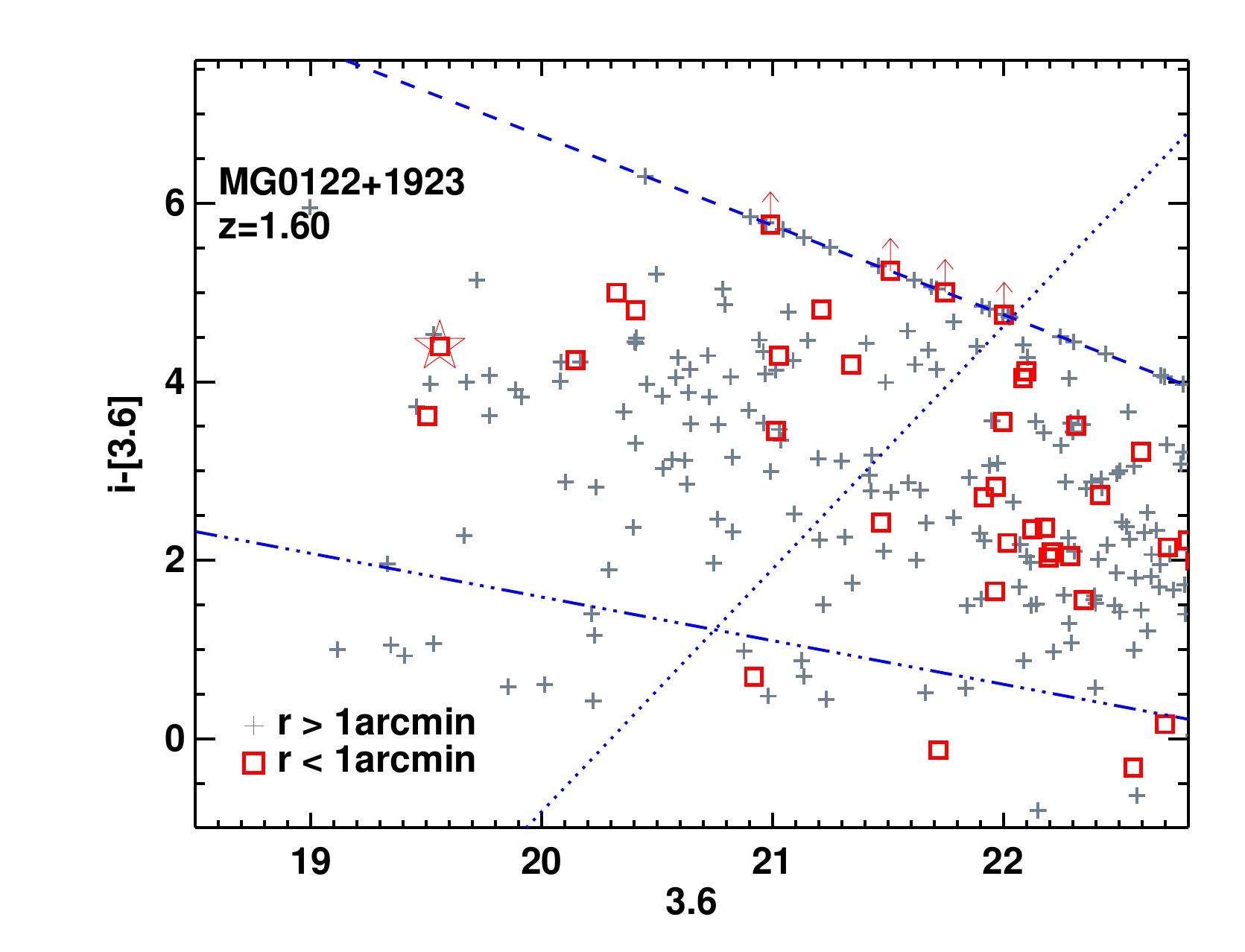}
\includegraphics[scale=0.3]{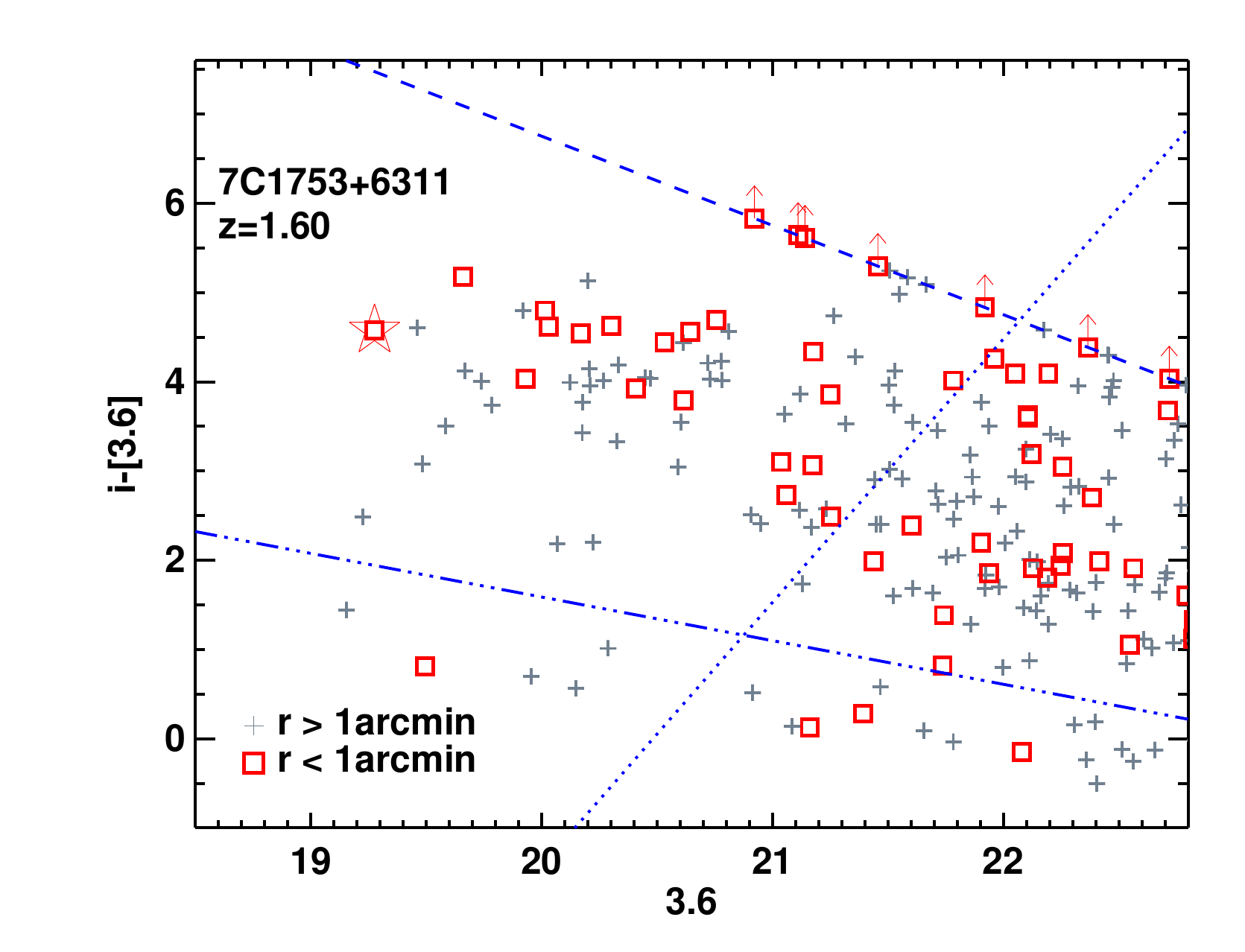}
\includegraphics[scale=0.3]{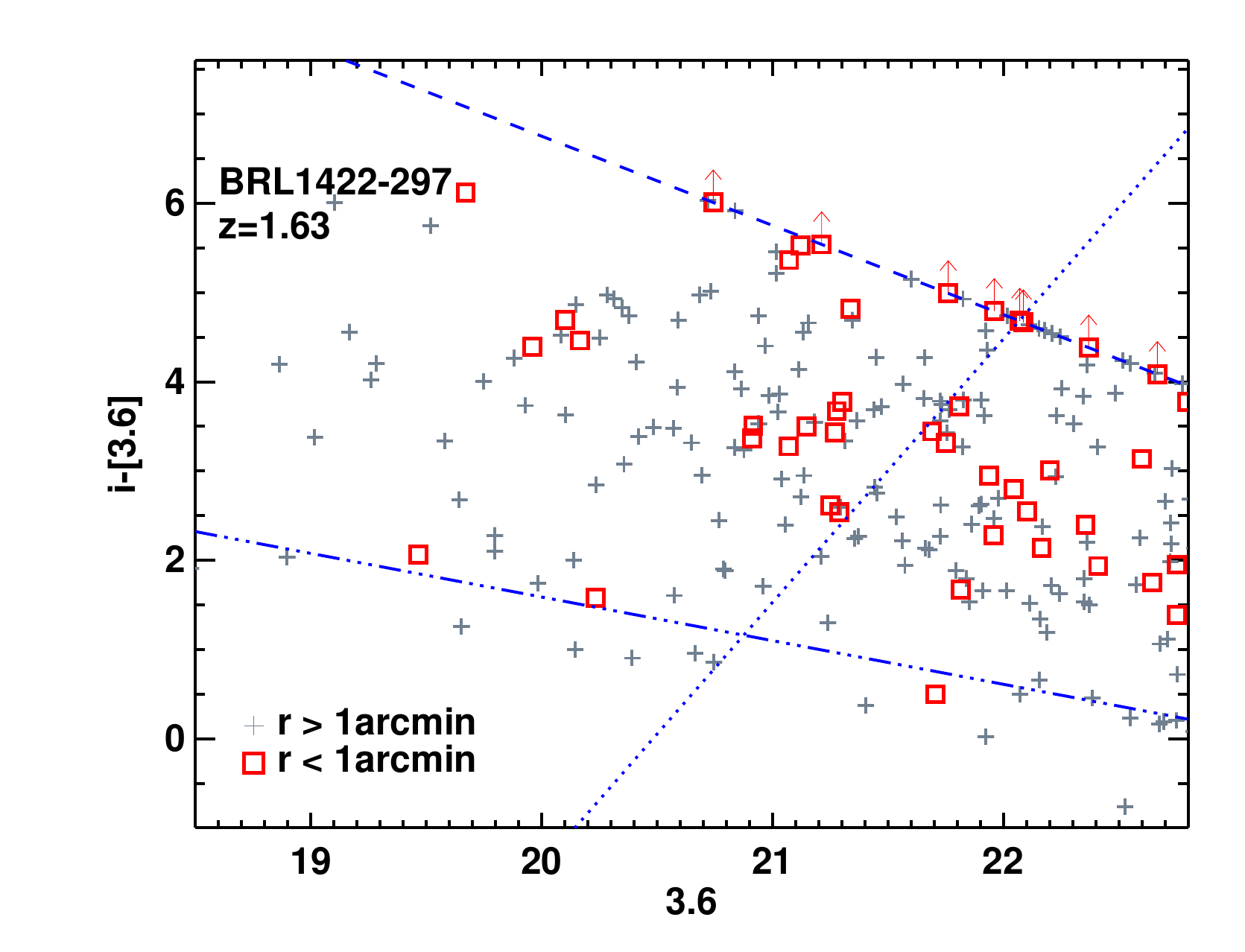}
\includegraphics[scale=0.3]{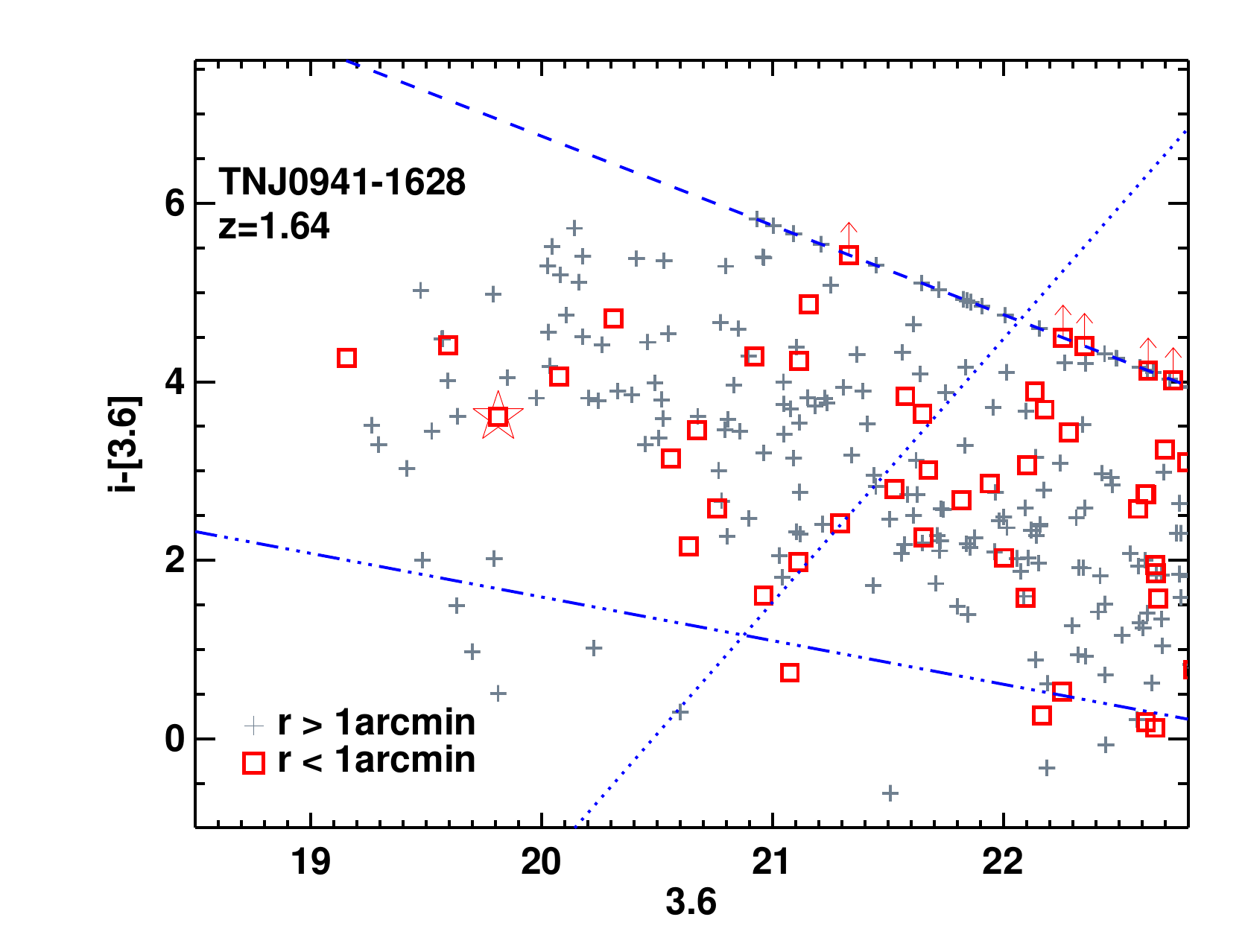}
\includegraphics[scale=0.3]{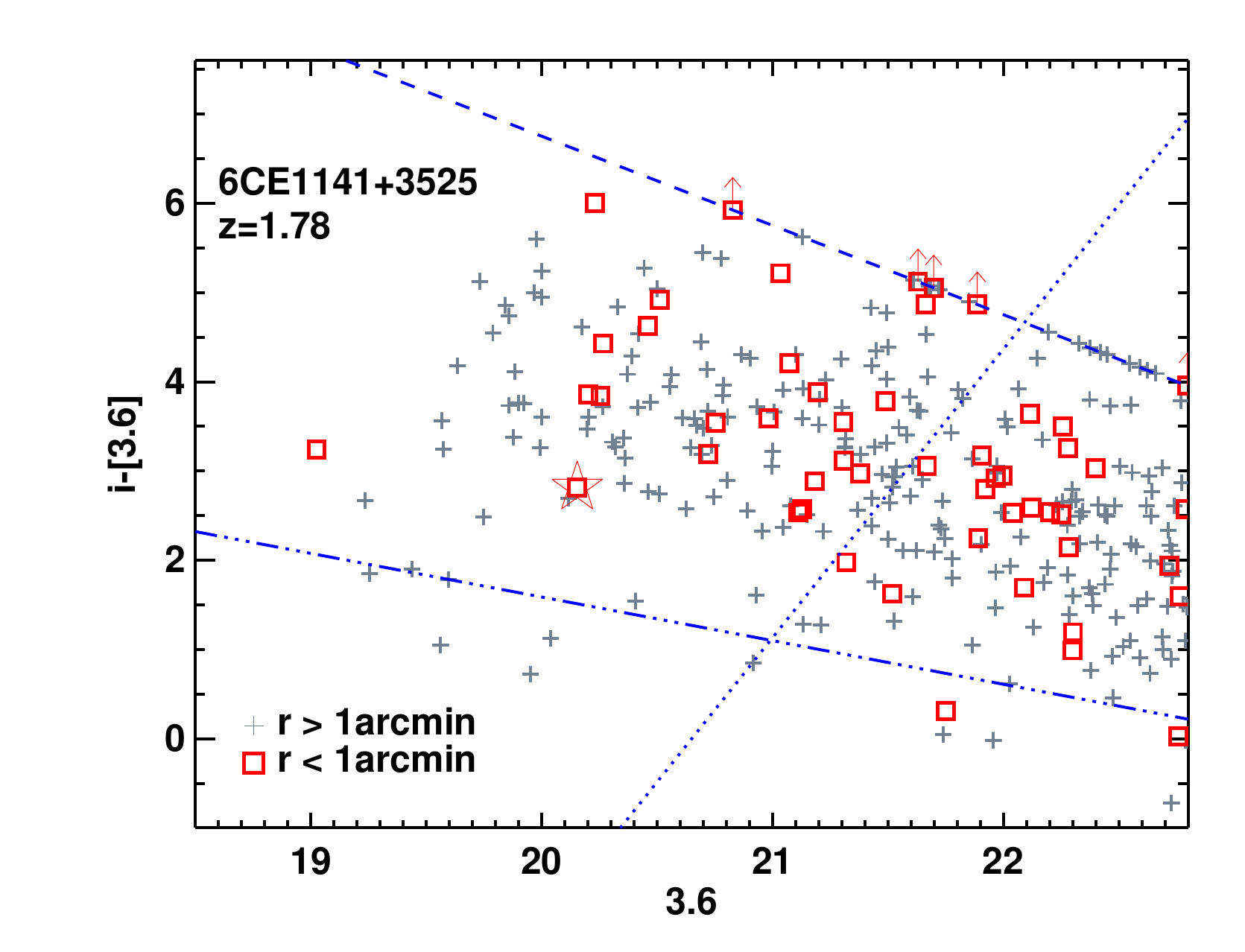} 
\includegraphics[scale=0.3]{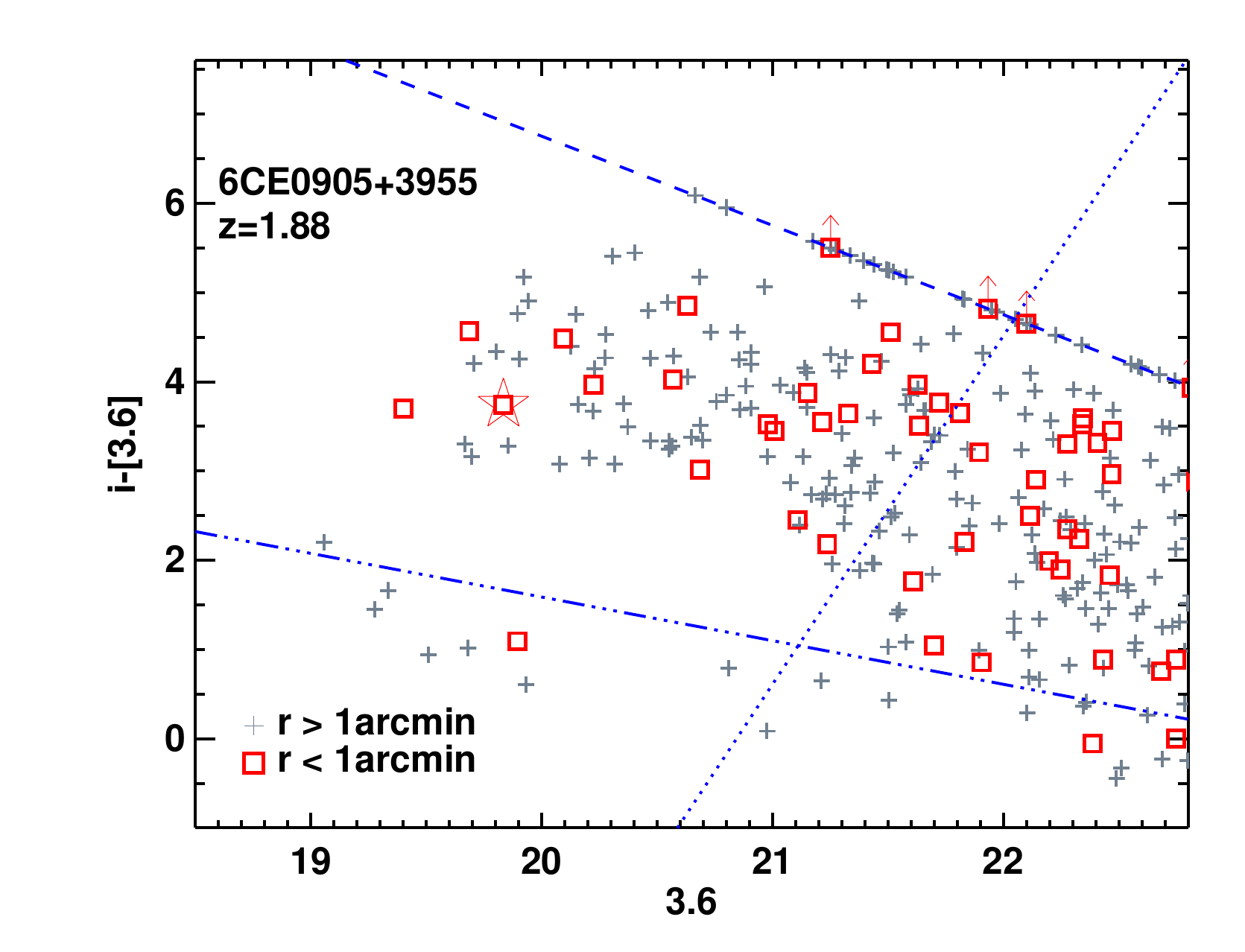}
\includegraphics[scale=0.3]{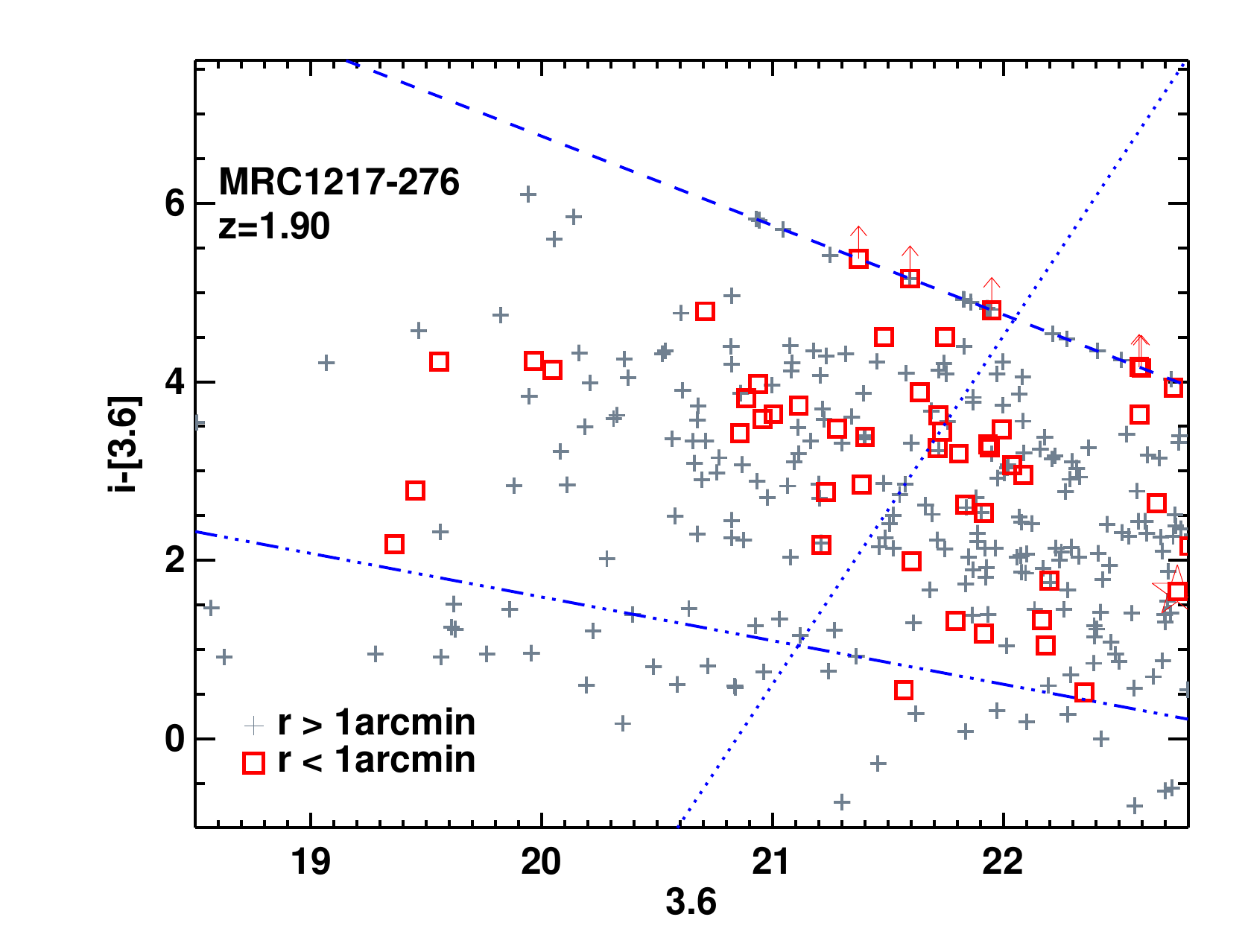}
\caption{$i'-[3.6]$ colour-magnitude diagrams. Red squares show sources within 1\,arcmin of the RLAGN. Grey plus symbols show those sources lying further than 1\,arcmin from the RLAGN, which are more likely to be contaminants. The blue dashed lines show our $1\sigma$ $i'$ median depth. Sources fainter than the $1\sigma$ median depth are set to $1\sigma_{med}$ and shown as lower limits. The dotted blue lines show a $M_*>10^{10.5}$\,M$_{\odot}$ mass cut and the dash-triple dotted blue lines show the cut used to remove low redshift contaminants. The RLAGN is shown by a red star. In some diagrams, the AGN is too bright to fit on the scale and so is not shown. For 7C1756$+$6520 confirmed cluster members from \citet{Galametz2010a} are shown by green circles.}
\label{fig:AppendCMDs}
\end{figure*}
\begin{figure*}
\centering
\includegraphics[scale=0.3]{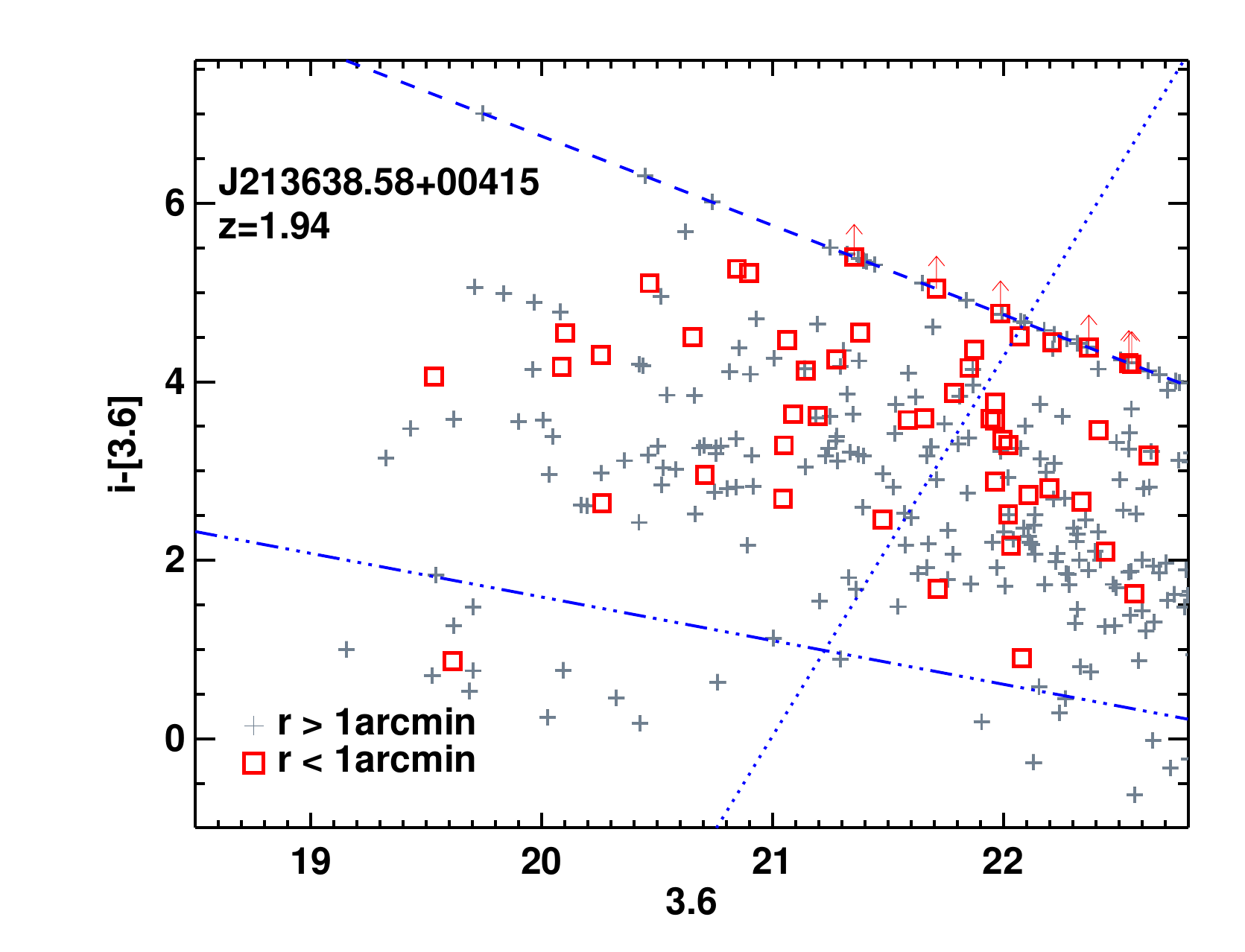}
\includegraphics[scale=0.3]{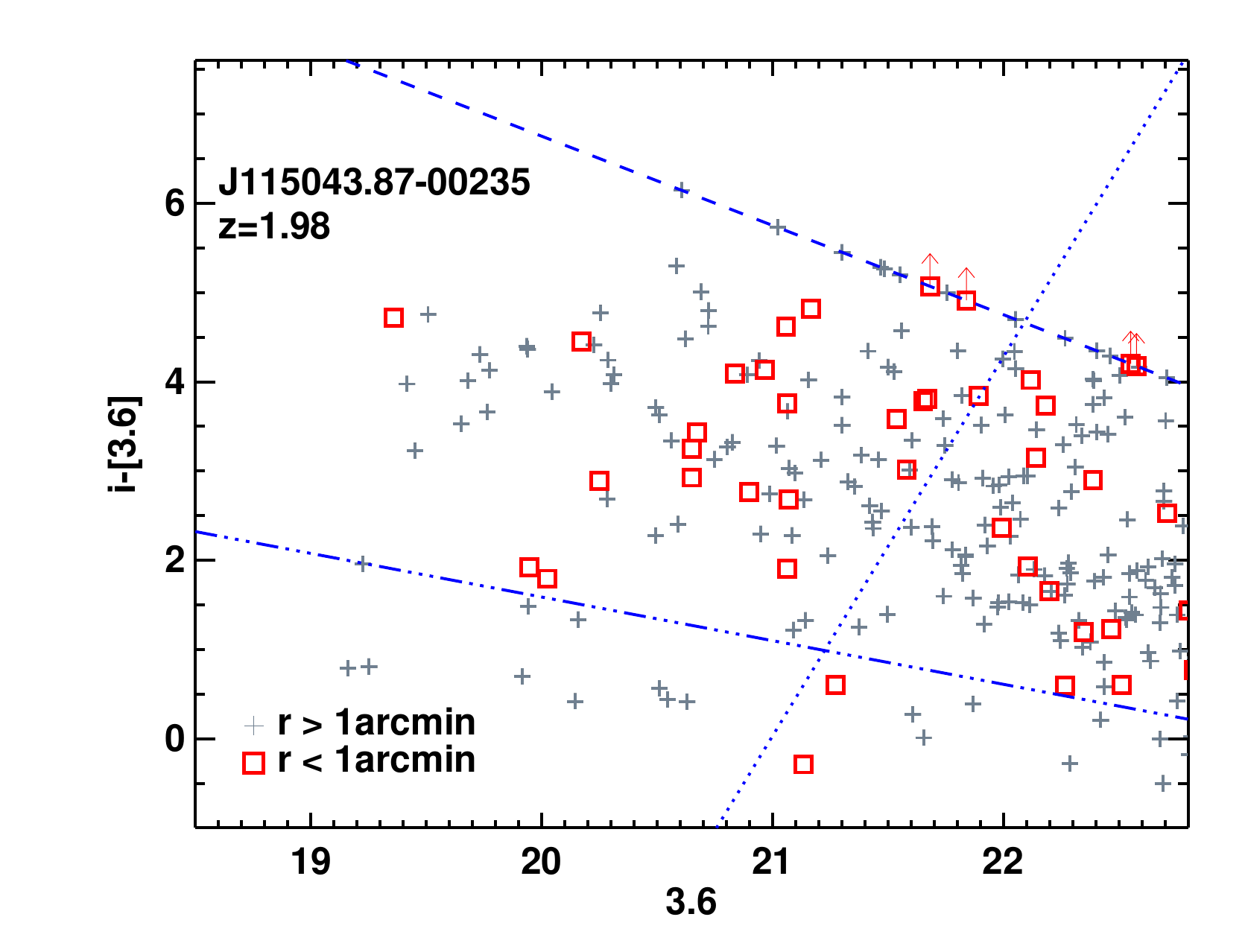}
\includegraphics[scale=0.3]{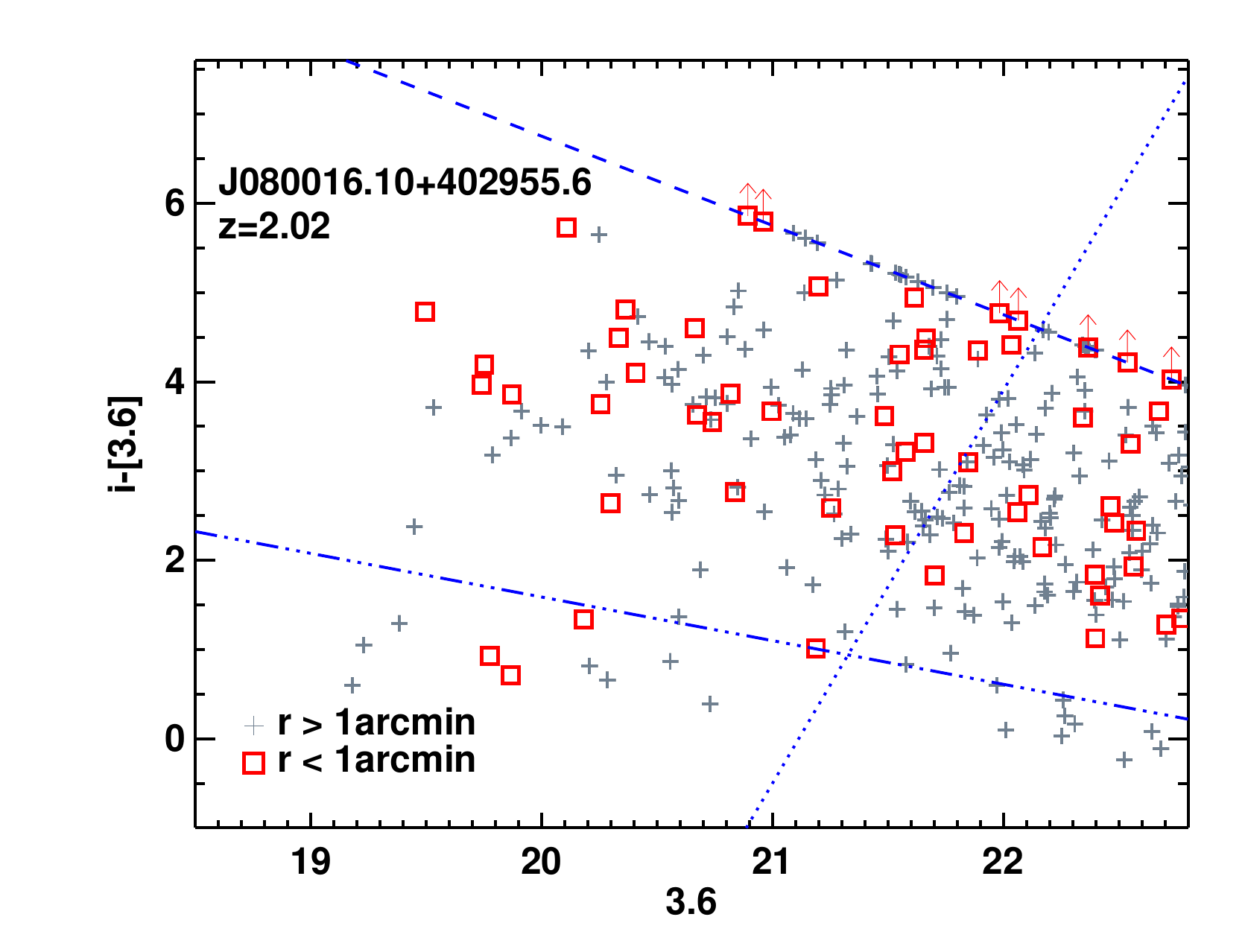}
\includegraphics[scale=0.3]{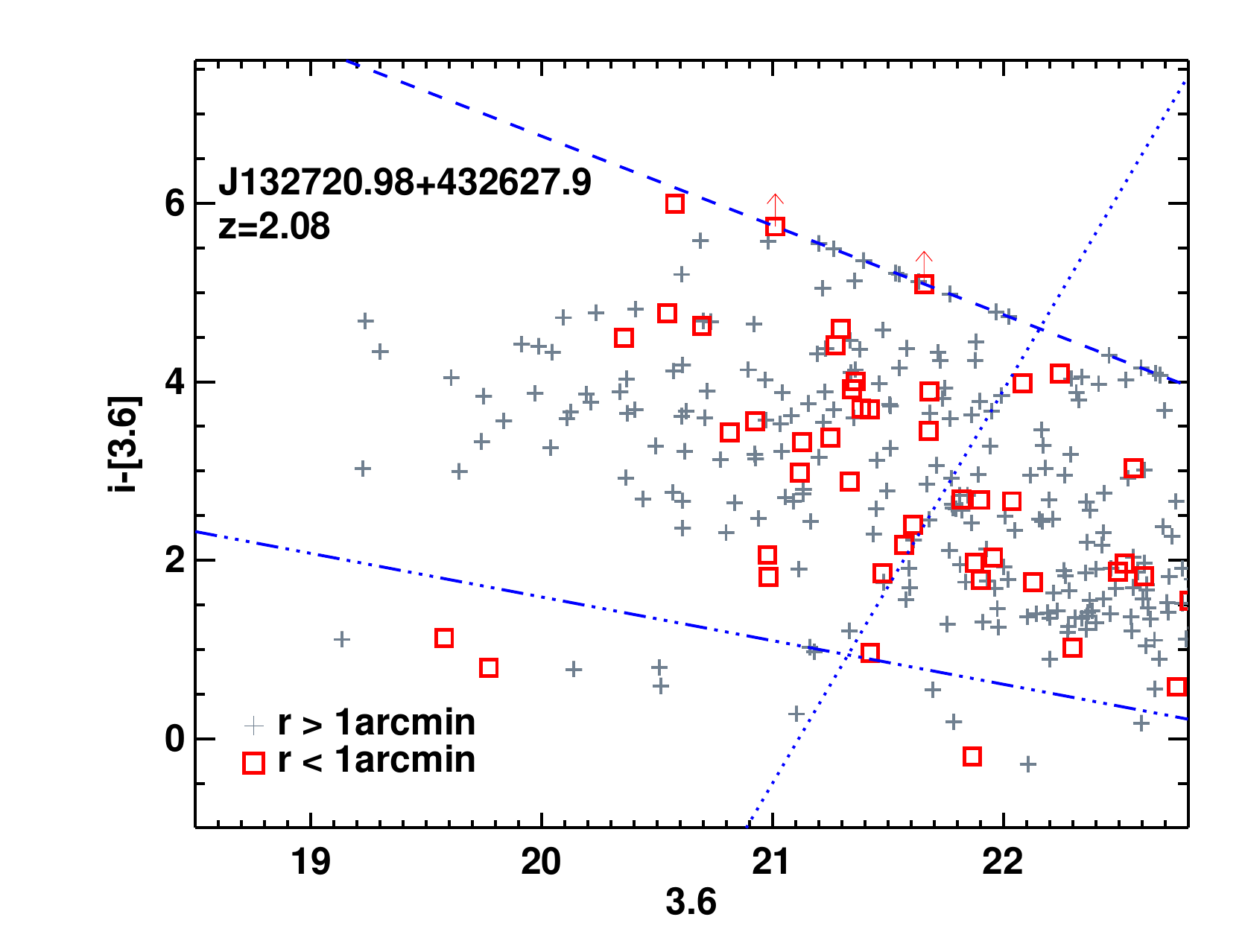}
\includegraphics[scale=0.3]{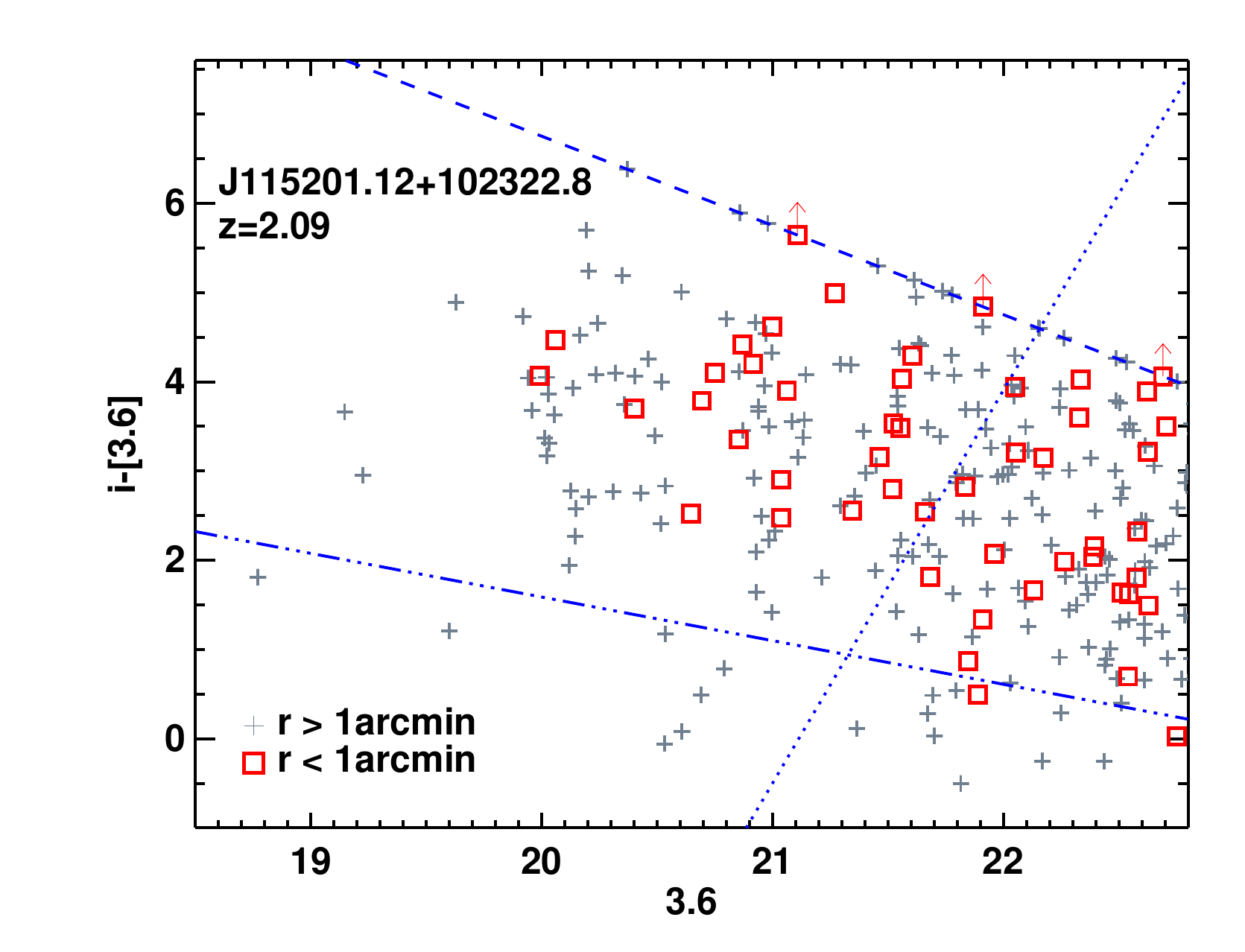}
\includegraphics[scale=0.3]{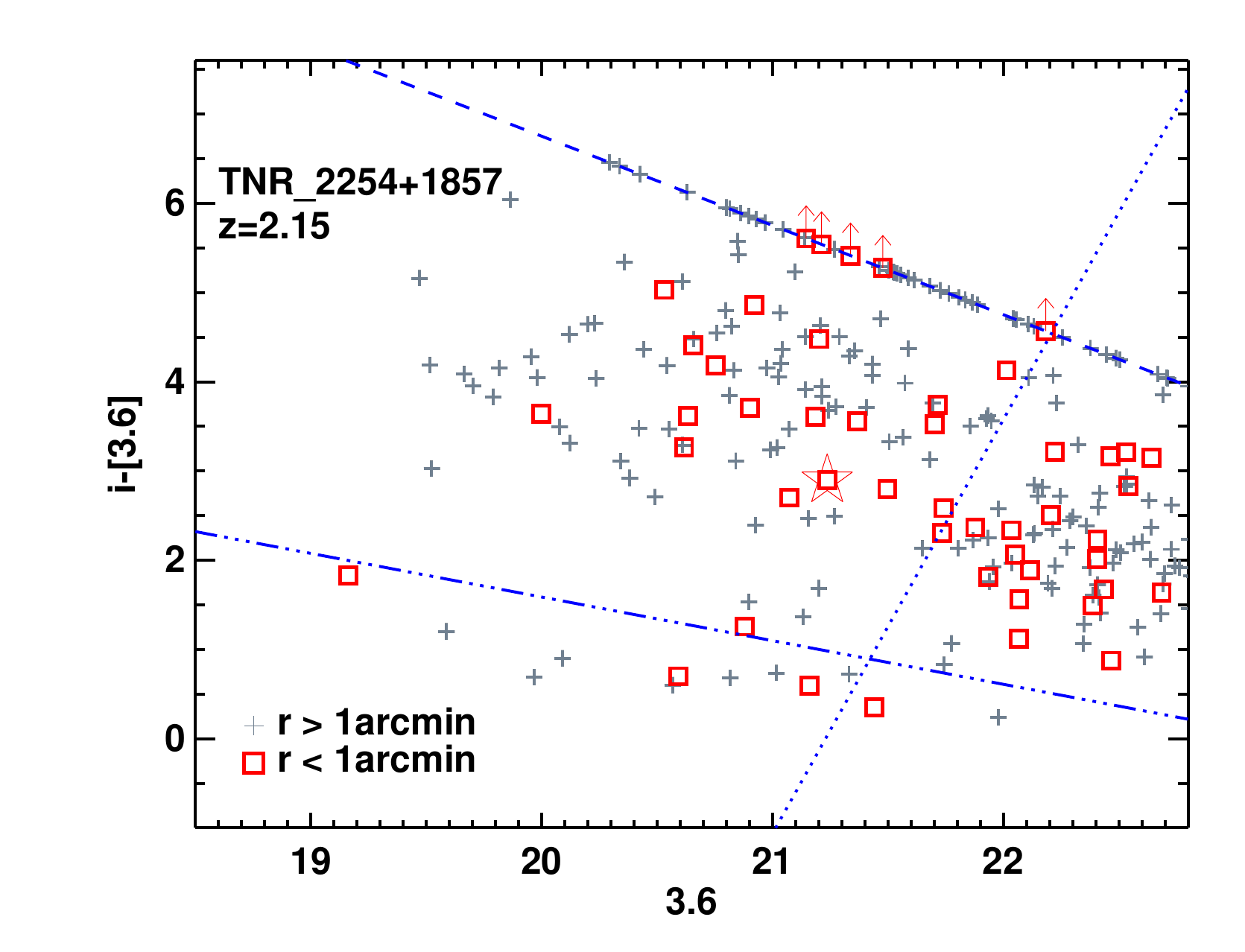}
\includegraphics[scale=0.3]{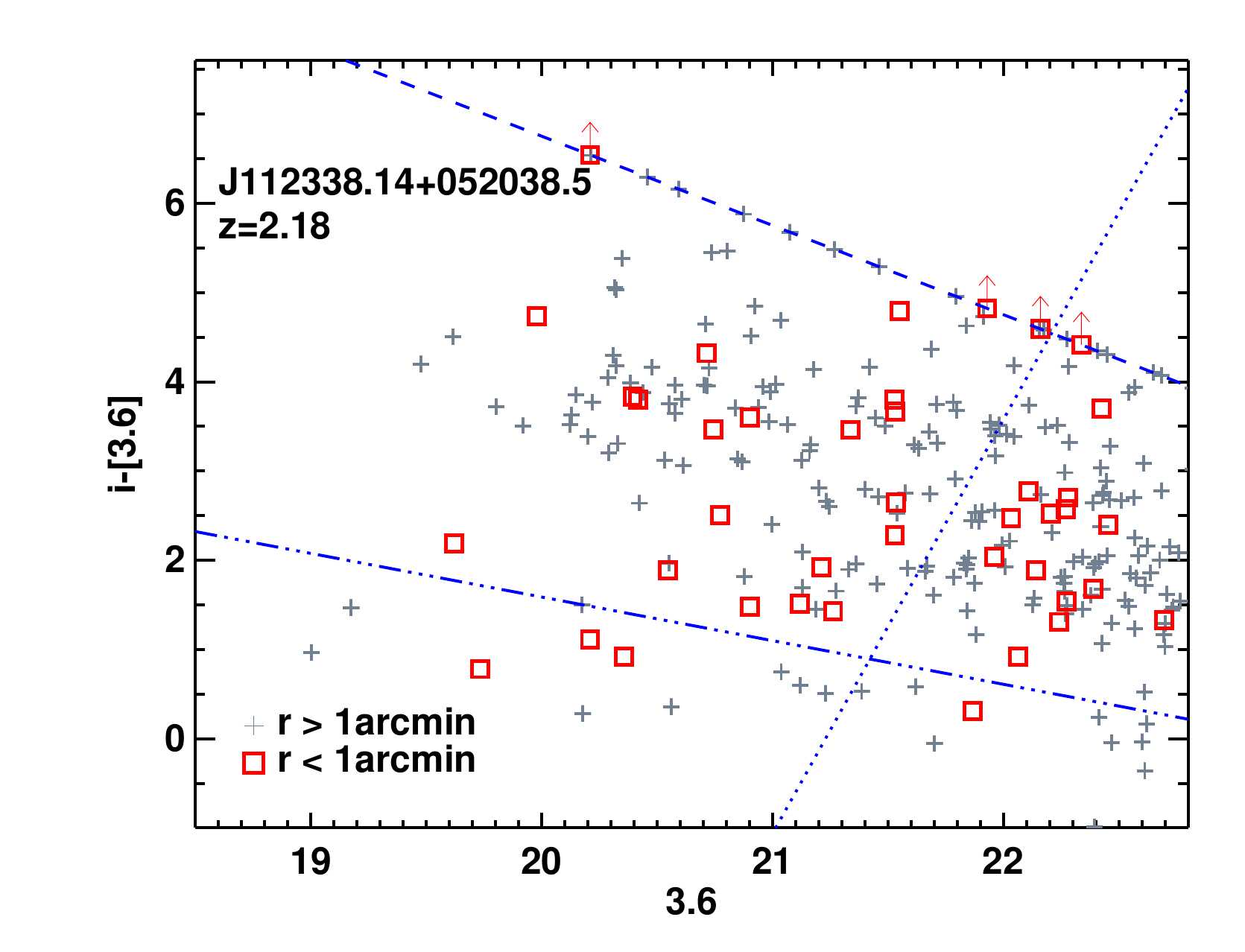}
\includegraphics[scale=0.3]{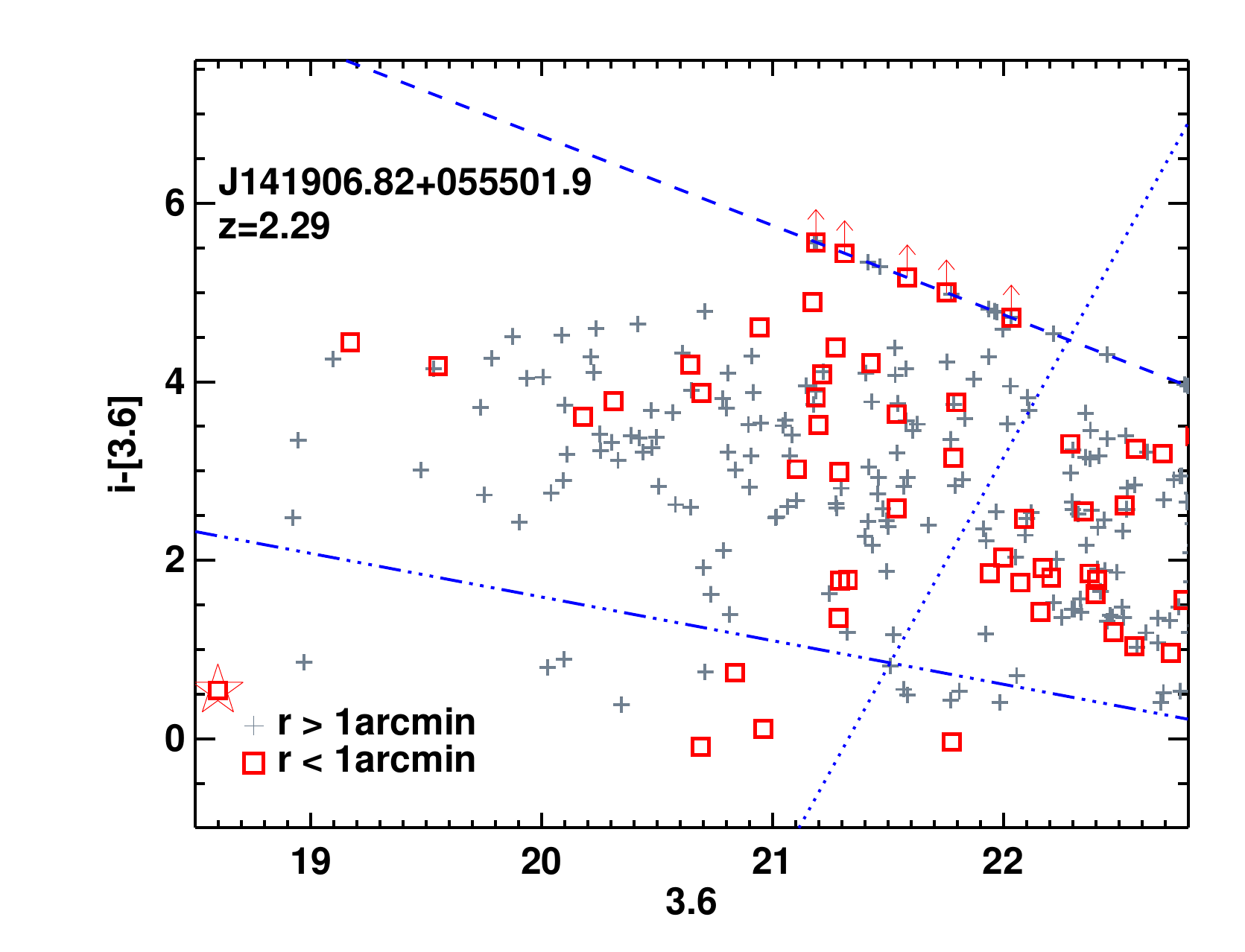}
\includegraphics[scale=0.3]{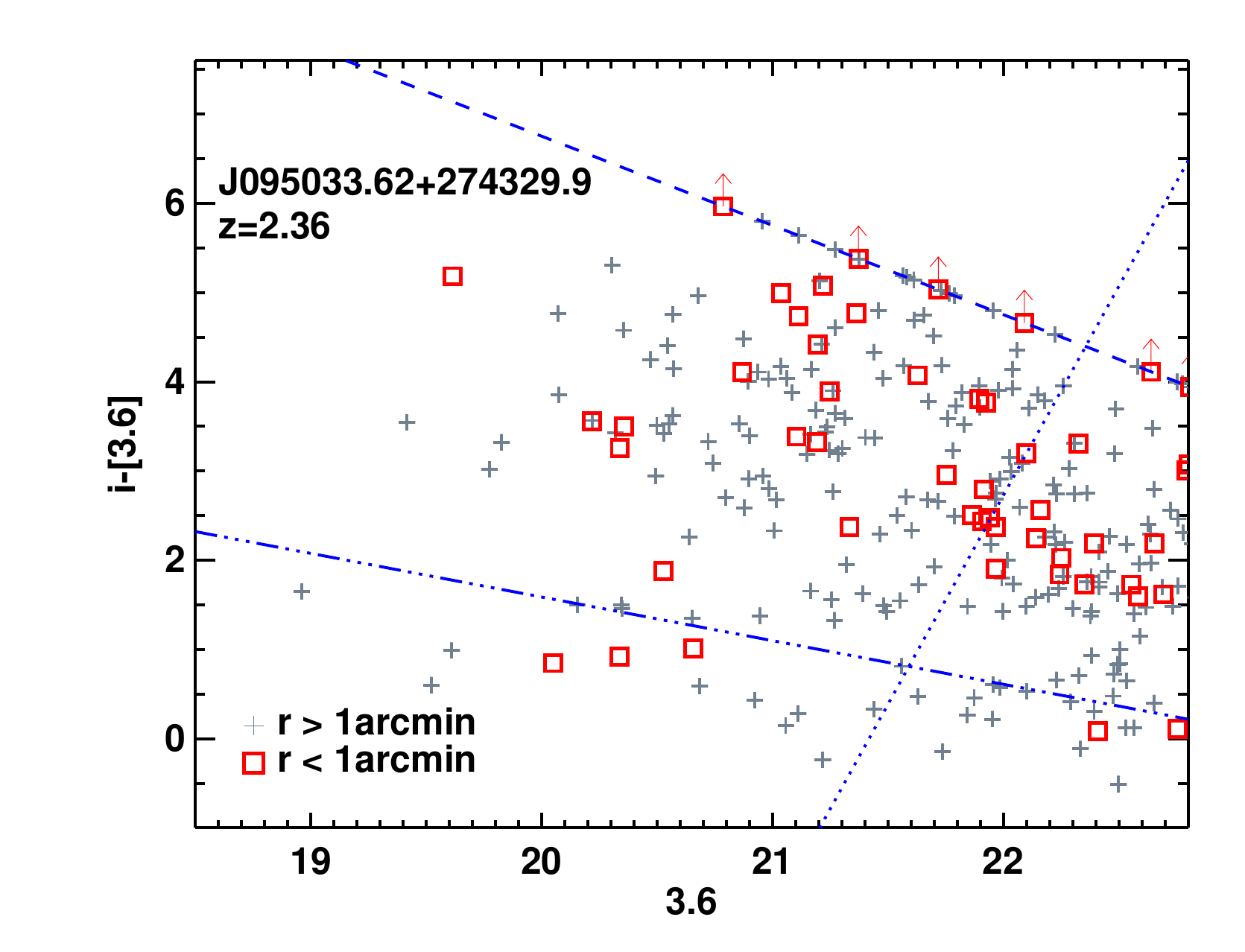}
\includegraphics[scale=0.3]{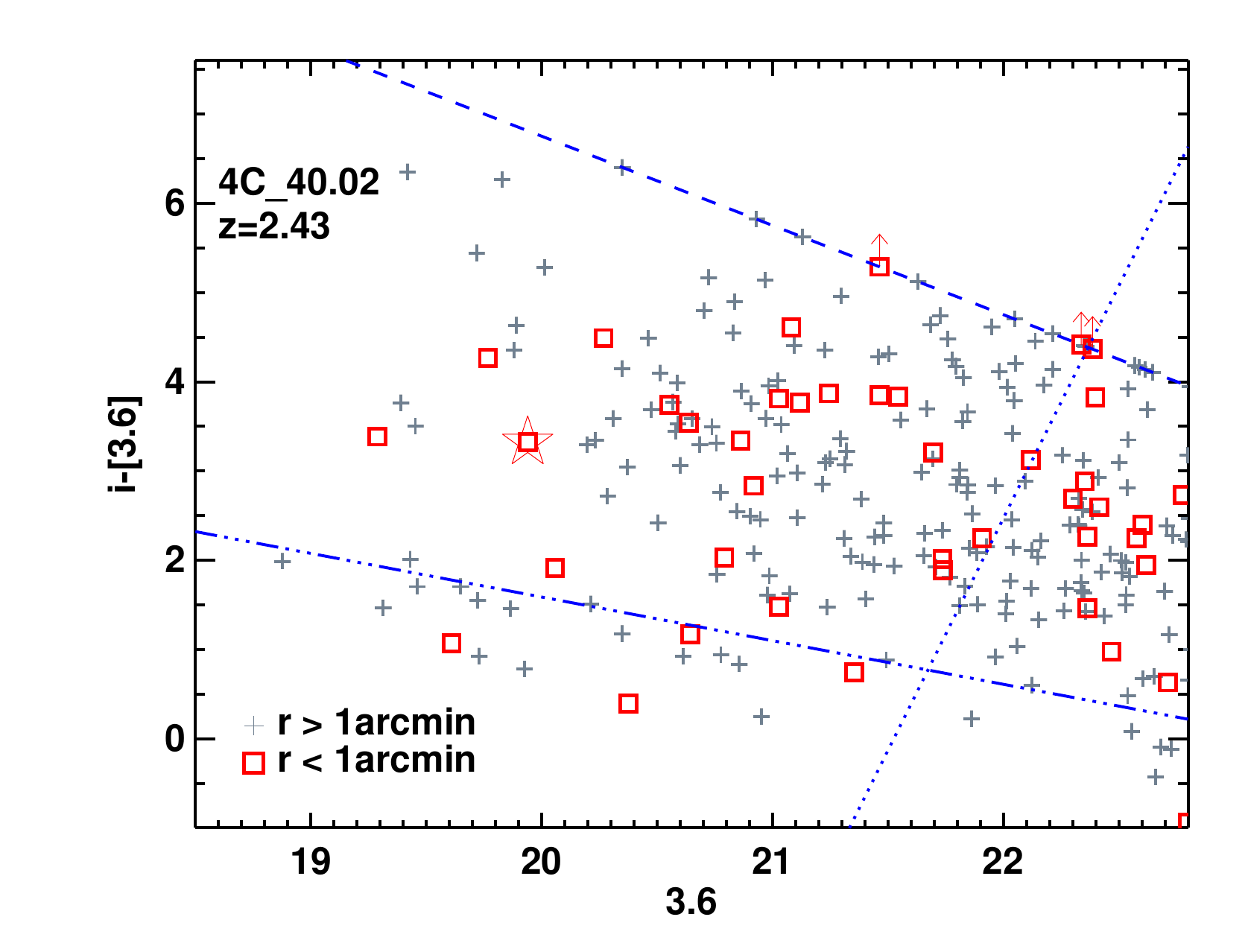}
\includegraphics[scale=0.3]{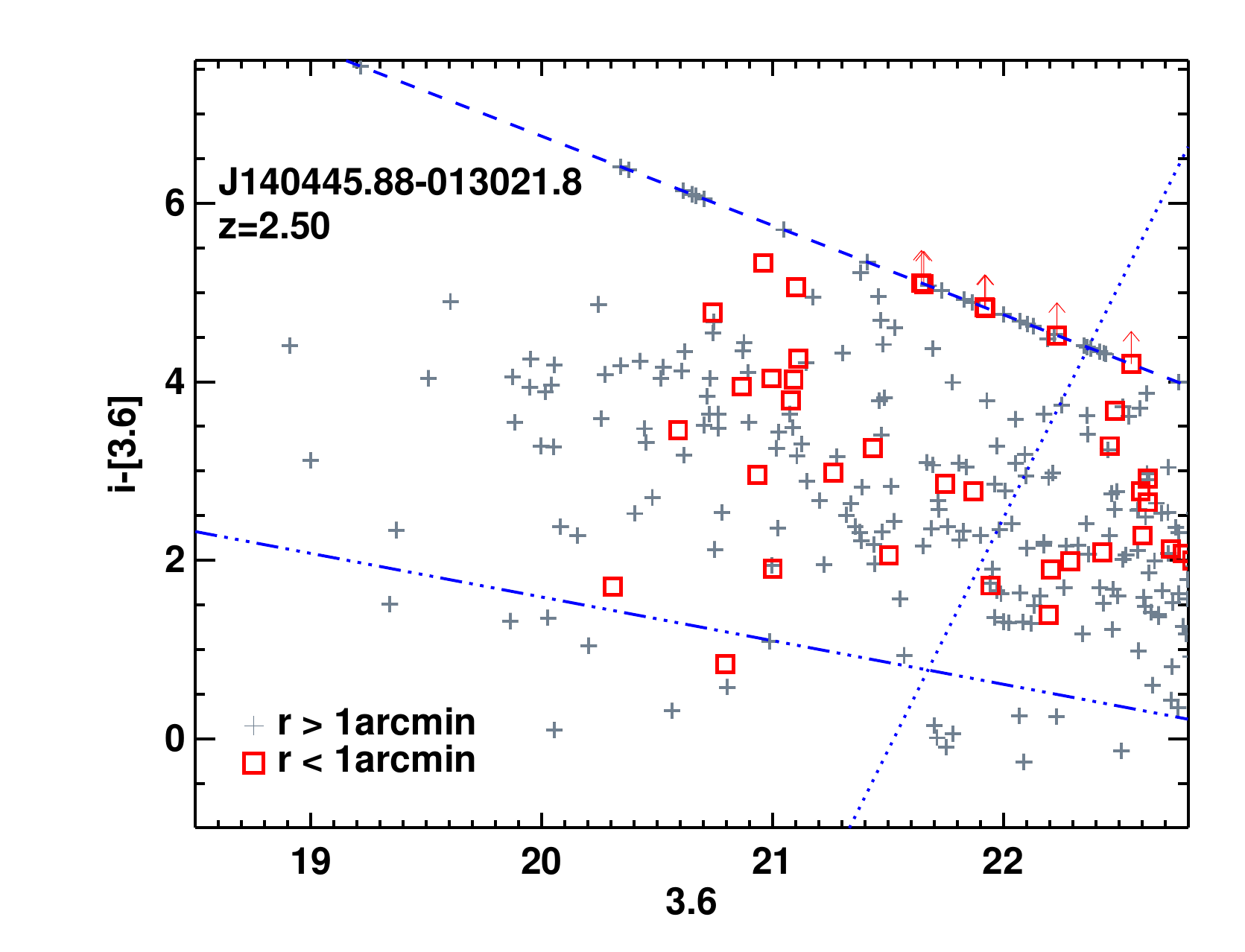}
\includegraphics[scale=0.3]{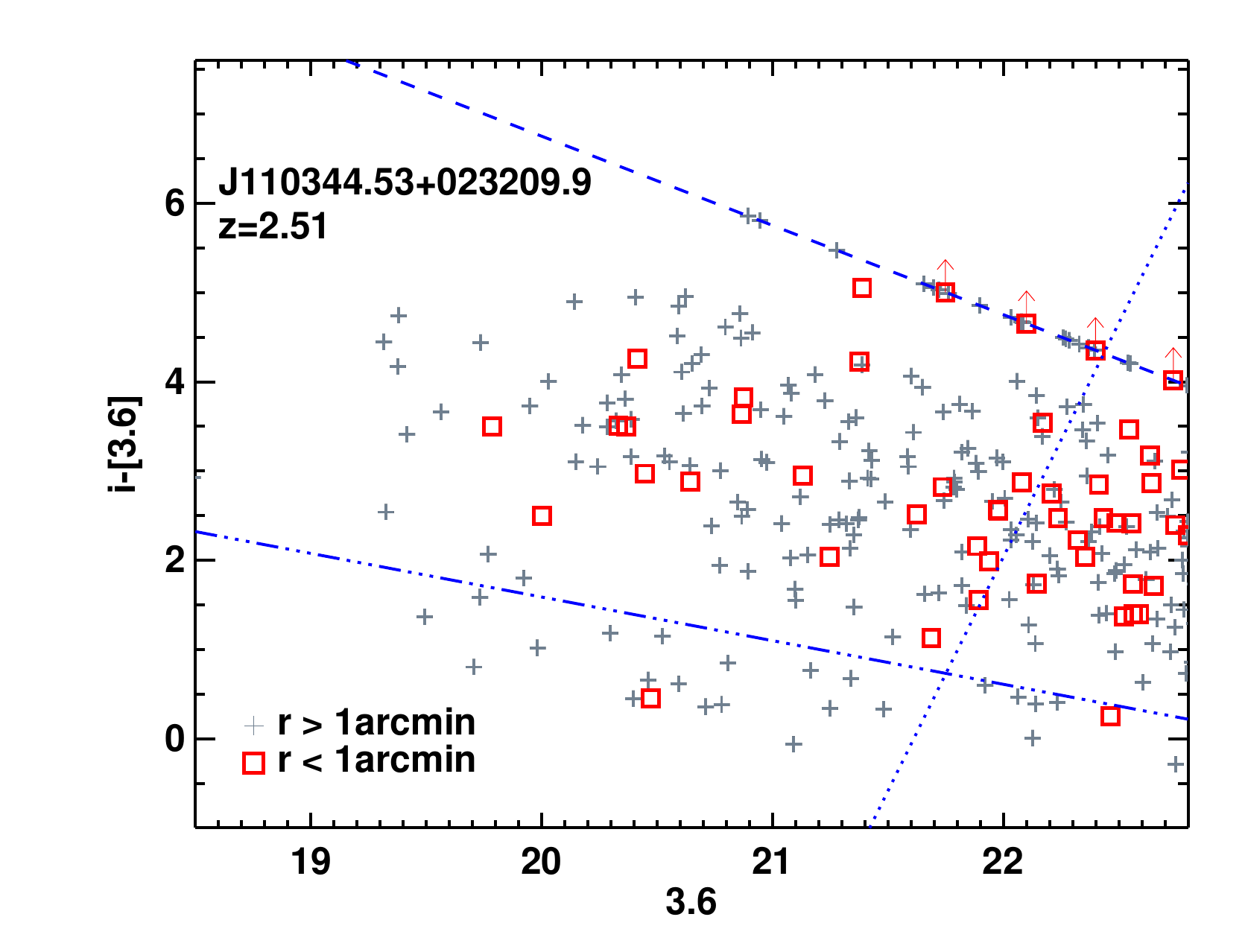}
\includegraphics[scale=0.3]{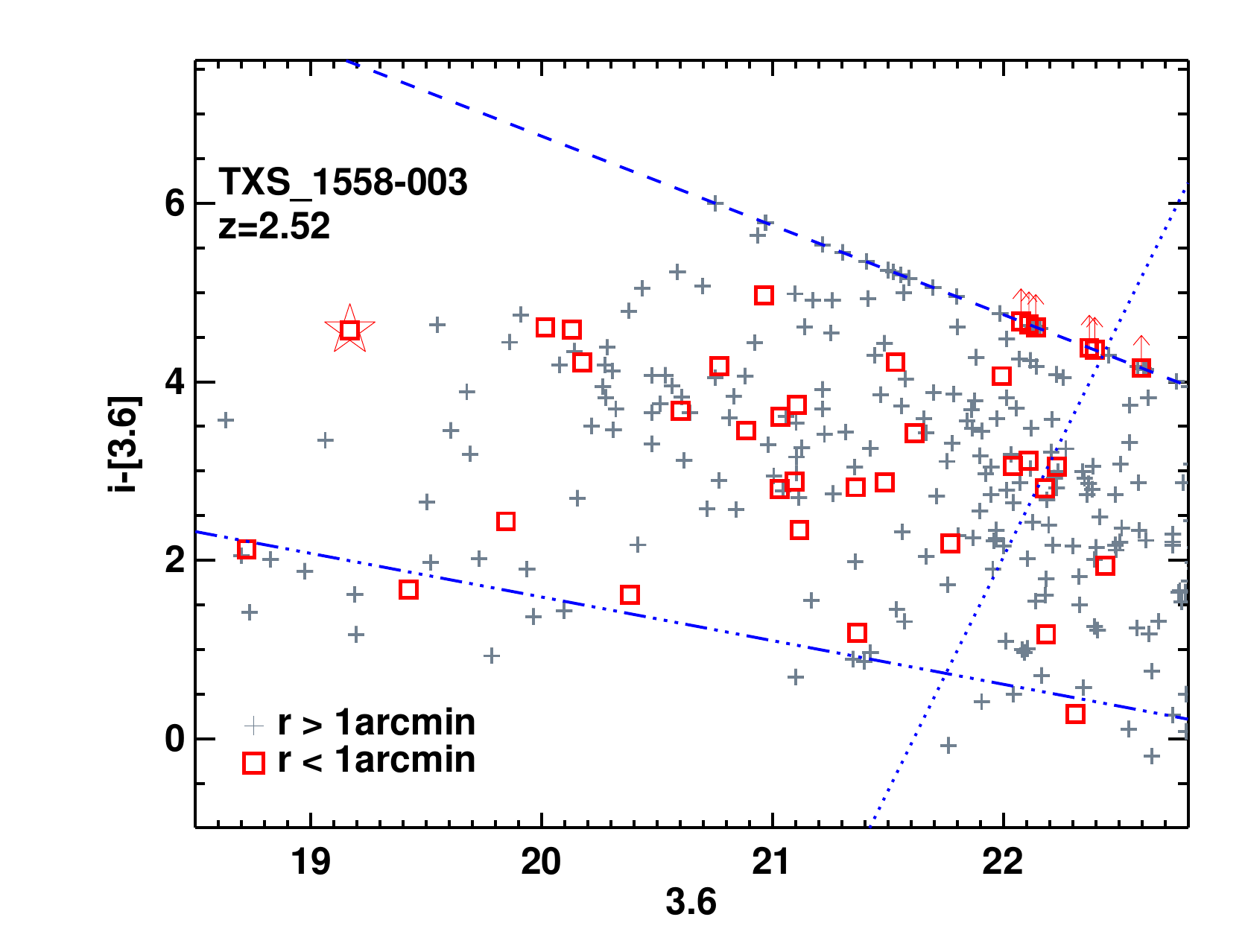}
\includegraphics[scale=0.3]{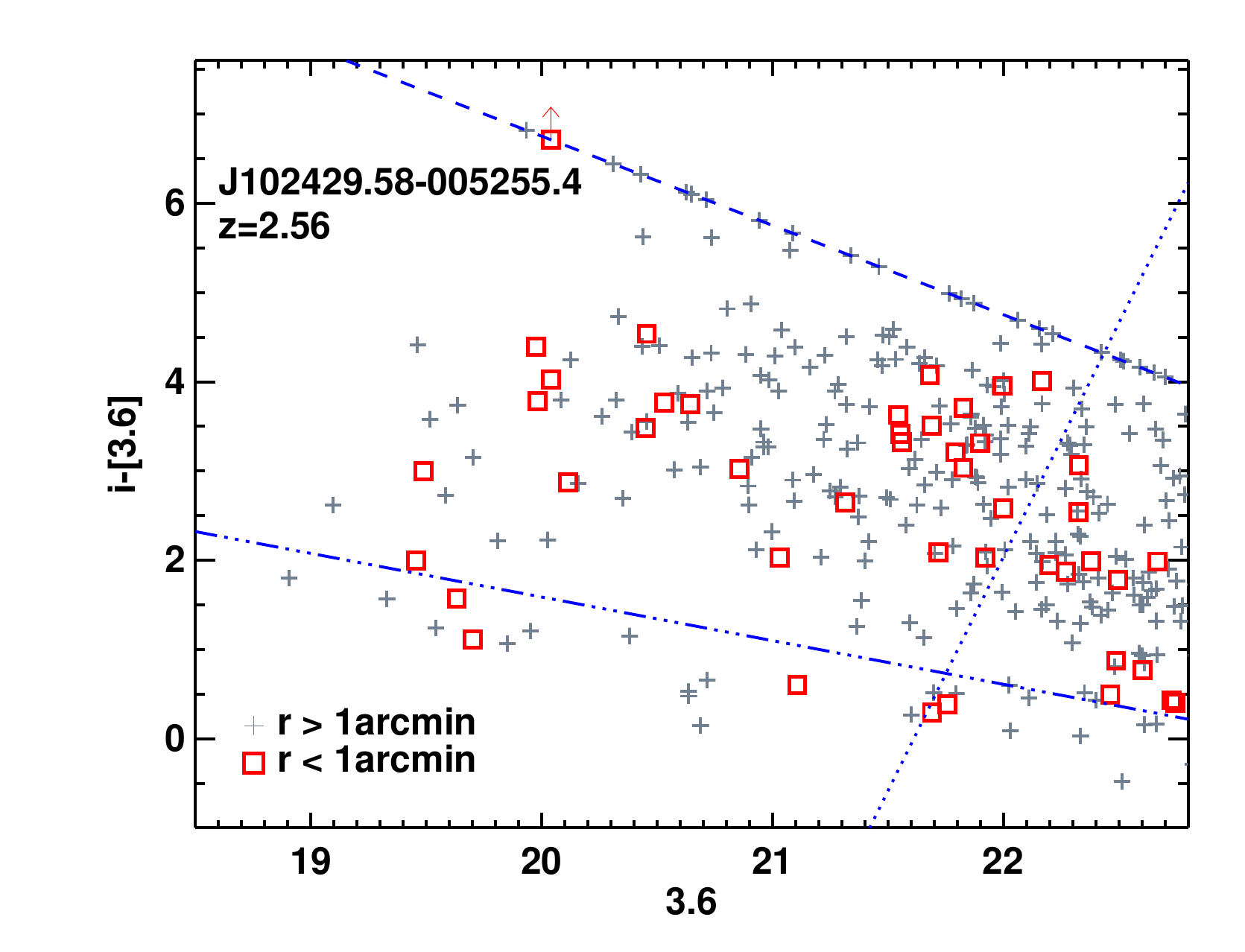}
\includegraphics[scale=0.3]{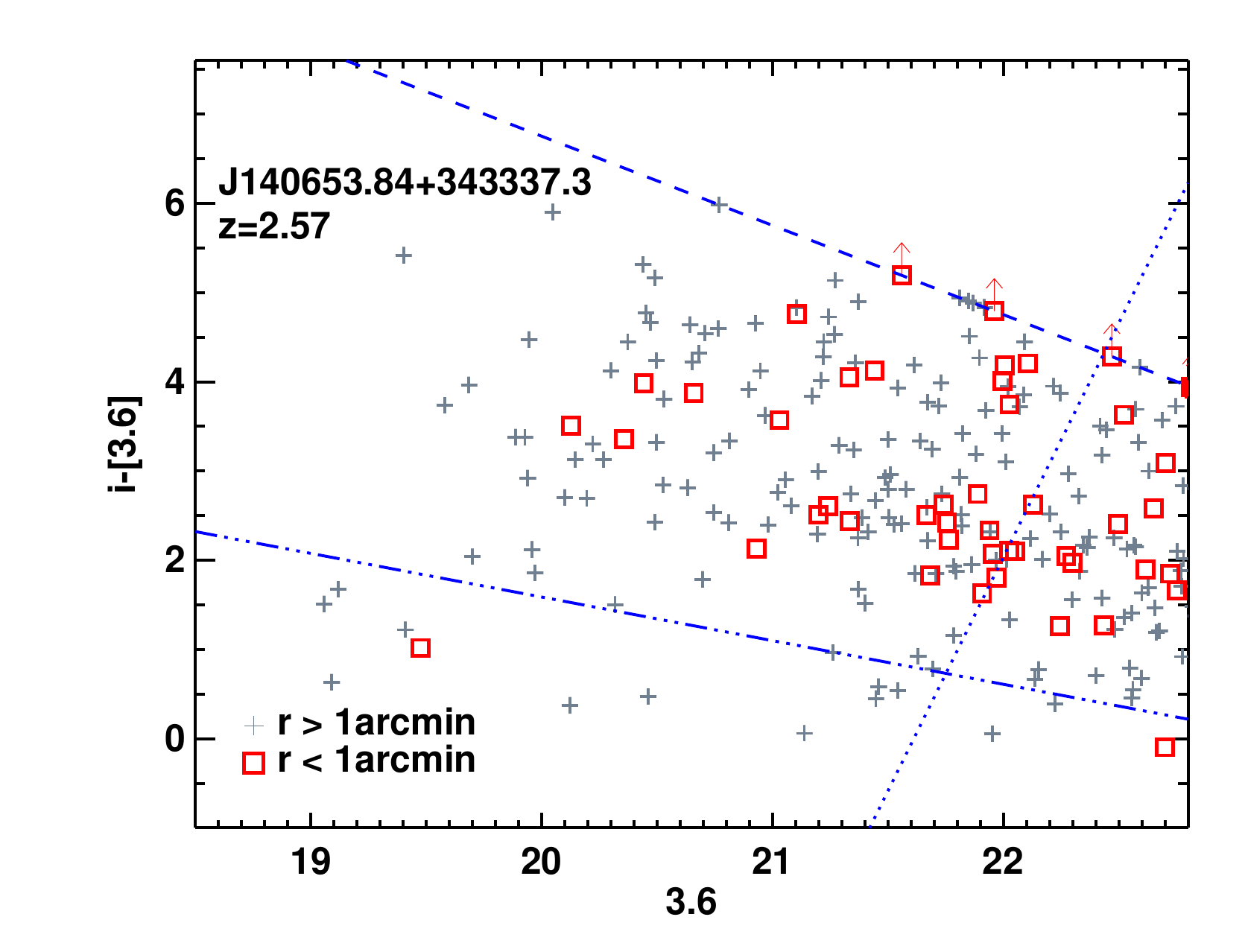}

\contcaption{}
\end{figure*}
\begin{figure*}

\includegraphics[scale=0.3]{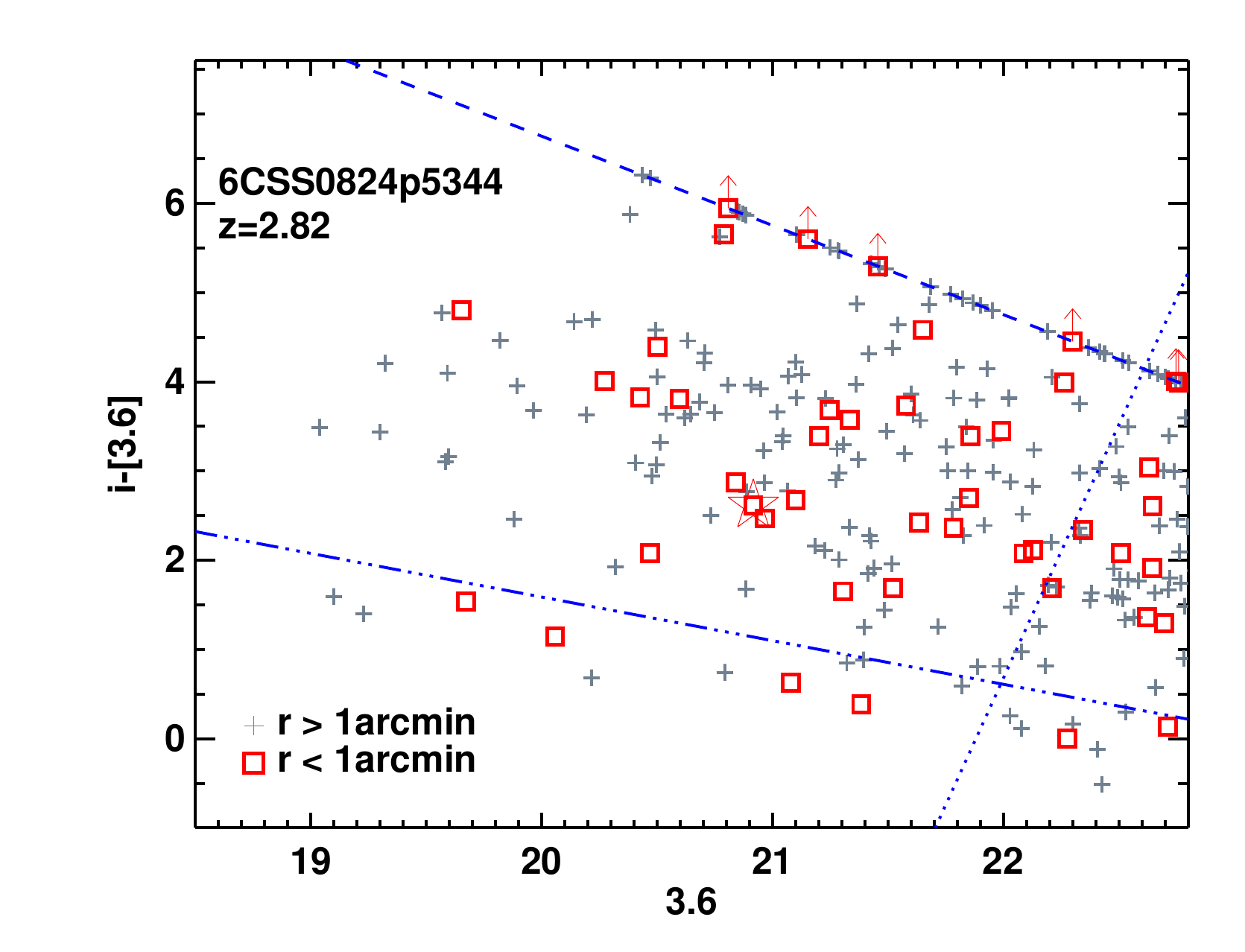} 
\includegraphics[scale=0.3]{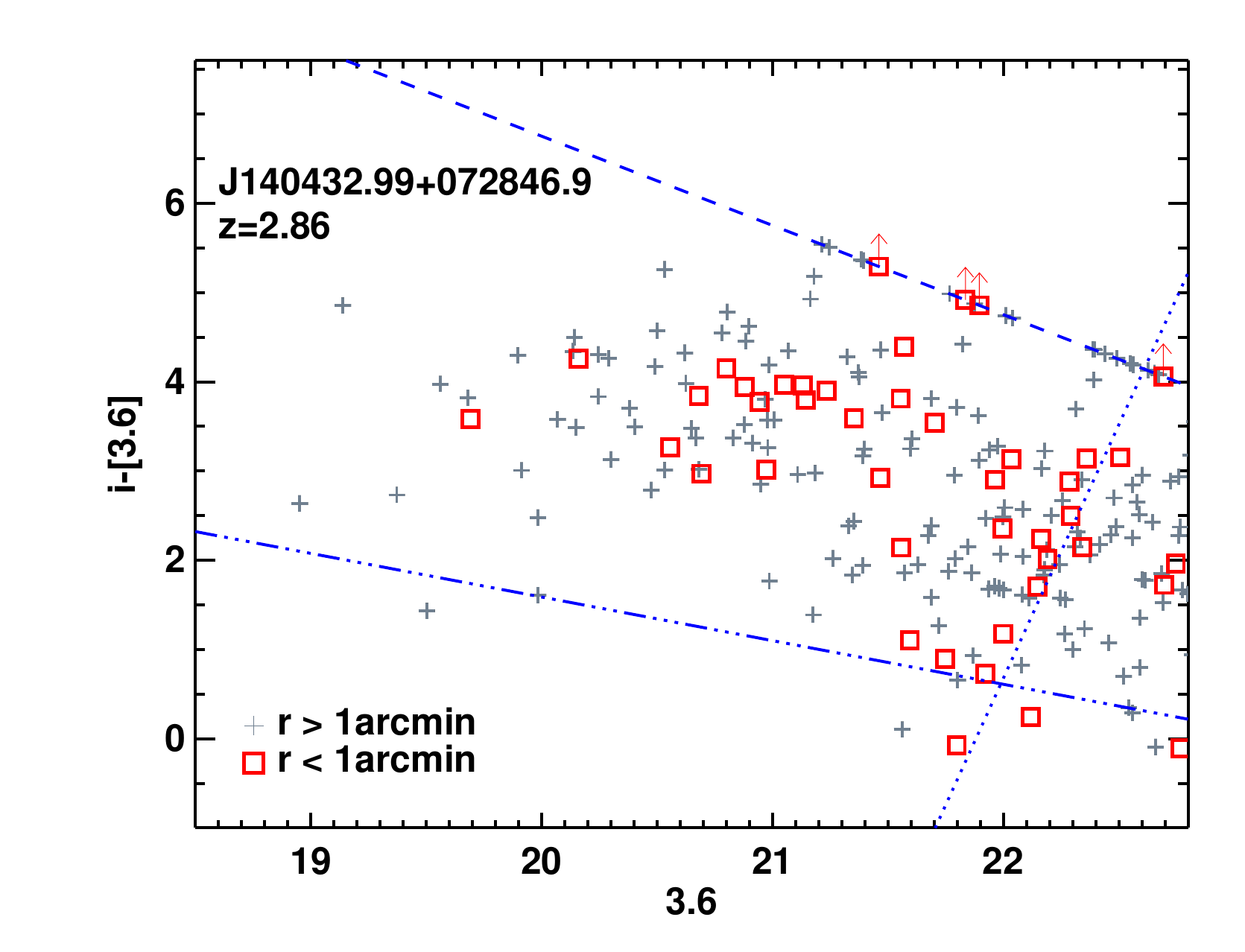} 
\includegraphics[scale=0.3]{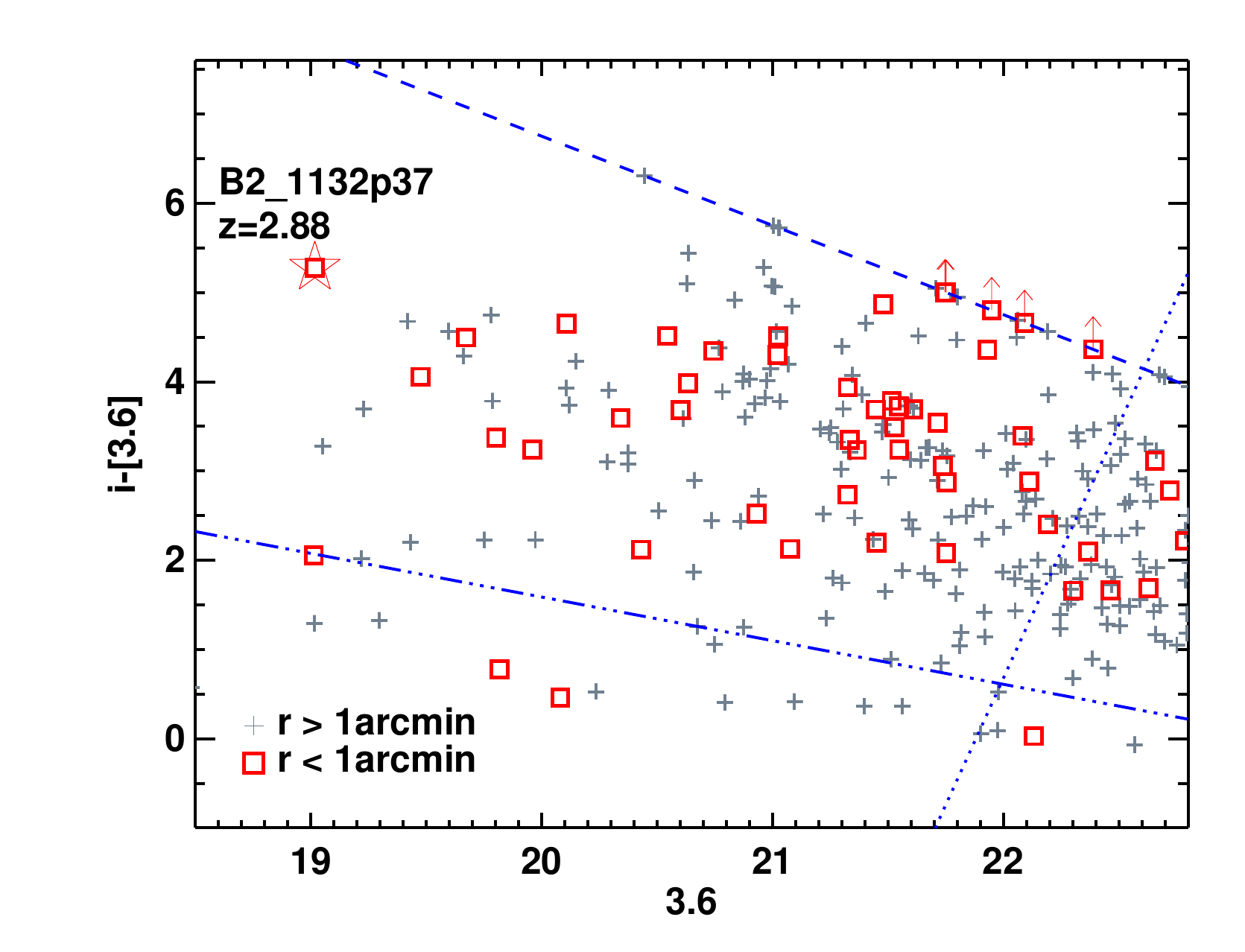}
\includegraphics[scale=0.3]{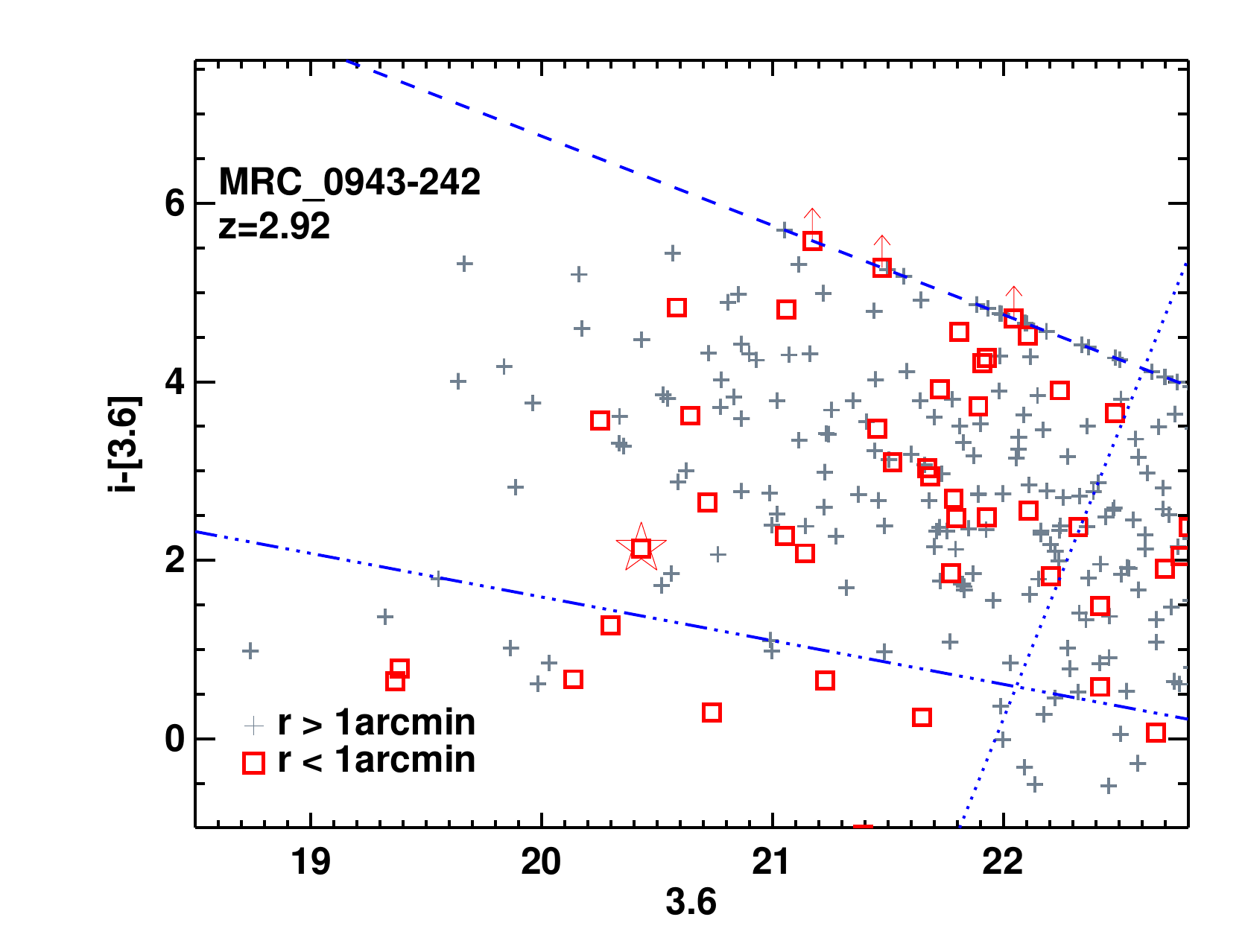}
\includegraphics[scale=0.3]{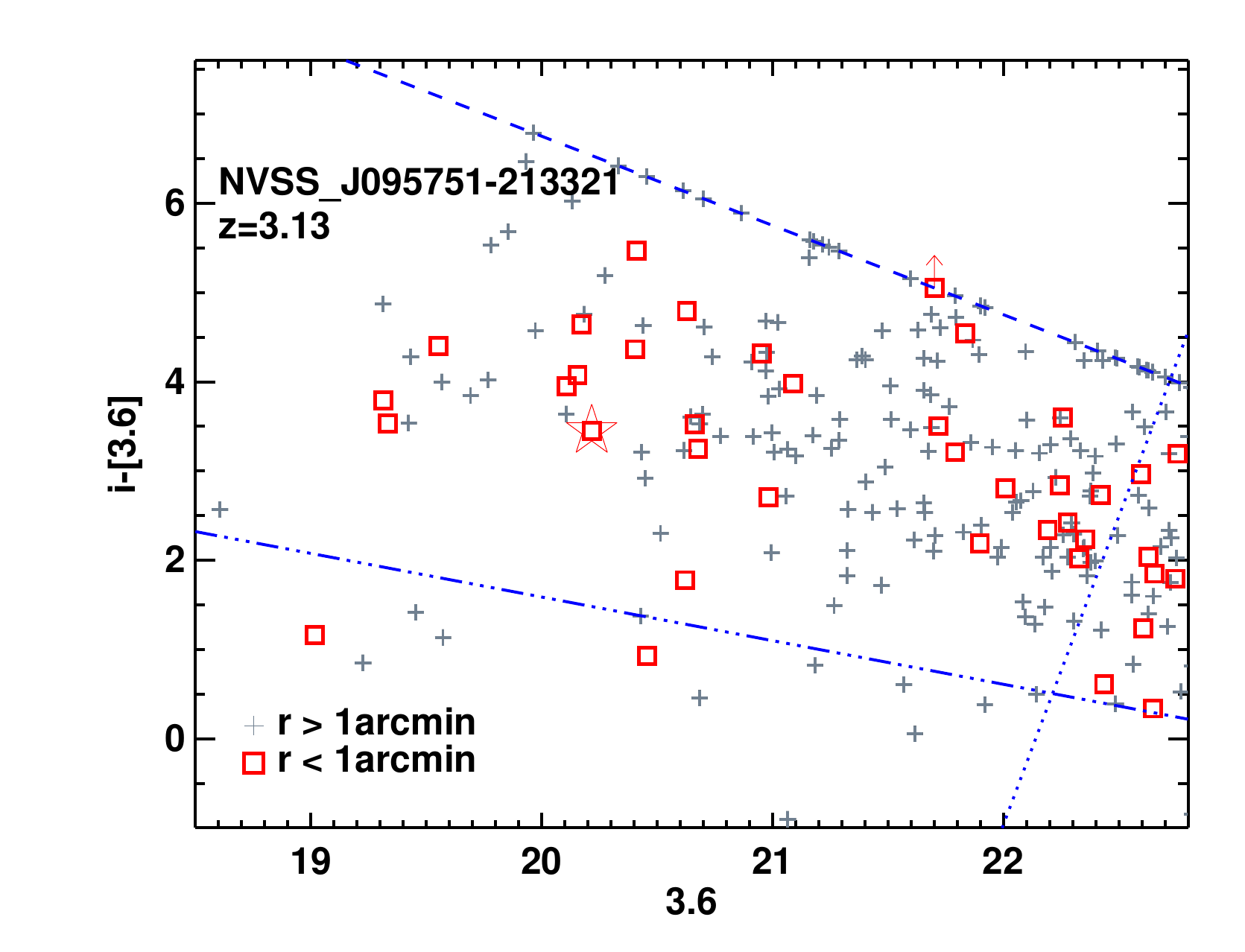}

\contcaption{}
\end{figure*}

\section{Maraston models} \label{Append:Maraston}
At high redshift ($1.4< z <2.7$), the treatment of the asymptotic giant branch (AGB) phase of stellar evolution becomes important in the \emph{Spitzer} wavebands. At these redshifts, the AGB effect is expected to be at a maximum. 
\citet{Maraston2006} showed that the AGB phase of stellar evolution can affect the measured age and mass of high redshift galaxies and produce systematically younger ages than \citet{BC03} models. This effect is unlikely to be significant, as \citet{Kriek2010} showed that \citet{BC03} provide better fits to post-starburst galaxy spectral energy distributions than \citet{Maraston2005} models which take into account the effects of AGB stars.

\begin{figure}
\centering
\includegraphics[scale=0.5]{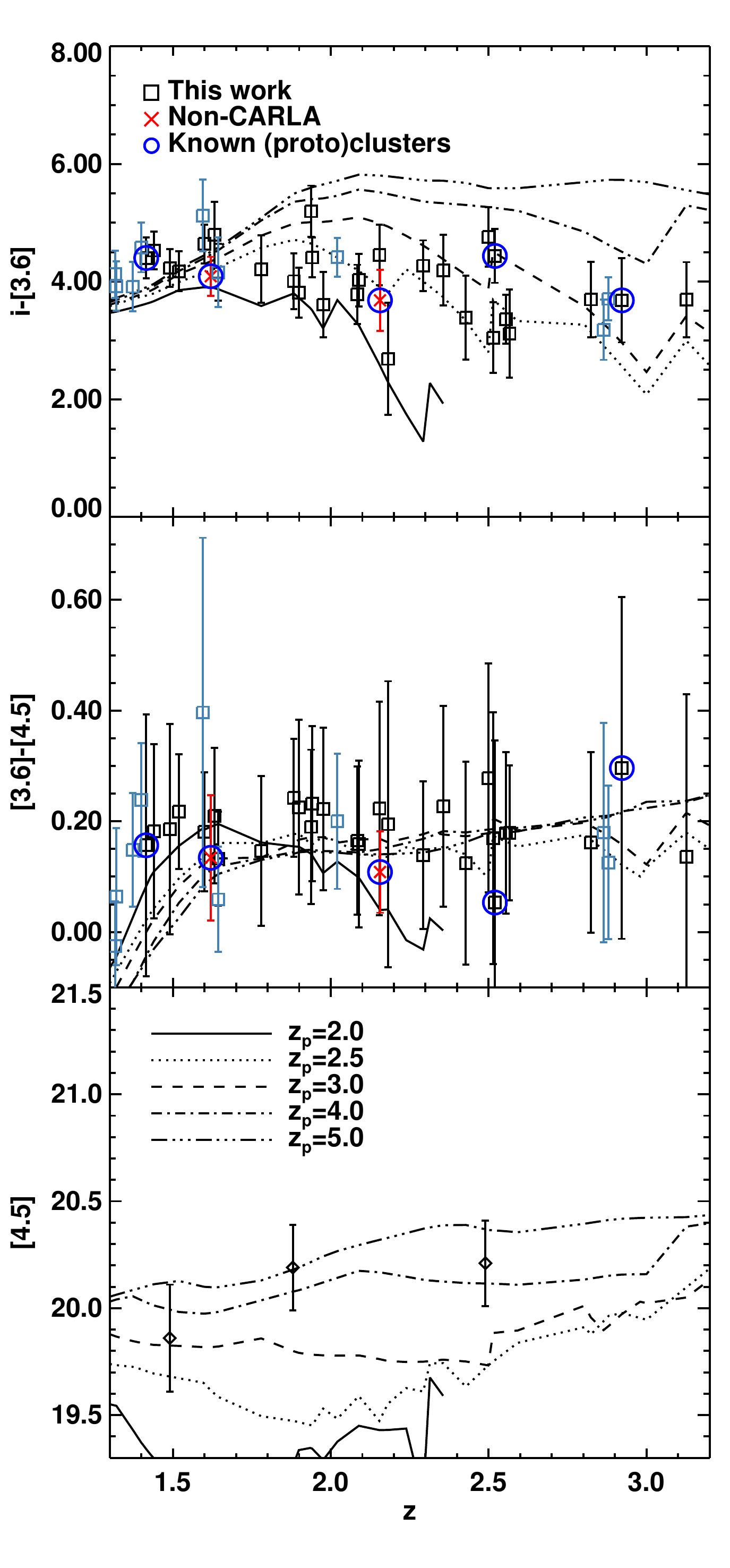}
\caption{The same as the right hand column of Figure \ref{fig:3x3}, with \citet{Maraston2005} mSSP models used instead of \citet{BC03} models.
}
\label{fig:mSSPMaraston}
\end{figure}

We reproduce the mSSP models (Section \ref{sec:mSSP}) using \citet{Maraston2005} models (Figure \ref{fig:mSSPMaraston}). Models with a \citet{Chabrier2003} IMF were not available for the \citet{Maraston2005} models so we use a \citet{Kroupa2001} IMF, which produces similar results.  Qualitatively the \citet{Maraston2005} models show the same trends as the \citet{BC03} models for the $i'-[3.6]$ colours. The CARLA IRAC colours are better fit by \citet{Maraston2005} models, however the \citet{BC03} models are also consistent within scatter in the colours and flux errors. We use \citet{BC03} models in our analysis as the models of the $[4.5]$ magnitudes and $i'-[3.6]$ colours give a consistent estimate of $z_{peak}$ for the CARLA cluster data, whereas the \citet{Maraston2005} models for the $[4.5]$ magnitudes suggest a much higher $z_{peak}$ than the $i'-[3.6]$ colours.

\label{lastpage}

\end{document}